\documentclass[twocolumn]{aastex61}
\usepackage{CJK} 

\usepackage{amsmath}
\usepackage{amssymb}
\usepackage{epsfig}
\usepackage{apjfonts}
\usepackage{natbib}
\usepackage{hyperref}
\usepackage{epstopdf}
\usepackage{verbatim}
\usepackage{enumitem}
\usepackage{color}
\setlist[enumerate]{itemsep=-1mm}
\usepackage{soul, xcolor}
\setstcolor{blue}

\usepackage{mathtools}

\usepackage{bigints}

\newcommand\ddfrac[2]{\frac{\displaystyle #1}{\displaystyle #2}}

\usepackage{tikz}

\makeatletter
\global\let\tikz@ensure@dollar@catcode=\relax
\makeatother

\received{August 8, 2020; Revised April 29, 2021; Accepted May 5, 2021}
\submitjournal{ApJ}

\begin{document}


\begin{CJK*}{UTF8}{gbsn}

\title{Uniform Forward-Modeling Analysis of Ultracool Dwarfs. \\II. Atmospheric Properties of 55 Late-T Dwarfs}

\author[0000-0002-3726-4881]{Zhoujian Zhang (张周健)}
\affiliation{Institute for Astronomy, University of Hawaii at Manoa, Honolulu, HI 96822, USA}
\affiliation{Visiting Astronomer at the Infrared Telescope Facility, which is operated by the University of Hawaii under contract 80HQTR19D0030 with the National Aeronautics and Space Administration.}

\author[0000-0003-2232-7664]{Michael C.\ Liu}
\affiliation{Institute for Astronomy, University of Hawaii at Manoa, Honolulu, HI 96822, USA}

\author[0000-0002-5251-2943]{Mark S.\ Marley}
\affiliation{NASA Ames Research Center, Mail Stop 245-3, Moffett Field, CA 94035, USA}
\affiliation{The University of Arizona, Tuscon AZ 85721, USA}

\author[0000-0002-2338-476X]{Michael R.\ Line}
\affiliation{School of Earth \& Space Exploration, Arizona State University, Tempe AZ 85287, USA}

\author[0000-0003-0562-1511]{William M. J.\ Best}
\affiliation{Department of Astronomy, University of Texas at Austin, Austin, Texas 78712, USA}

\begin{abstract}
We present a large uniform forward-modeling analysis for 55 late-T (T7$-$T9) dwarfs, using low-resolution ($R\approx50-250$) near-infrared ($1.0-2.5$~$\mu$m) spectra and cloudless Sonora-Bobcat model atmospheres. We derive the objects' effective temperatures, surface gravities, metallicities, radii, masses, and bolometric luminosities using our newly developed Bayesian framework, and use the resulting population properties to test the model atmospheres. We find (1) our objects' fitted metallicities are $0.3-0.4$~dex lower than those of nearby stars; (2) their ages derived from spectroscopic parameters are implausibly young ($10$~Myr$-0.4$~Gyr); (3) their fitted effective temperatures show a similar spread as empirical temperature scales at a given spectral type but are $\sim50-200$~K hotter for $\geqslant$T8 dwarfs; and (4) their spectroscopically inferred masses are unphysically small (mostly $1-8$~M$_{\rm Jup}$). These suggest the Sonora-Bobcat assumptions of cloudless and chemical-equilibrium atmospheres do not adequately reproduce late-T dwarf spectra. We also find a gravity- and a metallicity-dependence of effective temperatures as a function of spectral type. Combining the resulting parameter posteriors of our sample, we quantify the degeneracy between fitted surface gravity and metallicity such that an increase in $Z$ combined with a $3.4\times$ increase in $\log{g}$ results in a spectrum that has similar fitted parameters. We note the systematic difference between late-T dwarf spectra and Sonora-Bobcat models is on average $\approx 2\%-4\%$ of the objects' peak $J$-band fluxes over the $1.0-2.5$~$\mu$m range, implying modeling systematics will exceed measurement uncertainties when analyzing data with $J$-band S/N $\gtrsim50$. Using our large, high-quality sample, we examine the spectral-fitting residuals as a function of wavelength and atmospheric properties to discern how to improve the model assumptions. Our work constitutes the largest analysis of brown dwarf spectra using multi-metallicity models and the most systematic examination of ultracool model atmospheres to date.
\end{abstract}

\section{Introduction}
\label{sec:introduction}
Characterization of ultracool dwarf spectra is essential to understand the physical and chemical processes in the atmospheres of brown dwarfs and exoplanets. Such processes govern these objects' appearance and evolution, and the emergent spectra encode signatures of their formation pathways \citep[e.g.,][]{2001RvMP...73..719B, 2008ApJ...683.1104F, 2015ARA&A..53..279M}. However, spectroscopic analysis of substellar objects can be challenging given the numerous complex processes and interactions in ultracool atmospheres. 

As brown dwarfs evolve to cooler effective temperatures, substantial changes in atmospheric chemistry drive the emergence and disappearance of various atomic and molecular spectral features. The transition from M to L spectral type is marked by the increasing strengths of alkali lines (e.g., Na and K) and metal hydride bands (e.g., FeH and CrH), and the disappearance of TiO and VO due to the formation of titanium-bearing condensates \citep[e.g., CaTiO$_{3}$;][]{1999ApJ...512..843B, 2002ApJ...577..974L}. During the M/L transition, clouds of iron and silicate grains (e.g., MgSiO$_{3}$ and Mg$_{2}$SiO$_{4}$) form and cause the objects' emergent spectra to become redder toward later L subtypes \citep[][]{1999ApJ...512..843B, 2002ApJ...568..335M}. The hallmark of the transition from L to T spectral type is the emergence of CH$_{4}$ absorption, along with the strengthening of H$_{2}$O bands. The cool temperatures of the L/T transition ($T_{\rm eff} \approx 1200-1400$~K) apparently lead to cloud patchiness, condensation below the photosphere, and dissipation \citep[e.g.,][]{2008ApJ...689.1327S, 2010ApJ...723L.117M}, resulting in brighter $J$-band emission and bluer $J-K$ colors of early-T dwarfs as compared to late-L dwarfs \citep[e.g.,][]{2004AJ....127.3553K, 2006ApJ...647.1393L, 2012ApJS..201...19D}. The transition from T to Y spectral type is defined by the emergence of NH$_{3}$ absorption in the near-infrared, along with the formation of new condensates (e.g., Na$_{2}$S, KCl, and H$_{2}$O) given the cold atmosphere temperature \citep[$\lesssim 600$~K;][]{2011ApJ...743...50C, 2012ApJ...756..172M}. An alternative interpretation of the L-T-Y evolution does not rely on clouds but rather a thermo-chemical instability \citep[e.g.,][]{2015ApJ...804L..17T, 2016ApJ...817L..19T, 2017ApJ...850...46T, 2019ApJ...876..144T}. Such an instability can trigger local compositional convection, drive chemical abundances (e.g., CO/CH$_{4}$ and N$_{2}$/NH$_{3}$) out of equilibrium, and reduce the vertical temperature gradient in the atmospheres.

Efforts have been devoted to describe the above complicated processes via models of ultracool atmospheres \citep[e.g.,][]{1998A&A...337..403B, 2003A&A...402..701B, 2000ApJ...542..464C, 2002ApJ...573..394B, 2002ApJ...568..335M, 2010ApJ...723L.117M, 2017AAS...23031507M, 2006ApJ...640.1063B, 2008ApJ...689.1327S, 2012RSPTA.370.2765A, 2012ApJ...756..172M, 2015ApJ...804L..17T, 2016ApJ...817L..19T, 2019A&A...627A..67M, 2020A&A...637A..38P}. Characterization of brown dwarfs is commonly conducted by comparing grids of such pre-computed, forward models to observations. Synthetic model spectra have been quite successful in predicting the spectral morphology of brown dwarfs and giant planets, but inconsistencies between data and grid models have been long noticed \citep[e.g.,][]{2008ApJ...678.1372C, 2009ApJ...702..154S, 2017ApJ...842..118L, 2020ApJ...891..171Z}, indicating that the assumptions of these models (e.g., cloud properties, equilibrium chemistry, and radiative convective equilibrium) should be improved.  Recently, we have conducted a forward-modeling analysis using cloudless Sonora-Bobcat models (\citealt{2017AAS...23031507M}; Marley et al. submitted) for three benchmark late-T dwarfs, HD~3651B, GJ~570D, and Ross~458C (\citealt{2020arXiv201112294Z}; referred to as ``Paper~I'' hereinafter). Comparing our spectral-fitting results to those derived from these objects' bolometric luminosities, their primary stars' ages and metallicities, and the Sonora-Bobcat evolutionary models, we identified potential shortcomings of model predictions of these three objects. However, such analysis is hampered by the small census of known benchmarks with high-quality spectroscopy, meaning an insufficient diversity in surface gravity and metallicity at a given effective temperature (especially for $\lesssim 1000$~K).

The atmospheric retrieval technique is an alternative approach for inferring brown dwarf properties, which is not limited to the physical assumptions made by grid models and has the freedom to explore more physical (and unphysical) conditions of ultracool atmospheres by using many more free parameters \citep[e.g.,][]{2015ApJ...807..183L, 2017ApJ...848...83L, 2017MNRAS.470.1177B, 2019ApJ...877...24Z, 2020ApJ...905...46G}. Aiming to find models that almost exactly explain the observed spectra, retrieval can robustly test whether the physical assumptions made within grid models are reasonable. However, when an object has a peculiar spectrum, this data-driven method might converge to atmospheric compositions and thermal structures that are not physically self-consistent \citep[e.g.,][]{2019ApJ...877...24Z}.

Studying large samples of brown dwarfs can help surpass the aforementioned limitations of both forward modeling and retrieval methods, so that fundamental properties of the ultracool population can be robustly investigated. In addition, forward-modeling analysis of an ensemble of brown dwarfs can uncover the discrepancies between data and models, providing feedback about the strength and weakness of current modeling. Thanks to the wide coverage and depth of imaging sky surveys, including Pan-STARRS1 \citep[PS1;][]{2016arXiv161205560C}, 2MASS \citep{2006AJ....131.1163S}, the UKIRT Infrared Deep Sky Survey \citep[UKIDSS;][]{2007MNRAS.379.1599L}, and the Wide-field Infrared Survey Explorer ({\it WISE}, \citealt{2010AJ....140.1868W}; CatWISE, \citealt{2020ApJS..247...69E, 2020arXiv201213084M}), we now have a large census of brown dwarfs in the field and in young associations \citep[e.g.,][]{1999ApJ...519..802K, 2011ApJS..197...19K, 2006ApJ...645..676L, 2009ApJ...703..399L, 2012ApJ...758...31L, 2008MNRAS.391..320B, 2010MNRAS.406.1885B, 2013MNRAS.433..457B, 2015ApJ...814..118B, 2017ApJ...837...95B, 2018ApJS..234....1B, ultracoolsheet, 2007MNRAS.374..372L, 2013MNRAS.431.3222L, 2018ApJ...856...23G, 2018ApJ...858...41Z, 2020ApJ...899..123M}. Most notably, the sample of ultracool dwarfs with near-infrared spectra is abundant \citep[$\geqslant 20$~objects in each spectral subtype from M6 to T8; e.g.,][]{2014ASInC..11....7B, filippazzo_2016, 2019AJ....157..101M}, enabling ensemble analyses instead of the more common single-object studies.

In this paper, we apply our forward-modeling framework constructed and validated in Paper~I to 55 T7--T9 dwarfs ($T_{\rm eff} \approx 600-1200$~K). Late-T dwarfs likely lack optically thick clouds in their near-infrared photospheres \citep[though optically thin, sulfide clouds might exist; e.g.,][]{2012ApJ...756..172M}. Therefore, choosing late-T dwarfs helps simplify the model parameter space and can validate the cloud-free models, which serve as the starting point for the more complex cloudy models.  We compare our objects' low-resolution near-infrared spectra to cloudless Sonora-Bobcat models with both solar and non-solar metallicities. Our forward-modeling analysis uses the Bayesian framework Starfish (developed by \citealt{2015ApJ...812..128C} and updated by \citealt{2017ApJ...836..200G}; see Paper~I), which accounts for uncertainties from model interpolation, as well as correlated (data$-$model) residuals due to instrumental effects and modeling systematics, thereby providing robust physical parameters and more realistic error estimates than the traditional ($\chi^{2}$-based) methods adopted in most previous work. Our work is the largest analysis of brown dwarf spectra using multi-metallicity models and altogether the most systematic test of any set of ultracool model atmospheres to date. 

We aim to investigate the physical properties of late-T dwarfs and examine the model atmospheres through detailed comparisons between data and models. We start with a description of our sample (Section~\ref{sec:obs}) and present our methodology and results of the forward-modeling analysis (Section~\ref{sec:atm}). We identify known and candidate binaries in our sample and exclude them from our subsequent analysis (Section~\ref{sec:binaries}). We then study the inferred atmospheric properties of late-T dwarfs (Section~\ref{sec:atm_lateT}) and use them to investigate the performance of the cloudless Sonora-Bobcat models (Section~\ref{sec:comments_sonora}). Finally, we provide a summary and a brief discussion of further work (Section~\ref{sec:summary}).

\section{Observations}
\label{sec:obs}

\subsection{Sample of T7--T9 Dwarfs}
\label{subsec:sample}
We construct our sample of late-T dwarfs by using the catalog of ultracool dwarfs by \cite{2018ApJS..234....1B} and selecting all T7--T9 objects with prism-mode ($R\approx 50-250$) spectra taken from the SpeX spectrograph \citep{2003PASP..115..362R} mounted on the NASA Infrared Telescope Facility (IRTF). These low-resolution spectra are from our own observations (Section~\ref{subsec:irtf}), the SpeX Prism Library \citep[][]{2014ASInC..11....7B, 2017ASInC..14....7B}, and the literature \citep[e.g.,][]{2004AJ....127.2856B, 2006ApJ...639.1095B, 2010ApJ...710.1142B, 2015ApJ...814..118B}, leading to a total of 55 late-T dwarfs. In this sample, 54 objects have parallaxes \citep{2007AandA...474..653V, 2008ApJ...689L..53B, 2012ApJS..201...19D, 2012ApJ...748...74L, 2016AandA...595A...1G, 2016AandA...595A...2G, 2018AandA...616A...1G, 2019ApJS..240...19K, 2020AJ....159..257B} and the only remaining object, WISE~J024512.62$-$345047.8 \citep[WISE~0245$-$3450, T8;][]{2013ApJS..205....6M}, is part of our astrometric monitoring program. Four objects in our sample are co-moving companions to either stars or brown dwarfs: HD~3651B \citep[T7.5;][]{2006MNRAS.373L..31M, 2007ApJ...654..570L}, GJ~570D \citep[T7.5;][]{2000ApJ...531L..57B}, Ross~458C \citep[T8;][]{2010MNRAS.405.1140G, 2010AandA...515A..92S}, and ULAS~J141623.94+134836.3 \citep[ULAS~1416+1348; T7.5][]{2010AandA...510L...8S, 2010MNRAS.404.1952B}, for which we adopt the parallaxes of their primary hosts from {\it Gaia} DR2 \citep{2016AandA...595A...1G, 2018AandA...616A...1G}. The first three benchmark objects have been analyzed in Paper~I and here we simply keep them in our sample. Our sample contains $90\%$ of the known T7$-$T9 dwarfs that have distances $\leqslant 25$~pc, $J$-band magnitudes $\leqslant 17.5$~mag, and declinations from $-40^{\circ}$ to $+70^{\circ}$\footnote{The five late-T dwarfs with such properties not included in our sample are: Gl~229B \citep[T7;][]{1995Natur.378..463N}, WISEPA~J052844.51$-$330823.9 \citep[T7;][]{2011ApJS..197...19K}, WISE~J214706.78$-$102924.0 \citep[T7.5;][]{2013ApJS..205....6M}, WISEPA~J085716.25$+$560407.6 \citep[T8;][]{2011ApJS..197...19K}, and WISEPA~J143602.19$-$181421.8 \citep[T8;][]{2011ApJS..197...19K}.}. Information about our sample's spectroscopy, astrometry, and photometry is listed in Tables~\ref{tab:spectroscopy}, \ref{tab:astrometry}, and \ref{tab:photometry}, respectively.

\subsection{IRTF/SpeX Spectroscopy}
\label{subsec:irtf}
We obtained near-infrared spectra of 16 T7--T9 dwarfs in our sample using IRTF (Table~\ref{tab:spectroscopy}). We used SpeX in prism mode with either the $0.5''$ slit ($R \approx 80-250$) or the $0.8''$ slit ($R \approx 50-160$). For each target, we took at least six exposures in a standard ABBA pattern and contemporaneously observed a nearby A0V standard star within 0.1~airmass for telluric correction. We reduced the data using version 4.1 of the Spextool software package \citep[][]{2004PASP..116..362C}. Our resulting spectra are reported with vacuum wavelengths and have typical S/N $\approx 50$ per pixel in $J$ band.

Combining all available IRTF/SpeX spectra, we find 42 objects in our sample have spectra observed using the $0.5''$ slit and 15 objects using the $0.8''$ slit, with 2 objects having spectra observed using both slit widths. Therefore, we have in total 57 near-infrared spectra for 55 objects. We flux-calibrate all spectra using the objects' $H_{\rm MKO}$ magnitudes, which are either observed \citep[e.g.,][]{2004AJ....127.3553K, 2012yCat.2314....0L, 2019ApJS..240...19K} or synthesized from other bands by \cite{2021AJ....161...42B}. The WFCAM $H$-band filter and the corresponding zero-point flux are from \cite{2006MNRAS.367..454H} and \cite{2007MNRAS.379.1599L}, respectively.

\begin{figure*}[t]
\begin{center}
\includegraphics[height=3.5in]{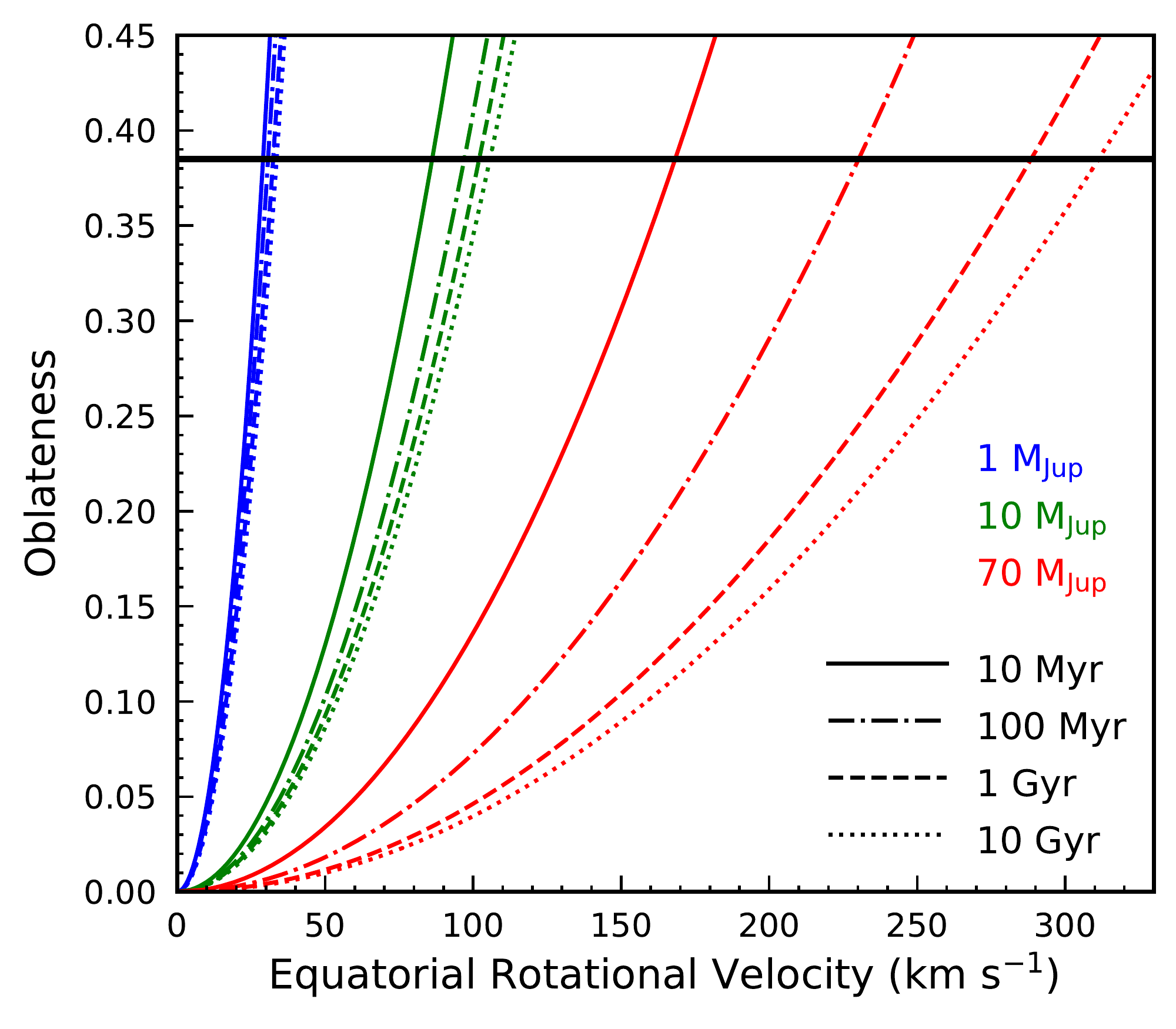}
\caption{Rotation-induced oblateness as a function of equatorial rotational velocity $v_{\rm rot}$ with different ages and masses, based on the solar-metallicity cloudless Sonora-Bobcat evolutionary models. We compute oblateness as $f = 2Cv_{\rm rot}^{2} / 3gR$ \citep[e.g.,][]{2003ApJ...588..545B}, where $\{g,\ R\}$ are derived from the evolutionary models and $C = 0.9669$ corresponds to the $n=1.5$ polytrope representative of fully convective brown dwarfs \citep[e.g.,][]{1993RvMP...65..301B}. We use the black horizontal line to mark the critical oblateness ($f_{\rm crit} = 0.385$) below which objects are rotationally stable and have $v_{\rm rot} \leqslant 300$~km~s$^{-1}$. These theoretical relations are only slightly changed using the Sonora-Bobcat models with non-solar metallicities (i.e., $Z = -0.5$ and $+0.5$). A $10$~Gyr-old $70$~M$_{\rm Jup}$ object can induce the same oblateness if it has a $0.5$~dex lower metallicity but $\approx 20$~${\rm km}\ {\rm s}^{-1}$ higher $v_{\rm rot}$. Such an effect due to non-solar metallicity diminishes with younger ages and lower masses.}
\label{fig:oblateness_vs_vrot}
\end{center}
\end{figure*}

\section{Forward-Modeling Analysis}
\label{sec:atm}

\subsection{Methodology}
\label{subsec:method}
In Paper~I, we constructed and validated a forward-modeling framework using the Bayesian inference tool Starfish \citep{2015ApJ...812..128C, 2017ApJ...836..200G} and the Sonora-Bobcat models (\citealt{2017AAS...23031507M}; Marley et al. submitted), which assume cloudless and chemical-equilibrium atmospheres. Our framework is customized for the parameter space relevant to late-T dwarfs: $[600,1200]$~K in effective temperature $T_{\rm eff}$, $[3.25,5.5]$~dex in logarithmic surface gravity $\log{g}$, and $[-0.5,+0.5]$~dex in bulk metallicity $Z$ (before removal of any species by condensation), and for near-infrared ($0.8-2.5$~$\mu$m) spectra taken using the prism mode of IRTF/SpeX. Here we summarize our methods and refer readers to Paper~I and \cite{2015ApJ...812..128C} for more details.

We start our analysis by training Starfish's spectral emulator to generate a probability distribution of model spectra for an arbitrary set of grid parameters. Starfish propagates the resulting interpolation uncertainties into the inferred posteriors and is thereby fundamentally different from linear interpolation adopted by traditional ($\chi^{2}$-based) forward-modeling analyses \citep[e.g.,][]{2010ApJS..186...63R, 2020ApJ...891..171Z}. We first trim the cloudless Sonora-Bobcat models over the aforementioned grid parameter space and wavelength range, and then downgrade their spectral resolution using two different Gaussian kernels, which correspond to the $0.5''$ slit and the $0.8''$ slit of SpeX. The convolution also accounts for the wavelength-dependent spectral resolution of the SpeX prism mode \citep[][]{2003PASP..115..362R}. We conduct principal component analysis and train Gaussian processes on these processed models, leading to spectral emulators tailored for spectra obtained with the $0.5''$ and $0.8''$ slits. 

We determine six physical parameters: effective temperature $T_{\rm eff}$, logarithmic surface gravity $\log{g}$, metallicity $Z$, radial velocity $v_{r}$, projected rotational velocity $v\sin{i}$, and logarithmic solid angle $\log{\Omega} = \log\ (R/d)^{2}$, where $i$, $R$, and $d$ is the inclination of the rotation axis, the radius, and the distance, respectively. As in Paper~I, we assume uniform priors of $[600,1200]$~K in $T_{\rm eff}$, $[3.25,5.5]$~dex in $\log{g}$, $[-0.5,+0.5]$~dex in $Z$, $(-\infty, +\infty)$ for both $v_{r}$ and $\log{\Omega}$, and $[0, v_{\rm max}]$ for $v\sin{i}$. We determine $v_{\rm max}$ using the objects' distances and fitted $\{\log{g},\ \log{\Omega}\}$ in each step of the spectral-fitting process by assuming the rotationally induced oblateness is within the stability limit. For the only object in our sample (WISE~$0245-3450$) without a parallax, we use the theoretical relation between the oblateness and the equatorial rotational velocity based on the cloudless Sonora-Bobcat models (see Figure~\ref{fig:oblateness_vs_vrot}) and adopt $v_{\rm max} = 300$~km~s$^{-1}$.

We also determine three hyper-parameters $\{a_{N},\ a_{G},\ \ell\}$ as part of our spectral-fitting process. These characterize the final covariance matrix which contains both diagonal and off-diagonal components. The latter accounts for correlated data$-$model residuals due to instrumental effects and modeling systematics, leading to more realistic error estimates than a covariance matrix with only diagonal components. We adopt the same hyper-parameter priors as in Paper~I, but we assume a uniform prior of $[425, 1115\times5]$~km~s$^{-1}$ in $\ell$ for the $0.8''$ spectra in our sample. This is slightly different from the $[820, 1840\times5]$~km~s$^{-1}$ we use for $0.5''$ spectra, given the different instrumental line spread functions for these two slit widths (see Appendix of Paper~I).

\begin{figure*}[t]
\begin{center}
\includegraphics[height=2.4in]{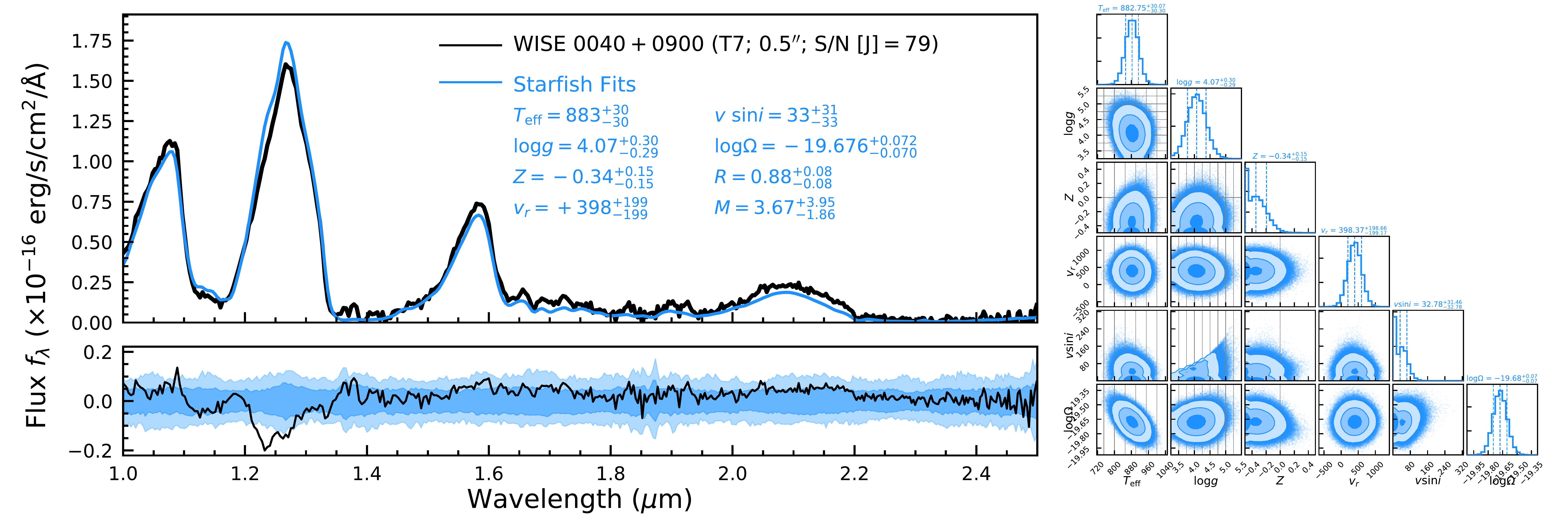}
\caption{Starfish-based forward-modeling results of our late-T dwarf sample. Left: The upper panel shows the observed spectrum (black) of each object and the median Sonora-Bobcat model spectra of those interpolated at parameters drawn from the MCMC chains (blue). The object's name, spectral type, the slit width and $J$-band S/N (median S/N in $1.2-1.3$~$\mu$m) of its spectrum, and inferred physical parameters are in the upper right corner.  The lower panel shows the residual (data$-$model; black), with the blue shadows being $1\sigma$ and $2\sigma$ dispersions of $5 \times 10^{4}$ draws from the Starfish's full covariance matrix. Right: Posteriors of the six physical parameters $\{T_{\rm eff},\ \log{g},\ Z,\ v_{r},\ v\sin{i},\ \log{\Omega}\}$ derived from the Starfish-based forward-modeling analysis. We use grey vertical and horizontal lines to mark the $\{T_{\rm eff},\ \log{g},\ Z\}$ grids points of the cloudless Sonora-Bobcat models. Figures of spectral-fitting results for our entire sample (55 late-T dwarfs with 57 spectra) are accessible online.}
\label{fig:emu_results}
\end{center}
\end{figure*}

We use {\it emcee} \citep{2013PASP..125..306F} to fit our objects' $1.0-2.5$~$\mu$m spectra with 48 walkers. We terminate the MCMC fitting process after $10^{5}$ iterations, since such number of iterations exceeds 50 times the autocorrelation length of all the fitted parameters. To complete our forward-modeling analysis, we add the following systematic uncertainties into the resulting parameter posteriors: $20$~K in $T_{\rm eff}$, $0.2$~dex in $\log{g}$, $0.12$~dex in $Z$, $180$~km~s$^{-1}$ in $v_{r}$, $40$~km~s$^{-1}$ in $v\sin{i}$, and $[0.05^{2} + (0.4\sigma_{H_{\rm MKO}})^{2}]^{1/2}$~dex in $\log{\Omega}$, where $\sigma_{H_{\rm MKO}}$ is the $H$-band magnitude uncertainty. These systematic errors were determined in Paper~I by (1) fitting the original Sonora-Bobcat model atmospheres themselves using our forward-modeling framework to quantify the uncertainties introduced by Starfish's spectral emulator, (2) incorporating the uncertainty in the wavelength calibration of the SpeX prism data into the systematics of  $v_{r}$, and (3) incorporating the uncertainty in the flux calibration of spectra (due to $H$-band magnitude errors) into the systematics of $\log{\Omega}$. We note all these added uncertainties are smaller than or comparable to the formal fitting errors of physical parameters (see Table~\ref{tab:typical_error}). 

In addition, we flux-calibrate the objects' spectra using $H$-band magnitudes in this work, and conducting this process using photometry in different bands will alter our resulting $\log{\Omega}$ posteriors \citep[e.g., Figure~8 of][]{2015ApJ...807..183L}. Each late-T dwarf in our sample has similar photometric uncertainties among $J$, $H$, and $K$ bands. Therefore, the objects' $\log{\Omega}$ posteriors, derived using spectra calibrated by $J_{\rm MKO}$ or $K_{\rm MKO}$, will have different median values but similar uncertainties compared to those inferred in this work using $H_{\rm MKO}$. As a reference, Table~\ref{tab:starfish_fitting_results} lists the shifts ($\Delta\log{\Omega}$) that our fitted $\log{\Omega}$ values would be changed by if our objects' spectra were instead flux-calibrated using $J$ or $K$ band. Our derived spectroscopic radii, masses, and bolometric luminosities (see Section~\ref{subsec:results}) would be accordingly increased by $0.5\Delta\log{\Omega}$, $\Delta\log{\Omega}$, and $\Delta\log{\Omega}$ in the logarithmic scale, respectively. We do not incorporate these small shifts (which are mostly smaller than the uncertainties from the spectral fitting) into the systematic errors, so our spectroscopically inferred properties are all tied to the objects' $H$-band photometry.

\begin{figure*}[t]
\begin{center}
\includegraphics[height=4.3in]{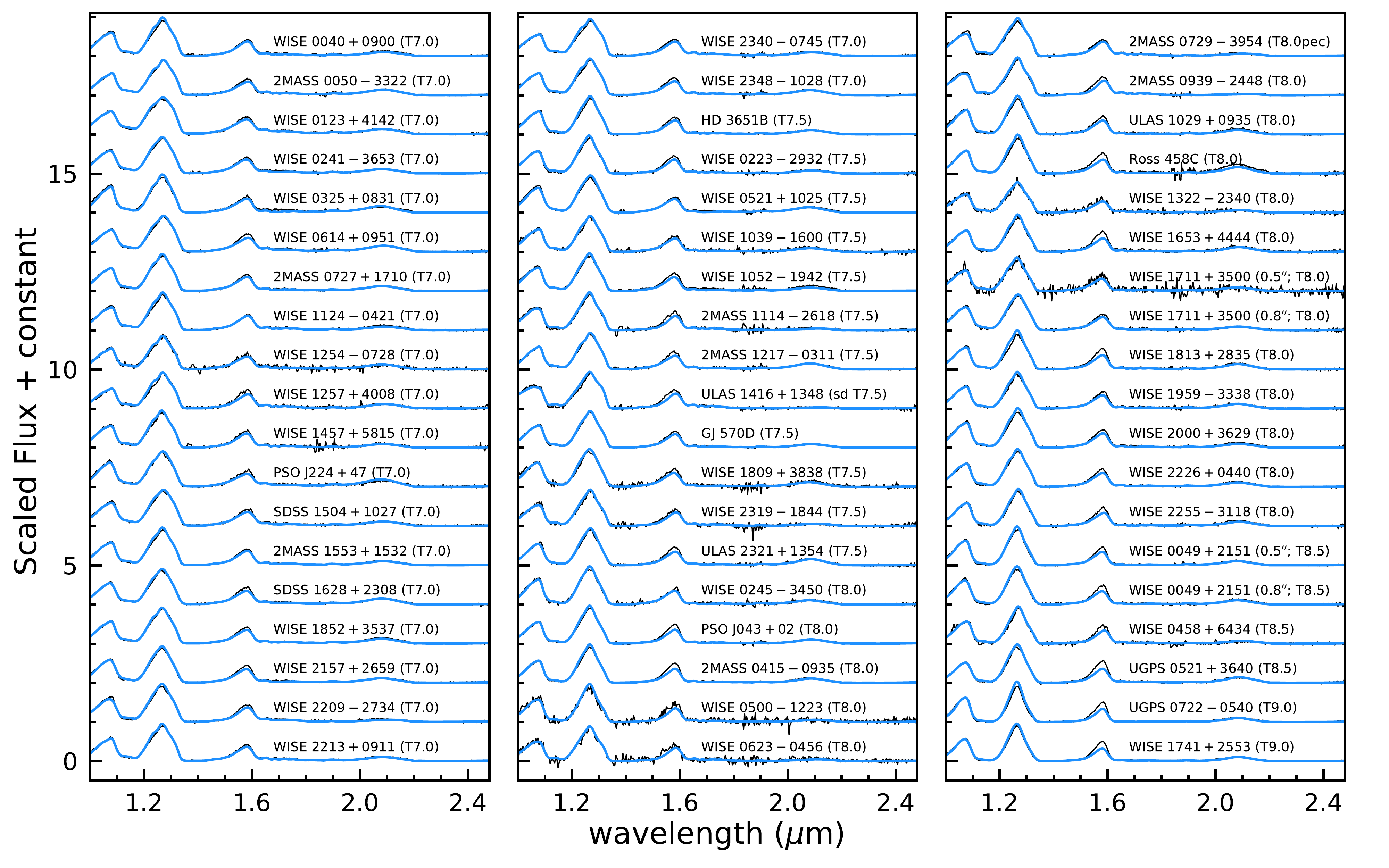}
\caption{Observed spectra (black) of our entire late-T dwarf sample and the median Sonora-Bobcat model spectra interpolated at parameters drawn from the MCMC chains (blue). }
\label{fig:spec_summary}
\end{center}
\end{figure*}

\begin{figure*}[t]
\begin{center}
\includegraphics[height=6.in]{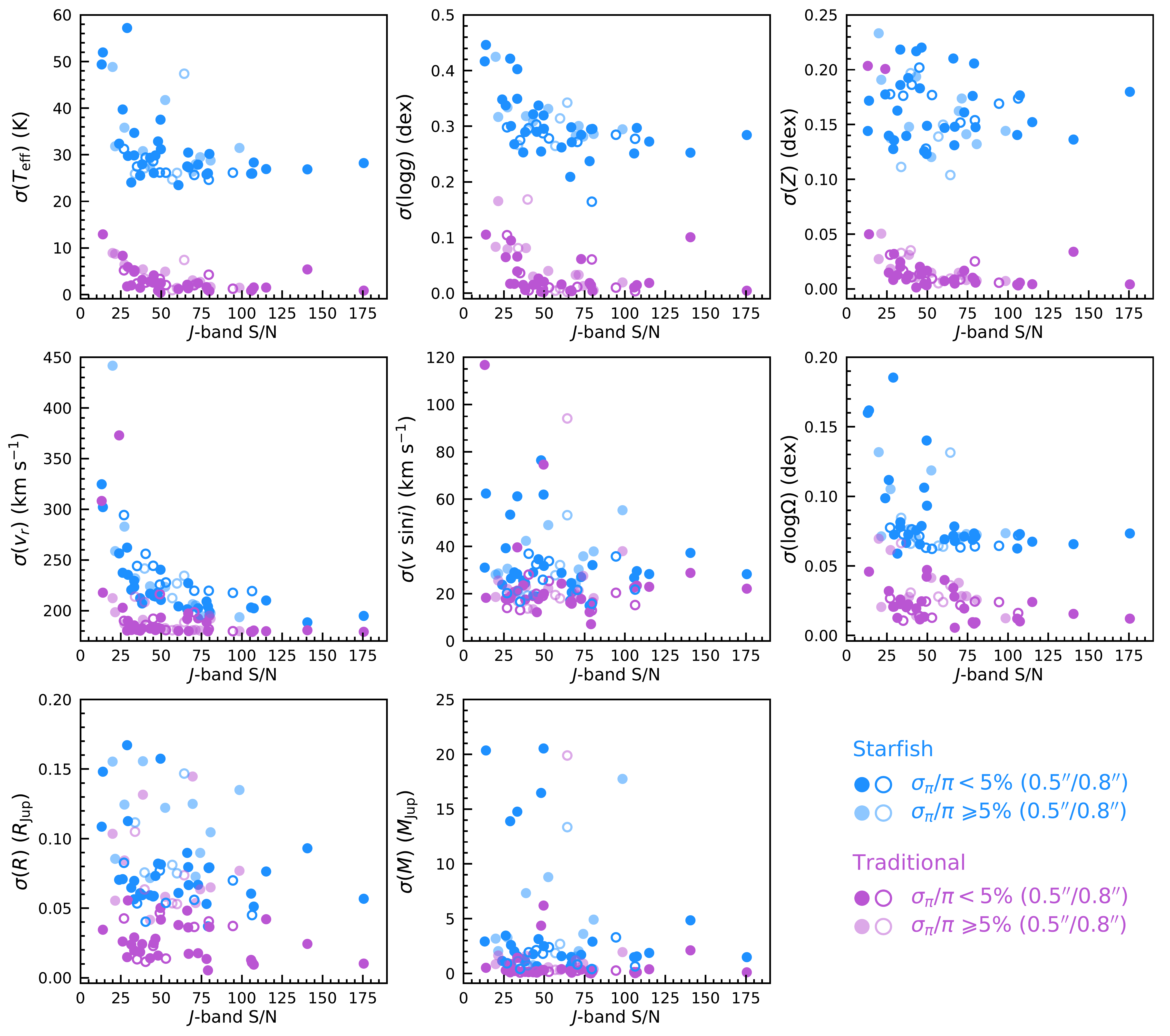}
\caption{Parameter uncertainties from Starfish (blue) and traditional (purple; Appendix~\ref{app:traditional}) forward-modeling analysis as a function of the $J$-band S/N, i.e., the median S/N in $1.2-1.3$~$\mu$m. We use solid circles for results using spectra taken with the $0.5''$ slit and open circles for the $0.8''$ slit. The uncertainties of $\{T_{\rm eff},\ \log{g},\ Z,\ v_{r},\ v\sin{i},\ \log{\Omega}\}$ have a significant spread at low S/N, but then shrink to lower values at S/N $\gtrsim 50$. Such trends also hold for $R$ and $M$ uncertainties of the objects with high-precision parallaxes (e.g., $\sigma_{\pi} / \pi <5\%$). The $v_{r}$ uncertainties derived from the lower-resolution spectra ($0.8''$ slit) are systematically larger than those from the higher-resolution ones as expected. Also, the $v_{r}$ uncertainties from both Starfish and traditional spectral-fitting methods are larger than the $180$~km~s$^{-1}$ uncertainty in the wavelength calibration of the SpeX prism data. }
\label{fig:comment_snr}
\end{center}
\end{figure*}

\subsection{Results}
\label{subsec:results}

\begin{figure*}[t]
\begin{center}
\includegraphics[height=6.3in]{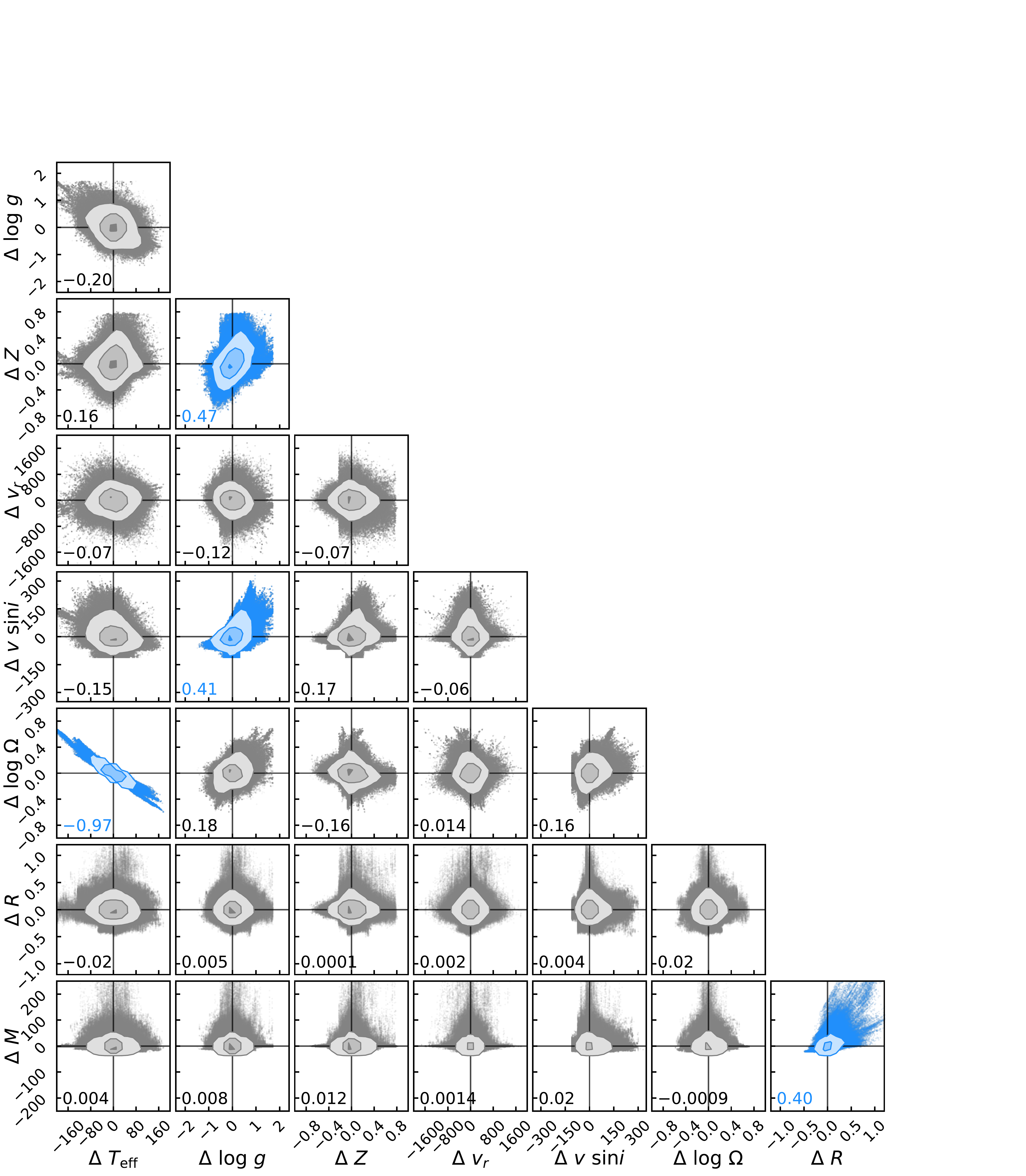}
\caption{Stacked posteriors of the eight physical parameters $\{T_{\rm eff},\ \log{g},\ Z,\ v_{r},\ v\sin{i},\ \log{\Omega},\ R,\ M\}$ from all the MCMC chains of our late-T dwarf sample. These chains have been subtracted by their medians, so they represent the deviations of parameter posteriors from median values, which we denote as $\Delta T_{\rm eff}$, $\Delta \log{g}$, $\Delta Z$, etc. We stack all the modified chains, generate stacked posteriors, and compute the Pearson correlation coefficients between each pair of parameters and give the coefficient values in the lower left corner of each panel. Most parameter pairs have very weak correlations (grey) with the absolute values of their coefficients below $0.2$. We find four parameter pairs (blue) with large correlation coefficients (absolute values above 0.4): $(\Delta T_{\rm eff},\ \Delta \log{\Omega})$, $(\Delta R,\ \Delta M)$, $(\Delta \log{g},\ \Delta v\sin{i})$, and $(\Delta \log{g},\ \Delta Z)$. The strong correlations within the first three pairs are as expected, while the fourth one suggests a $\log{g}-Z$ degeneracy (Equation~\ref{eq:corr_logg_z}) in our forward-modeling analysis. }
\label{fig:stack_posterior}
\end{center}
\end{figure*}

Figure~\ref{fig:emu_results} presents the resulting parameter posteriors of our late-T dwarfs and compares the observed data with Sonora-Bobcat model spectra interpolated at the physical parameters drawn from the MCMC samples. These observed data and fitted model spectra of our entire sample are summarized in Figure~\ref{fig:spec_summary}. We list the fitted physical parameters and uncertainties in Table~\ref{tab:starfish_fitting_results}.\footnote{The spectroscopically inferred radial velocities and projected rotational velocities in our analysis cannot be well constrained due to the low spectral resolution of the data. These physical parameters ($v_{r}$ and $v\sin{i}$) are in fact coupled with the model systematics but have no correlations with, and thereby no impact on, the other physical parameters (Figure~\ref{fig:stack_posterior}; also see Section~4.2 of Paper~I).} We further use the objects' parallaxes and their fitted $\log{g}$ and $\log{\Omega}$ to derive their radii ($R$) and masses ($M$). We also compute the bolometric luminosity ($L_{\rm bol}$) of each object by integrating its observed $1.0-2.5$~$\mu$m SpeX spectrum combined with the fitted model spectra at shorter and longer wavelengths spanning $0.4-50$~$\mu$m. We have incorporated uncertainties in observed spectral fluxes, parallaxes, and fitted physical parameters into our resulting $L_{\rm bol}$ values in a Monte Carlo fashion. Such computed $L_{\rm bol}$'s are not necessarily equal to those derived from the objects' $T_{\rm eff}$ and $R$ posteriors via the Stefan-Boltzmann law, given the mismatch between observed spectra and fitted models in $1.0-2.5$~$\mu$m wavelengths. We confirm our objects' bolometric luminosities computed from these two approaches are all consistent within uncertainties. For HD~3651B, GJ~570D, and Ross~458C, we adopt the more accurate $L_{\rm bol}$ derived in our Paper~I using their primary stars' ages and metallicities and the Sonora-Bobcat evolutionary models. We compare our computed $L_{\rm bol}$ with literature in Appendix~\ref{app:compare_lbol} and calculate bolometric corrections for T7$-$T9 dwarfs in Appendix~\ref{app:bc}. The derived physical properties ($R$, $M$, $L_{\rm bol}$) of our objects are summarized in Table~\ref{tab:starfish_derived_results}. 

We summarize the typical parameter uncertainties in Table~\ref{tab:typical_error} and find the resulting $\{T_{\rm eff},\ \log{g},\ Z\}$ uncertainties of our sample are about $1/3-1/2$ of the Sonora-Bobcat model grid spacing, the same as our finding in Paper~I for the three late-T benchmarks, HD~3651B, GJ~570D, and Ross~458C. The fitted models generally match well the observed spectra, and we discuss the data-model comparison in Section~\ref{sec:comments_sonora}. 

Some of our spectroscopically inferred physical parameters of late-T dwarfs can be under- or over-estimated relative to their true values, given that our set of models assume cloudless and chemical-equilibrium atmospheres. In Paper~I, we analyzed three late-T benchmarks and compared the fitted parameters to those derived from their bolometric luminosities, their primary stars' metallicities and ages, and the Sonora-Bobcat evolutionary models. Assuming the evolutionary-based parameters are more robust, we found the accuracy of our forward-modeling results exhibited two outcomes. For HD~3651B and GJ~570D, our spectral fits produced reliable $T_{\rm eff}$ and $R$, but underestimated $\log{g}$ and $Z$ by $\approx 1.1-1.3$~dex and $\approx 0.3-0.4$~dex, respectively. For Ross~458C, our spectral fit produced reliable $\log{g}$ and $Z$, but overestimated $T_{\rm eff}$ by $\approx 120$~K and underestimated $R$ by $\approx 0.4$~R$_{\rm Jup}$ (or a factor of $\approx 1.6$). Underestimation of the spectroscopically inferred $\log{g}$ and/or $R$ further led to unphysically small masses of these objects. The late-T dwarfs in our sample might have their fitted parameters biased by similar amount as these benchmarks, and in Sections~\ref{sec:atm_lateT} and \ref{sec:comments_sonora}, we demonstrate that the spectroscopically inferred parameters of this ensemble of late-T dwarfs are useful to assess the model atmospheres.

We also conducted a forward-modeling analysis following the traditional approach, where we use linear interpolation to generate the model spectrum in between grid points and adopt a diagonal covariance matrix (defined by observed flux uncertainties) to evaluate free parameters. We describe details in Appendix~\ref{app:traditional} and present results in Figure~\ref{fig:emulin_results} and Table~\ref{tab:traditional_results}. Our Starfish analysis produces generally consistent physical parameters but with more realistic error estimates than the traditional approach. We therefore adopt the Starfish-based properties as final results of our atmospheric model analysis and use them for subsequent discussions. 

In the rest of this section, we study the impact of the data S/N and spectral resolution on our forward-modeling analysis (Section~\ref{subsubsec:comment_snr}) and investigate the correlations among the inferred physical parameters (Section~\ref{subsubsec:param_corr}).

\subsubsection{Impact of S/N and Spectral Resolution}
\label{subsubsec:comment_snr}
Our near-infrared spectra span a wide range in $J$-band S/N\footnote{The median S/N around the $J$-band peak, i.e., $1.2-1.3$~$\mu$m, of the object's spectra.} from $15-180$ (Table~\ref{tab:spectroscopy}), and here we examine the impact of the S/N on the precision of our fitting results. Figure~\ref{fig:comment_snr} plots uncertainties of the eight physical parameters $\{T_{\rm eff},\ \log{g},\ Z,\ v_{r},\ v\ \sin{i},\ \log{\Omega},\ R,\ M\}$ as a function of $J$-band S/N. The uncertainties of the first six parameters have a significant spread at low S/N, but then shrink to small value at S/N$\gtrsim 50$. We note the Starfish-derived parameter errors at high S/N ($\gtrsim 50$) are in fact dominated by modeling systematics (instead of measurement uncertainties), which constitute $\approx 2\%-4\%$ of the peak $J$-band flux of late-T dwarf spectra (see Section~\ref{subsec:hyper_param}) and cause the parameter precision to be independent of the data S/N in this regime. The precision of $R$ and $M$ depends not only on the S/N of spectra but also on the parallax uncertainties. The $R$ and $M$ uncertainties of the objects with higher-precision parallaxes (e.g., relative uncertainties $\lesssim 5\%$) have the same dependence on S/N as the first six physical parameters.

We also investigate the impact of spectral resolution, since our sample contains spectra observed with both the $0.5''$ ($R \approx 80-250$) and $0.8''$ ($R \approx 50-160$) slits. As shown in Figure~\ref{fig:comment_snr}, the radial velocity uncertainties derived from the lower-resolution spectra are systematically larger than those from the higher-resolution ones, as expected, but such distinction between the two resolutions is not seen for the other parameters. 

Two objects in our sample, WISEPA~J$171104.60+350036.8$ \citep[WISE~$1711+3500$;][]{2011ApJS..197...19K} and WISE~J$004945.61+215120.0$ \citep[WISE~$0049+2151$;][]{2013ApJS..205....6M}, have spectra taken with both slits. Comparing results from each object with the two resolutions, we find almost all parameters are consistent within $0.6\sigma$. The only exception is the radial velocity of WISE~$1711+3500$, where the inferred values between the two slits differ by $2.1\sigma$. WISE~$1711+3500$ is a resolved $0.78''$ T8+T/Y binary (\citealt{2012ApJ...758...57L}; also see Section~\ref{sec:binaries}), so the two slits will exclude different amounts of the integrated light from the system and may cause the discrepant radial velocities. Also, the $J$-band S/N of this object's $0.5''$ spectrum is quite low ($\approx 20$) compared to the $0.8''$ spectrum (S/N $\approx 35$), so the radial velocity inferred from the $0.5''$ spectrum might be less accurate.\footnote{The $J$-band S/N of WISE~$0049+2151$ (which does not show the discrepant atmospheric parameters between two slits) is $\approx 80$ for its $0.5''$ spectrum and $\approx 35$ for its $0.8''$ spectrum.} 

\begin{figure*}[t]
\begin{center}
\includegraphics[height=5in]{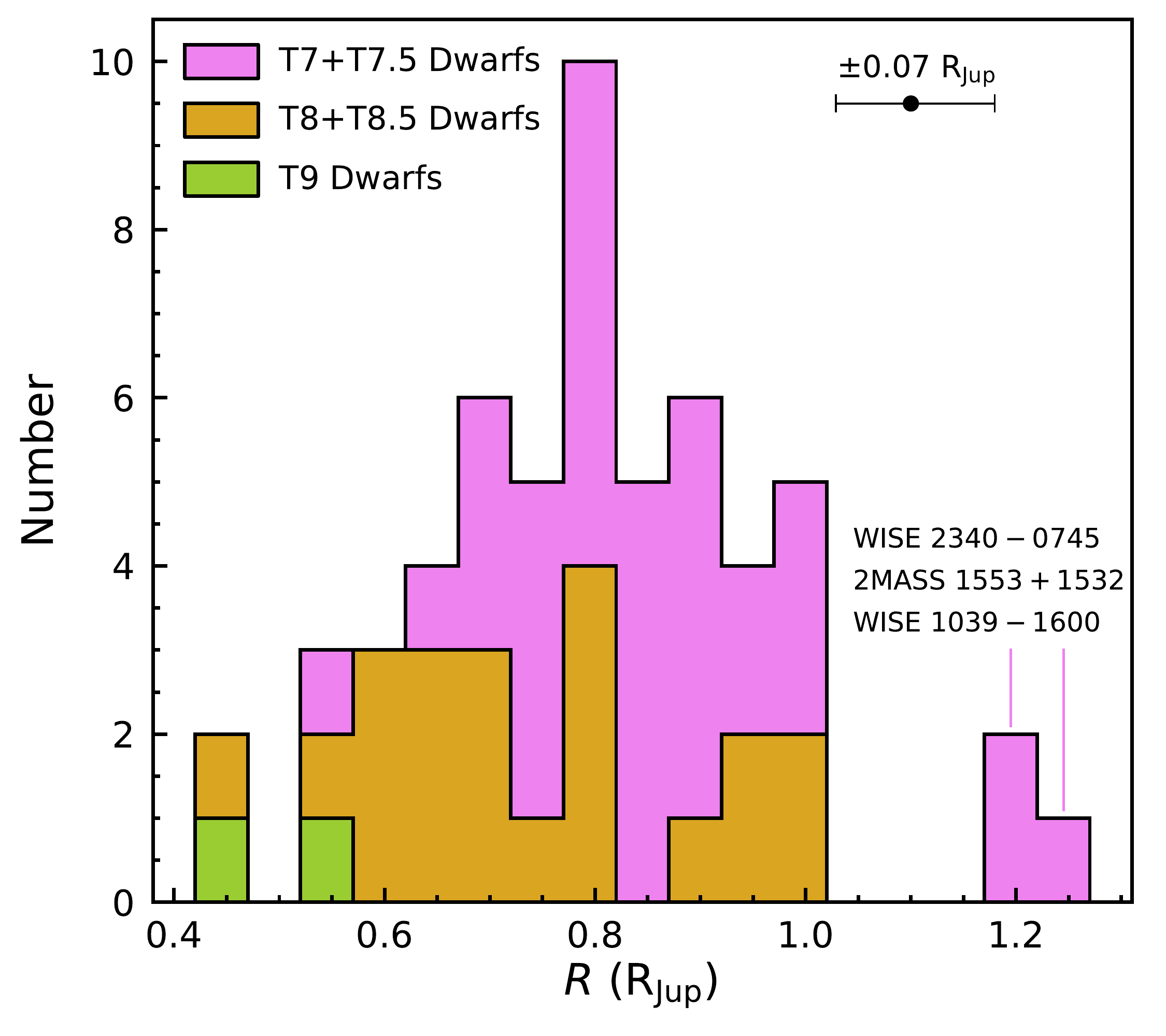}
\caption{Distribution of our derived atmospheric radii for late-T dwarfs (violet for T7 and T7.5, orange for T8 and T8.5, and green for T9). We label the three objects whose radii stand out with large values, among which 2MASS~$1553+1532$ is a resolved binary and the other two are candidate binaries (Section~\ref{sec:binaries}). }
\label{fig:radii}
\end{center}
\end{figure*}

\subsubsection{Parameter Correlations via Stacked Posteriors}
\label{subsubsec:param_corr}
We now investigate the correlations between the physical parameters $\{T_{\rm eff},\ \log{g},\ Z,\ v_{r},\ v\sin{i},\ \log{\Omega},\ R,\ M\}$. The correlations can shed light on how parameters change when one is under- or over-estimated. This information is useful when we examine the results from previous forward-modeling analyses \citep[e.g.,][]{2008ApJ...678.1372C, 2009ApJ...702..154S}, which relied on only solar-metallicity model atmospheres to analyze ultracool dwarfs which might have sub-solar or super-solar metallicities.

To investigate the parameter correlations, we stack the parameter posteriors of all our late-T dwarfs which are directly from the formal spectral fitting and have no systematic errors incorporated.\footnote{We note the parameter correlation involves the change of parameters' median values (due to the intrinsic parameter degeneracy). The systematic errors incorporated to our results only slightly inflate the parameter uncertainties (instead of their median values; Section~\ref{subsec:method}) and thus have no impact to the parameter correlations that we investigate here. } For each object's chain, we first subtract the median of each posterior so that the chain then represents the deviations of the posteriors from the median values, which we denote as $\Delta T_{\rm eff}$, $\Delta \log{g}$, $\Delta Z$, etc. Then we concatenate all the modified chains to generate the stacked parameter posteriors (Figure~\ref{fig:stack_posterior}). We use these stacked posteriors to compute the Pearson correlation coefficients between all pairs of parameters. We find that four pairs exhibit significant correlations: $(\Delta T_{\rm eff},\ \Delta \log{\Omega})$, $(\Delta R,\ \Delta M)$, $(\Delta \log{g},\ \Delta v\sin{i})$, and $(\Delta \log{g},\ \Delta Z)$. The correlations within the first three pairs are expected: (1) effective temperature and solid angle are naturally correlated due to the Stefan-Boltzmann law; (2) the mass is computed from radii and surface gravity and has a stronger dependence on radii; and (3) the inferred $v\sin{i}$ value of each object is primarily constrained by the prior, which is determined by surface gravity, solid angle, and distance (Section~\ref{subsec:method} and Equation~2 of Paper~I). 

The fourth correlation between $\log{g}$ and $Z$ is interesting. Conducting an orthogonal distance regression to the modified chains of $\Delta \log{g}$ and $\Delta Z$ using a straight line across the origin, we derive the following relation:
\begin{equation} \label{eq:corr_logg_z}
\Delta\ \log{g} = 3.42 \times \Delta Z
\end{equation}
The qualitative form of such $\log{g} - Z$ relation has been noted from previous spectroscopic analyses using other model atmospheres \citep[e.g.,][]{2006ApJ...639.1095B, 2007ApJ...667..537L, 2007ApJ...660.1507L, 2009MNRAS.395.1237B}. The cause of this degeneracy is that these two parameters have similar effect on the spectral morphology of cloudless model atmospheres, as either a high (low) $\log{g}$ or a low (high) $Z$ leads to the same suppressed (enhanced) $K$-band flux in late-T dwarf spectra. In addition, we find that $Z$ has the noticeable correlation only with the $\log{g}$ parameter in our analysis, suggesting that applying the solar-metallicity cloudless Sonora-Bobcat models to late-T dwarfs whose true metallicities are non-solar can bias $\log{g}$ but not the other parameters.

\section{Binary Systems}
\label{sec:binaries}
Now we examine binary systems in our sample, for which our derived atmospheric model properties will be not reliable. This is because some of the integrated light from these systems might be missing from our prism spectra (depending on their angular separations relative to the $0.5''$ and $0.8''$ slit widths), and our spectral-fitting process assumes all objects are single. We first discuss known resolved binaries (Section~\ref{subsec:resolved}) and then identify candidate binaries based on their spectrophotometric properties (Section~\ref{subsec:candidate}). In the end, we exclude six known or likely binaries from our subsequent discussions.

\begin{figure*}[t]
\begin{center}
\includegraphics[height=5.in]{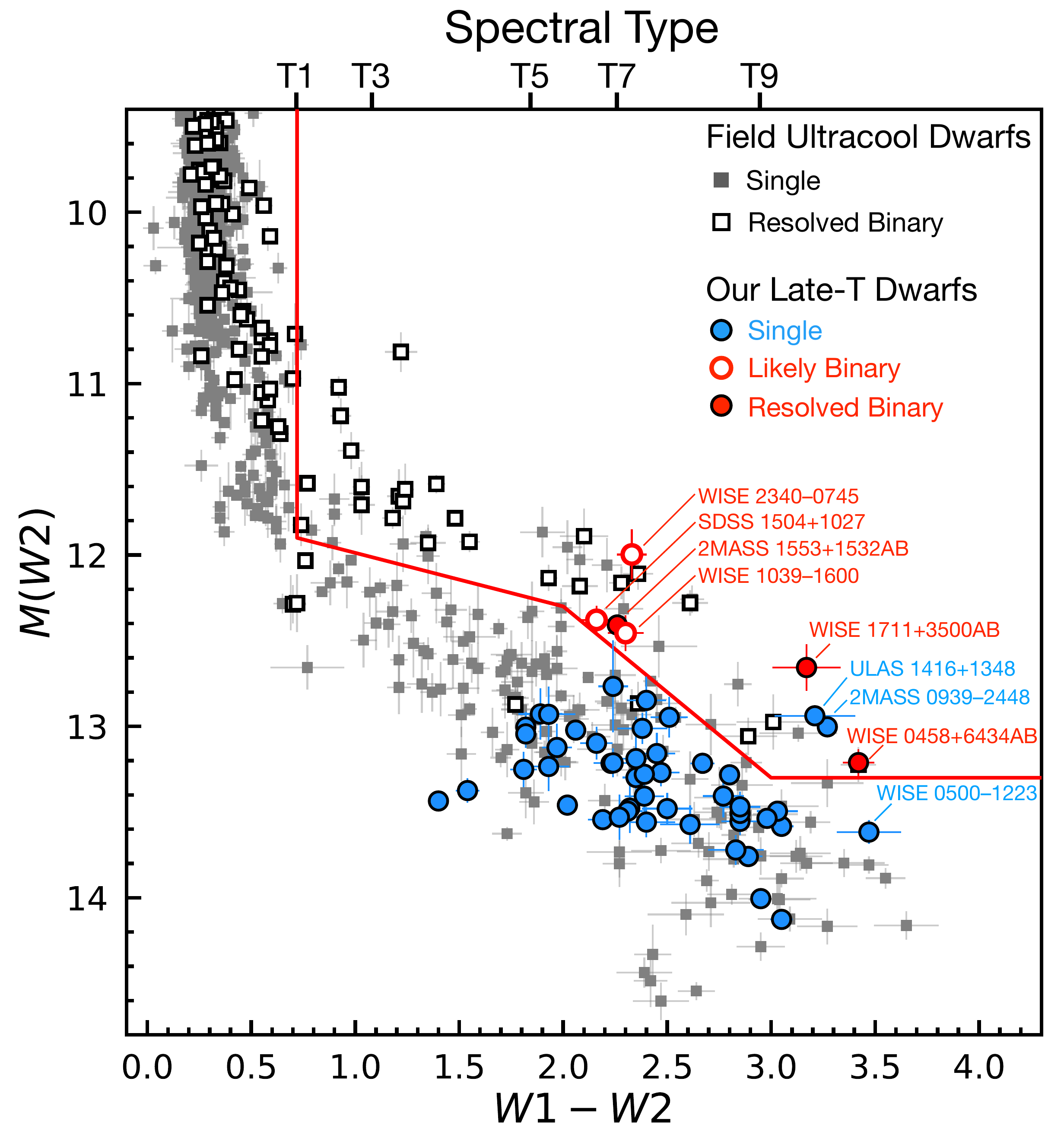}
\caption{$W2$-band absolute magnitudes as a function of $W1-W2$ colors for our late-T dwarfs and field ultracool dwarfs \citep[][]{ultracoolsheet} whose absolute magnitudes and colors both have S/N$>5$. We use red solid circles for resolved late-T binaries in our sample (Section~\ref{subsec:resolved}), red open circles for likely binaries (Section~\ref{subsec:candidate}), and blue solid circles for the remaining. For the field objects, we use white squares for resolved ultracool binaries and solid squares for those which are single or have unknown binarity. The upper x-axis shows corresponding T spectral types based on median $W1-W2$ colors of field dwarfs from \cite{2018ApJS..234....1B}. We draw a line in the diagram to help distinguish the T-type resolved binaries from the others, with the boundary defined by points of $(W1-W2,\ M(W2))\ =\ (0.72, 9.4),\ (0.72, 11.9),\ (2, 12.3),\ (3,13.3),\ (4.5,13.3)$. T-type binaries tend to have brighter magnitudes and redder colors than such boundary in this diagram, but we caution that low-metallicity, high-gravity single objects might occupy the same area as the resolved binaries (Section~\ref{subsec:candidate}). }
\label{fig:w1w2}
\end{center}
\end{figure*}

\subsection{Resolved Binaries}
\label{subsec:resolved}
Three late-T dwarfs in our sample are known to be resolved binaries: 2MASSI~J$1553022+153236$ (2MASS~$1553+1532$), WISE~$1711+3500$, and WISEPA~J$045853.89+643452.9$ (WISE~$0458+6434$). 2MASS~$1553+1532$ (T7) is a resolved $0.35''$ T6.5+T7 binary \citep{2006ApJS..166..585B}, with a bolometric luminosity ratio of $0.31 \pm 0.12$~mag and a mass ratio of $0.90 \pm 0.02$\footnote{\cite{2006ApJS..166..585B} estimated the bolometric luminosity of binary components using their HST/NICMOS resolved photometry and a bolometric correction, which they derived as a function of near-infrared colors using the photometry of unresolved T dwarfs in their sample and the $K$-band bolometric correction from \cite{2004AJ....127.3516G}. \citeauthor{2006ApJS..166..585B} estimated the mass of the binary components using the objects' effective temperatures based on the \cite{2004AJ....127.3516G} $T_{\rm eff}$-SpT relation, an assumed age of $0.5 - 5.0$~Gyr, and the \cite{1997ApJ...491..856B} evolutionary models.}. Our fitted model spectra of this object match the observed spectrum, but the fitted $R = 1.21^{+0.10}_{-0.09}$~R$_{\rm Jup}$ stands out from our sample (Figure~\ref{fig:radii}) and is larger than the evolutionary model predictions ($\approx 0.75-1.20$~R$_{\rm Jup}$; e.g., \citealt{1997ApJ...491..856B, 2008ApJ...689.1327S}; Marley et al. submitted), as well as the directly measured radii of transiting brown dwarfs, e.g., KELT-1b \citep[$1.116^{+0.038}_{-0.029}$~R$_{\rm Jup}$;][]{2012ApJ...761..123S}, KOI-$205$b \citep[$0.81\pm 0.02$~R$_{\rm Jup}$;][]{2013A&A...551L...9D}, and LHS~$6343$~C \citep[$0.783 \pm 0.011$~R$_{\rm Jup}$;][]{2015ApJ...800..134M}.  The large radii of 2MASS~$1553+1532$ is in accord with its nearly equal-brightness binarity. A similar large spectroscopic radii ($R = 1.59^{+0.14}_{-0.09}$~R$_{\rm Jup}$) was inferred by \cite{2017ApJ...848...83L} based on their retrieval analysis which also did not account for binarity.

WISE~$1711+3500$ (T8) is a resolved $0.78''$ T8+T/Y binary (\citealt{2012ApJ...758...57L}). The binary has near-infrared ($YJHK$ bands) flux ratios of $2.6-3.1$~mag, indicating the integrated-light spectra of this system is dominated by the T8 primary. This might explain why its fitted radii $R = 0.93^{+0.16}_{-0.12}$~R$_{\rm Jup}$ is not as large as the nearly equal-brightness binary 2MASS~$1553+1532$. 

WISE~$0458+6434$ (T8.5) is a resolved $0.51''$ T8.5+T9 binary \citep[][]{2011AJ....142...57G, 2012ApJ...745...26B, 2019ApJ...882..117L} with near-infrared ($JHK$ bands) flux ratios of $1.0-1.1$~mag. The fitted model spectra from our analysis do not match the observed spectrum and predict too faint $H$-band fluxes. Also, this object's parameter posteriors are bimodal based on the traditional method (Appendix~\ref{app:traditional}), with one peak consistent with the Starfish results and the other peak being $160$~K cooler in $T_{\rm eff}$, $1.5$~dex higher in $\log{g}$, and $0.2$~dex higher in $Z$. While the bimodal posteriors might be due to the object's binarity, they are likely caused by the low $J$-band S/N$=24$ of the data, which prevents the spectral-fitting process from converging on a unique set of solutions.

We exclude all these three known binaries from our subsequent analysis (Sections~\ref{sec:atm_lateT} and \ref{sec:comments_sonora}). 

\subsection{Candidate Binaries}
\label{subsec:candidate}
Here we choose three criteria to identify candidate binaries in our sample based on their fitted model parameters and observed photometry: 
\begin{enumerate}
\item[(1)] The fitted $R$ are outliers in the derived radii distribution of our sample (Figure~\ref{fig:radii}) with very large values (e.g., $\gtrsim 1.2$~R$_{\rm Jup}$),
\item[(2)] The parameter posteriors derived from either Starfish or traditional forward-modeling analysis are bimodal, or
\item[(3)] The AllWISE photometry is brighter and redder than most other objects in our sample, with critical values shown in Figure~\ref{fig:w1w2}.
\end{enumerate}

Figure~\ref{fig:w1w2} compares the integrated-light photometry of our late-T dwarfs to known field dwarfs from \cite{2020AJ....159..257B}, which shows that T-type resolved binaries have distinctly brighter $W2$-band absolute magnitudes and redder $W1-W2$ colors than the majority of field dwarfs which are single (or not known to be binaries). We draw a line in Figure~\ref{fig:w1w2} to help distinguish the resolved binaries from the others. We caution that this line can also select single objects with high-gravity, metal-poor atmospheres, as demonstrated by 2MASS~$0939-2448$ and ULAS~$1416+1348$ (discussed further below). Either high gravity or low metallicity can cause the objects' photospheres to reside at higher pressures, favoring the formation of CH$_{4}$ (with strong absorption in $W1$) over CO \citep[with strong absorption in $W2$; e.g.,][]{2007ApJ...655..522L, 2014ApJ...797...41Z, 2017ApJ...842..118L}. This leads to redder $W1-W2$ colors and might make such objects have similar AllWISE photometry as binaries.

We select candidate binaries whose properties satisfy at least one of our criteria and end up with six objects: WISE~J$103907.73-160002.9$ (WISE~$1039-1600$), WISEPC~J$234026.62-074507.2$ (WISE~$2340-0745$), WISEPA~J$050003.05-122343.2$ (WISE~$0500-1223$), 2MASS~J$09393548-2448279$ (2MASS~$0939-2448$), ULAS~$1416+1348$, and SDSS~J$150411.63+102718.4$ (SDSS~$1504+1027$). Based on the recent near-infrared adaptive optics imaging (M. Liu, private communication), all these candidates are unresolved down to $0.1''$, making their potential binarity intriguing.

WISE~$1039-1600$ (T7.5) and WISE~$2340-0745$ (T7) are selected by both the first and third criteria, with their spectroscopically inferred radii among the largest in our sample (Figure~\ref{fig:radii}) and their AllWISE photometry being anomalous (Figure~\ref{fig:w1w2}). The fitted model spectra of these two objects match the observed spectra, so their large radii are unlikely caused by poorly fitted models. Also, their kinematics do not support membership in any young associations (based on BANYAN~$\Sigma$ [\citealt{2018ApJ...856...23G}] or LACEwING [\citealt{2017AJ....153...95R}]; see \citealt{2021ApJ...911....7Z}) and their fitted $\log{g}$ are consistent with the other late-T dwarfs in our sample. Therefore, we find no evidence of youth that would lead to their large radii \citep[e.g.,][]{2001RvMP...73..719B, 2008ApJ...689.1295K, 2013ApJ...772...79A}. In addition, their near-infrared spectra do not exhibit any peculiarities that might result from high-gravity, metal-poor atmospheres. Therefore, we conclude their large atmospheric radii and atypical AllWISE photometry are likely caused by binarity.

WISE~$0500-1223$ (T7) is selected by the second criterion, i.e., bimodal posteriors from atmospheric model analysis. Similar to WISE~$0458+6434$, the parameter posteriors of this object based on the traditional forward-modeling analysis have two peaks, with one consistent with the Starfish results, and the other being 120~K cooler in $T_{\rm eff}$, $1.5$~dex higher in $\log{g}$, and $0.4$~dex higher in $Z$. Its $W2$-band absolute magnitude is fainter than our critical line in Figure~\ref{fig:w1w2}, although this line is not well-established around the $W1-W2 = 3.47 \pm 0.16$~mag of WISE~$0500-1223$, given that there are too few known binaries with such red colors. While the bimodal posteriors of this object could be caused by binarity, we note the low $J$-band S/N$\approx 13$ of data might also cause this anomaly. We thus suggest a re-analysis of this system using higher-quality data.

\begin{figure*}[t]
\begin{center}
\includegraphics[height=6.in]{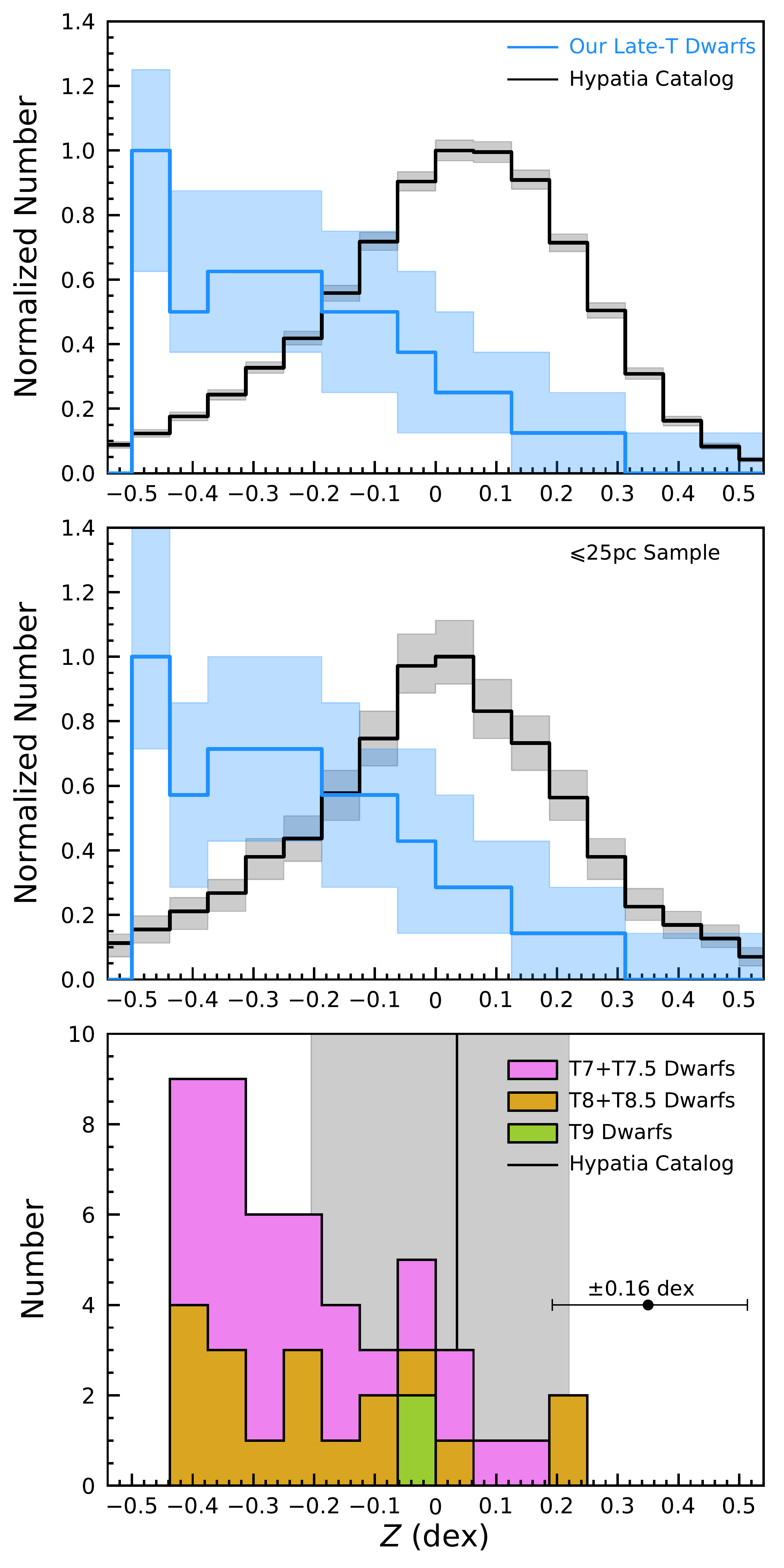}
\caption{Top: Metallicity distributions for late-T dwarfs (binaries removed) from our atmospheric model analysis and for thin-disk stars from the Hypatia catalog \citep[][]{2014AJ....148...54H, 2016ApJS..226....4H, 2017ApJ...848...34H}. For our late-T dwarfs, we generate $10^{4}$ metallicity distributions with each constructed by a random draw from each object's $Z$ posterior. We then plot the median and the $1\sigma$ confidence interval of these distributions using the blue line and shadow, respectively. Distributions for Hypatia stars (black) are produced in the same approach except that we assume the metallicity of each star follows a Gaussian distribution. We normalize these two types of distributions so that their peaks correspond to one. Middle: Similar format as the top but with late-T and stellar samples constrained to $\leqslant 25$~pc. Bottom: The distribution of our late-T dwarfs, with the violet color for T7 and T7.5, orange for T8 and T8.5, and green for T9. Our typical $Z$ uncertainty is shown as the black error bar. We use the vertical black line and grey shadow to mark the median ($0.04$~dex) and the $1\sigma$ spread ($-0.21$ to $0.22$~dex) of the Hypatia stars. }
\label{fig:z}
\end{center}
\end{figure*}

The remaining three candidates are all selected by the third criterion, i.e., anomalous AllWISE photometry. 2MASS~$0939-2448$ (T8) has a metal-poor atmosphere given its broad $Y$-band and faint $K$-band spectra. We derive a low metallicity, $Z = -0.41^{+0.15}_{-0.09}$~dex, which is consistent with the past forward-modeling \citep[][]{2006ApJ...639.1095B, 2008ApJ...689L..53B, 2009ApJ...695.1517L} and retrieval analyses \citep[][]{2017ApJ...848...83L}. Combining near-infrared and mid-infrared spectra, both \cite{2008ApJ...689L..53B} and \cite{2009ApJ...695.1517L} have suggested 2MASS~$0939-2448$ is an unresolved binary, primarily due to its uncommon brightness in the mid-infrared. However, \cite{2012ApJS..201...19D} suggested that 2MASS~$0939-2448$ might be a single object with a very red $W1-W2$ color due to its low metallicity and/or high gravity, given that its near-infrared absolute magnitudes are not brighter than field dwarfs with similar spectral types. 

ULAS~$1416+1348$ (T7.5) is a $9''$ ($85$~au) companion \citep[][]{2010MNRAS.404.1952B, 2010AandA...510L...8S} to a L6 dwarf SDSS~J141624.08+134826.7 \citep[][]{2010ApJ...710...45B, 2010AJ....139.1045S}.\footnote{The primary has weak TiO and CaH absorption features in optical, an unusually blue $J-K$ color, enhanced FeH 0.99~$\mu$m absorption, deep H$_{2}$O absorption features, and thin-disk kinematics \citep[e.g.,][]{2010ApJ...710...45B, 2010AJ....139.1045S}. \cite{2010ApJ...710...45B} suggested these features are caused by the object's low metallicity and its less-opaque condensate clouds than the majority of L6 dwarfs \citep[similar to  the blue L4.5 dwarf 2MASS~J$11263991-5003550$ as discussed by][]{2008ApJ...674..451B}. The metallicity and age of the system are not well constrained as compared to the three benchmark companions (HD~3651B, GJ~570D, and Ross~458C) studied in Paper~I.} \cite{2010MNRAS.404.1952B} noted the unusually red $H-W2$ and $W1-W2$ colors of this object and connected these properties to its metal-poor atmosphere. The very blue $Y-K$ color of this companion suggests high gravity and low metallicity, in accord with our derived $\log{g} = 5.21^{+0.26}_{-0.25}$~dex and $Z = -0.39^{+0.14}_{-0.11}$~dex, which are consistent with past spectral analyses \citep[][]{2010AJ....139.2448B, 2017ApJ...848...83L, 2020ApJ...905...46G}.\footnote{The best-fit parameters derived by \cite{2010AJ....139.2448B} using the \cite{2008ApJ...689.1327S} spectral models also favors a weak-mixing atmosphere with the vertical diffusion coefficients of $K_{zz} = 10^{4}$~cm$^{2}$~s$^{-1}$.} \cite{2019ApJS..240...19K} suggested ULAS~$1416+1348$ is an unresolved binary, given that three known late-T subdwarfs (BD~$+$01$^{\circ}$~2920B [\citealt{2012MNRAS.422.1922P}], Wolf~1130C [\citealt{2013ApJ...777...36M}], and WISE~J083337.82+005214.1 [\citealt{2014MNRAS.437.1009P}]) have similar positions as field dwarfs in the $M_{H}$~vs.~$H-W2$ diagram whereas ULAS~$1416+1348$ is an apparent outlier.

SDSS~$1504+1027$ (T7) has a normal near-infrared spectrum given its spectral type, and our fitted model spectra match the data ($J$-band S/N of 50 per pixel). \cite{2014AJ....148..129A} obtained high spatial-resolution images using HST/WFC3 and did not resolve this object down to a resolution of $0.096''-0.142''$ and a magnitude contrast of $1.5-2.0$~mag.

To conclude, we flag WISE~$1039-1600$, WISE~$2340-0745$, and SDSS~$1504+1027$ as likely binaries, and exclude them from our subsequent analysis (Sections~\ref{sec:atm_lateT} and \ref{sec:comments_sonora}) along with the three known binaries discussed in Section~\ref{subsec:resolved}. For the remaining candidates, WISE~$0500-1223$, 2MASS~$0939-2448$, and ULAS~$1416+1348$, we keep them in our analysis, given that their anomalous spectrophotometry is likely caused by low-S/N data or abnormal atmospheres. However, we do not rule out the possibility that these objects are tight unresolved binaries.

\section{Atmospheric Properties of Late-T Dwarfs}
\label{sec:atm_lateT}
Here we examine our forward-modeling results on the distributions of metallicities (Section~\ref{subsec:metallicity}) and ages (Section~\ref{subsec:age}) of late-T dwarfs, as well as their effective temperature--spectral type relation (Section~\ref{subsec:teff_spt}). We demonstrate spectral-fitting results for an ensemble of objects can provide useful diagnostics about the physical assumptions made within model atmospheres, by virtue of having a large sample of spectra over a focused spectral type range (as opposed to a smaller sample of objects spanning a wide spectral type range).

\subsection{Metallicity Distribution}
\label{subsec:metallicity}
We investigate the metallicity of 49 late-T dwarfs\footnote{Among these objects, WISE~$0049+2151$ has two sets of atmospheric model parameters, derived from $0.5''$ and $0.8''$ spectra (Section~\ref{subsubsec:comment_snr}). These results are consistent, and we use those determined from the higher-S/N $0.5''$ spectrum.} in our sample (with the other 6 objects excluded due to their known or likely binarity; Section~\ref{sec:binaries}) and compare them to metallicities of nearby stars. This analysis can in principle assess whether these two populations have similar formation histories but can also shed light on the accuracy of our fitted atmospheric properties.

Spectroscopic analysis of stellar metallicities and elemental abundances involves (1) measuring equivalent widths of individual absorption lines and then converting them into abundances using the curve of growth and theoretical line lists \citep[e.g.,][]{1993A&A...275..101E, 2004AandA...415.1153S, 2006MNRAS.370..163B, 2013ApJ...764...78R, 2018A&A...615A..76S}, or (2) fitting the entire observed spectra using stellar model atmospheres with a range of abundances \citep[e.g.,][]{2005ApJS..159..141V, 2008A&A...485..571J, 2011ApJ...735...41P, 2016ApJS..225...32B}. The abundances of several thousands of nearby ($\lesssim 100$~pc) stars have been studied via high-resolution spectroscopy ($R \gtrsim 3 \times 10^{4}$). The vast majority of these stars have similar iron content ($[$Fe/H$]$) as our Sun \citep[e.g., Figure~4 of][]{2014AJ....148...54H}. While it has been known that different methods, line lists, and spectral resolutions can lead to notable spreads in the derived metallicities and elemental abundances \citep[e.g.,][]{2012ApJ...757..161T, 2014A&A...570A.122S, 2016ApJS..226....4H, 2017MNRAS.468.4151I}, much effort has been made to construct large, homogeneous stellar abundance catalogs \citep[e.g.,][]{2014AJ....148...54H, 2017ApJ...848...34H, 2016ApJS..225...32B}.

In contrast, systematic metallicity analysis of substellar objects has been lacking in most previous spectroscopic studies. For late-type ultracool dwarfs, direct metallicity measurements from atomic/molecular absorption features are particularly challenging, given that (1) the spectral continuum required for computing absorption depths is hard to determine due to the presence of very broad molecular bands (e.g., H$_{2}$O and CH$_{4}$) and (2) elemental abundances cannot be entirely tracked by the observed absorption features, as some portions of species might have been perturbed by vertical mixing and/or sequestered into condensates and thereby sinking to below the photosphere \citep[e.g.,][]{1994Icar..110..117F, 2017ApJ...848...83L}. Metallicity estimates are viable using model atmosphere analyses. While there has been previous such studies for one to a few objects \citep[e.g.,][]{2007ApJ...656.1136S, 2010ApJ...723..850B, 2017ApJ...842..118L}, many analyses have used models with only the solar metallicity \citep[e.g.,][]{2008ApJ...678.1372C, 2009ApJ...702..154S, 2009AandA...501.1059D, 2011ApJ...740..108L, 2015ApJ...804...92S, 2017MNRAS.467.1126D}. 

Recently, some studies have begun characterizing metallicities for large samples of ultracool dwarfs. \citeauthor{2017MNRAS.464.3040Z} (\citeyear{2017MNRAS.464.3040Z, 2018MNRAS.479.1383Z}) studied 28 M7$-$L7 subdwarfs with thick-disk or halo membership. They determined the objects' physical properties by visually comparing their optical and near-infrared spectra to BT-Settl model atmospheres \citep[][]{2011ASPC..448...91A, 2014IAUS..299..271A} and found low $[$Fe/H$]$ values, spanning $-2.5$ to $-0.5$~dex. \cite{2017ApJ...842..118L} visually compared near-infrared spectra of 20 Y dwarfs to the \cite{2015ApJ...804L..17T} models, which assume cloudless atmospheres with dis-equilibrium chemistry and reduced vertical temperature gradient, and found solar-like metallicities with a spread of $\pm0.3$~dex. \cite{2017ApJ...848...83L} conducted a retrieval analysis for 11 late-T dwarfs and found notably lower metallicities (spanning $-0.4$ to $0.1$~dex) and higher carbon-to-oxygen ratios (C/O) than nearby FGK stars. These results are likely related to the sequestration of oxygen into condensates in late-T atmospheres, which then rain out of the photosphere \citep[e.g.,][]{1994Icar..110..117F, 2015ARA&A..53..279M}, resulting in retrieved metallicities and C/O that are systematically too low and high, respectively. Applying the \cite{2017ApJ...848...83L} methodology to 14 late-T and Y dwarfs, \cite{2019ApJ...877...24Z} derived metallicities that span $-0.2$ to $0.6$~dex, which are consistent with (if not slightly higher than) those of nearby stars.

\begin{figure*}[t]
\begin{center}
\includegraphics[height=5.in]{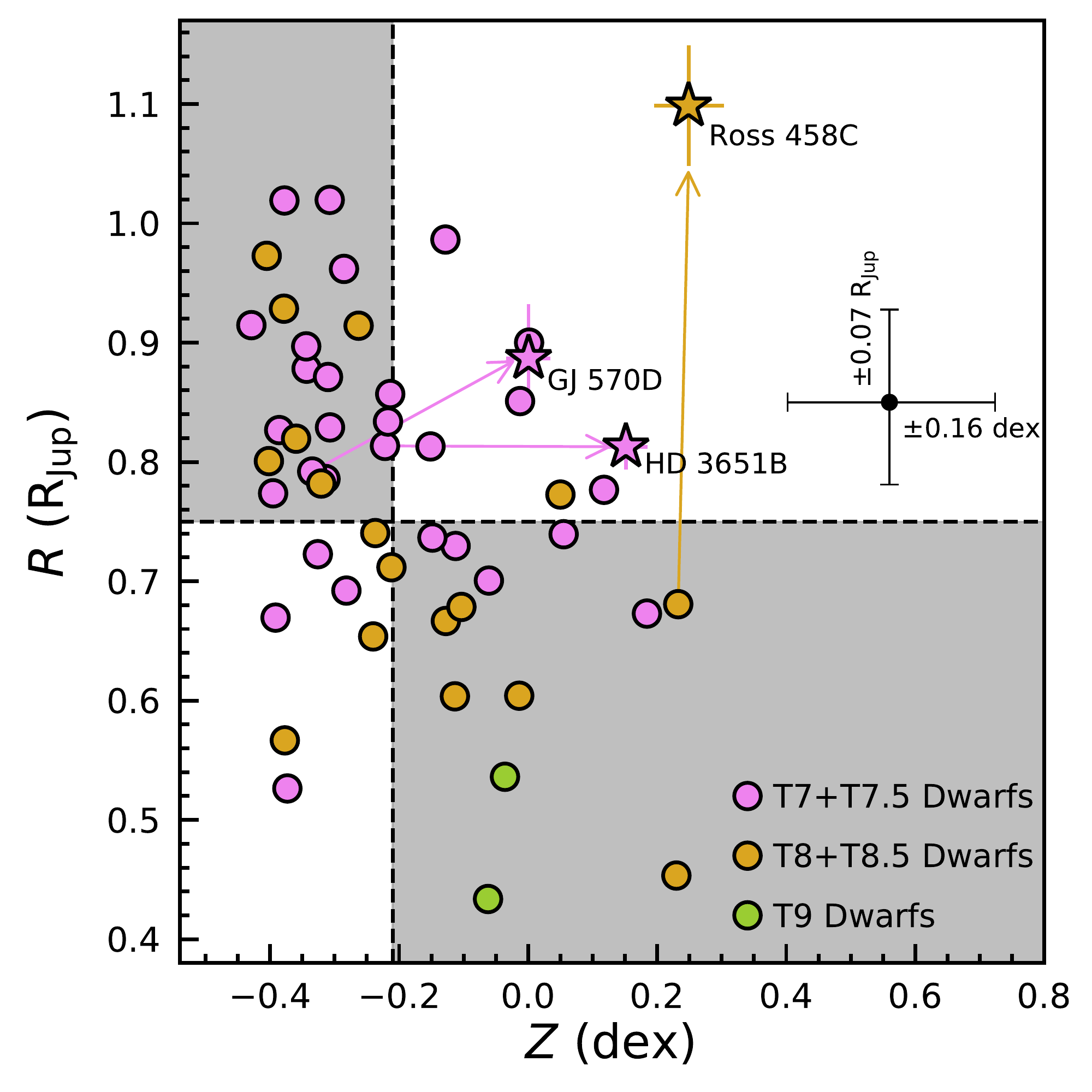}
\caption{Our spectroscopically inferred $R$ and $Z$ for late-T dwarfs (binaries removed; violet for T7 and T7.5, orange for T8 and T8.5, and green for T9), with typical uncertainties shown as black error bars. We use two dashed lines for $Z=-0.21$~dex and $R = 0.75$~R$_{\rm Jup}$ to divide the space into four parts. Objects located in the upper left area (i.e., $Z < -0.21$~dex and $R \geqslant0.75$~R$_{\rm Jup}$) have metallicities that are $>1\sigma$ lower than Hypatia stars but their radii appear reasonable. They might be in similar situations as GJ~570D and HD~3651B, whose fitted $Z$ are underestimated by $\approx 0.35$~dex but their $R$ are reliable, as noted in Paper~I. Objects located in the lower right area (i.e., $Z \geqslant -0.21$~dex and $R < 0.75$~R$_{\rm Jup}$) have solar-like metallicities but unphysically small radii. Some of them might be in a similar situation as Ross~458C whose fitted $Z$ is reliable but $R$ is underestimated by a factor of $\approx 1.6$. The evolutionary model parameters of three benchmark companions (derived from Paper~I) are shown as stars and connected to their spectroscopically inferred parameters by arrows. }
\label{fig:Rz}
\end{center}
\end{figure*}

Our work analyzes 49 late-T dwarfs and thus constitutes the largest homogeneous analysis of brown dwarf metallicities to date. Figure~\ref{fig:z} compares the derived metallicities of our late-T dwarfs to those of F0--M4 stars from the Hypatia catalog\footnote{\url{https://www.hypatiacatalog.com/hypatia}.} \citep[][]{2014AJ....148...54H, 2016ApJS..226....4H, 2017ApJ...848...34H}. The latest version of this catalog (as of June 2019, compiling nearly 200 literature studies) contains 6196 stars within 150~pc that have spectroscopically determined $[$Fe/H$]$ and at least one other element. This compilation of stellar abundance is homogeneous in a sense that \citeauthor{2014AJ....148...54H} carefully examined the results reported by different studies for the same element within the same star and normalized all abundances to the same solar scale by \cite{2009LanB...4B..712L}. We only include thin-disk Hypatia stars (5021 objects) in our analysis and find these objects have a median metallicity of $Z = 0.04$~dex with a $1\sigma$ confidence interval (i.e., 16th-to-84th percentiles) of $-0.21$ to $0.22$~dex. In comparison, our 49 late-T dwarfs have much lower metallicities, spanning $-0.43$ to $0.23$~dex and with a median and mode of $-0.24$~dex and $-0.40$~dex, respectively. We run the Kolmogorov-Smirnov (K-S) test for metallicities of Hypatia stars and our late-T dwarfs in a Monte Carlo fashion\footnote{We generate $10^{4}$ distributions of Hypatia stellar metallicities, with each distribution constructed by a random draw from each object's measured $Z$ and errors by assuming a Gaussian distribution. We generate $10^{4}$ distributions of our late-T dwarfs' metallicities in a same approach except that we draw each object's metallicities from its atmospheric-based $Z$ posterior. We then run the K-S test for each pair of Hypatia and late-T distributions and compute the corresponding $p$-values.} and obtain $p$-values that are all smaller than $2 \times10^{-3}$, thereby indicating they are not drawn from the same distribution. These two samples' metallicity distributions are largely unchanged when we limit our comparison to $\leqslant 25$~pc for our late-T dwarfs (47 objects) and the Hypatia stars (590 objects; Figure~\ref{fig:z}), with the K-S test provideing similarly small $p$-values of $< 0.02$ (with a median of $2\times10^{-7}$).

\begin{figure*}[t]
\begin{center}
\includegraphics[height=6.in]{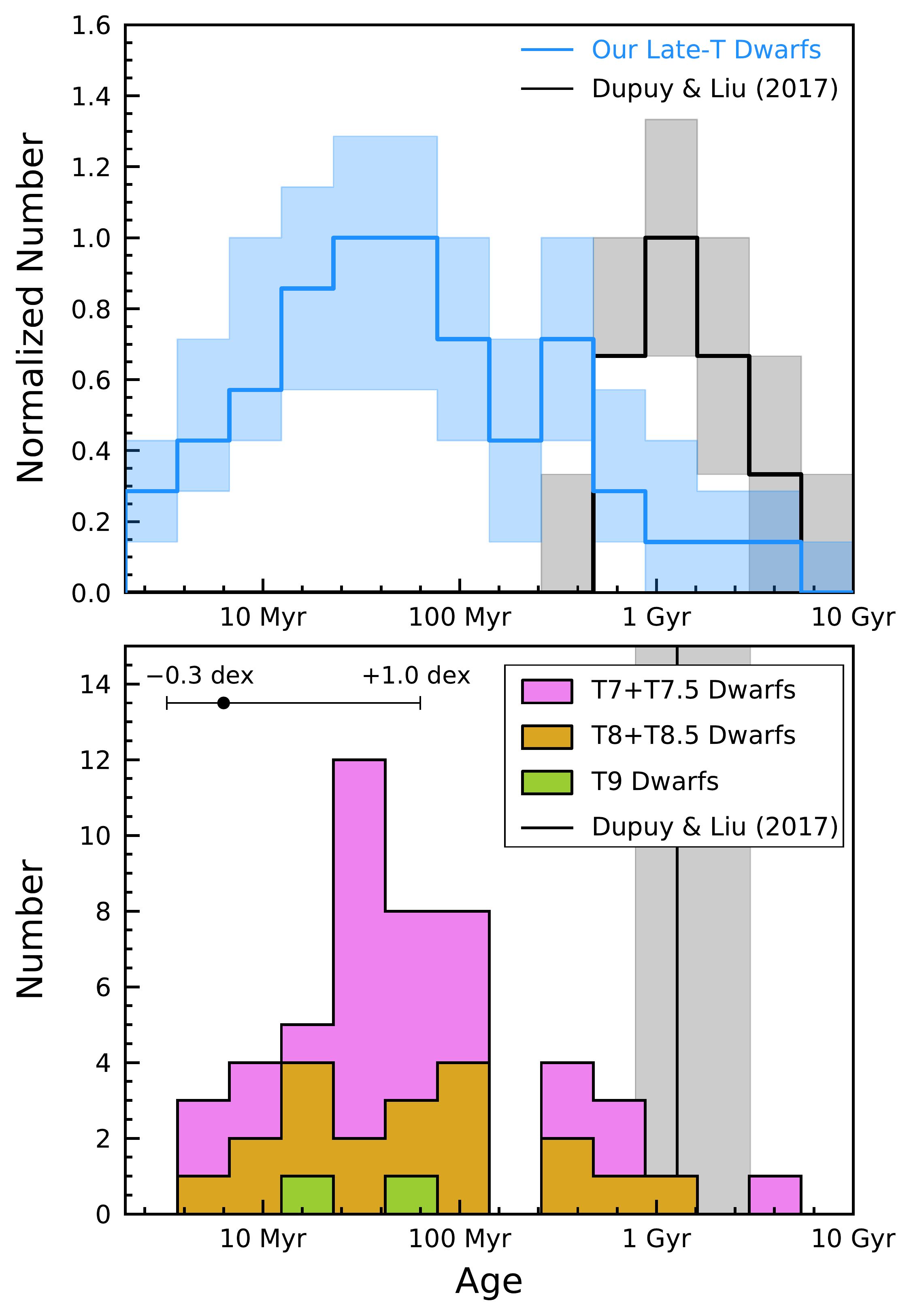}
\caption{Top: Age distributions for our late-T dwarfs (binaries removed) and for benchmark binaries with dynamical masses from \cite{2017ApJS..231...15D}. We compute age posteriors for our late-T dwarfs using the spectroscopically inferred $\{T_{\rm eff},\ \log{g},\ Z\}$ values at each step of the MCMC chains and the interpolated Sonora-Bobcat evolutionary models. We then generate $10^{4}$ distributions with each constructed by a random draw from each object's age posterior. We plot the median and the $1\sigma$ confidence interval of these distributions using the blue line and shadow. Distributions for the \cite{2017ApJS..231...15D} sample are produced in the same approach using the age posteriors of the mass benchmarks (T. Dupuy, private communication). We normalize these two types of distributions so that their peaks correspond to one. Bottom: The distribution of our late-T dwarfs, with the violet color for T7 and T7.5, orange for T8 and T8.5, and green for T9. Our typical logarithmic age uncertainty is shown as the black error bar. We use the vertical black line and shadow to mark the median ($1.3$~Gyr) and the $1\sigma$ spread ($0.8$ to $3.0$~Gyr) of the \cite{2017ApJS..231...15D} age distribution. }
\label{fig:age}
\end{center}
\end{figure*}

To examine whether the metallicity discrepancy between late-T and stellar populations is physical, we first analyze the accuracy of our derived $Z$ values. As shown in Figure~\ref{fig:Rz}, 30 late-T dwarfs have $Z$ of $-0.43$ to $-0.21$~dex and are thus $>1\sigma$ lower than the Hypatia stellar metallicity. Most of these objects have $R = 0.75-1.2$~R$_{\rm Jup}$ from our atmospheric modeling, consistent with the evolutionary model predictions and directly measured radii of transiting brown dwarfs (see Section~\ref{subsec:resolved}). The remaining 19 ($= 49 - 30$) late-T dwarfs have higher $Z$ of $-0.15$ to $0.23$~dex that are consistent with those of Hypatia stars, but most of these objects have unphysically small radii of $<0.75$~R$_{\rm Jup}$. Overall, we find a correlation between $Z$ and $R$, where the objects with higher $Z$ in our sample tend to have smaller $R$ from our atmospheric model analysis.

This $Z-R$ correlation is in contradiction to the cloudless Sonora-Bobcat evolutionary models, which predict that objects with higher metallicities should have increased atmospheric opacity, slower cooling, and thereby larger radii at a given mass and age \citep[also see][]{2011ApJ...736...47B}. Also, this correlation is not found in stacked posteriors of atmospheric parameters (Figure~\ref{fig:stack_posterior}). Therefore, it is likely not due to the parameter degeneracy within the Sonora-Bobcat models or our spectral-fitting machinery. 

This $Z-R$ trend might be because the spectroscopically inferred $Z$ or $R$ of our late-T dwarfs are underestimated. As shown in Paper~I, the fitted metallicities of the two benchmark companions, HD~3651B and GJ~570D, are underestimated by $\approx 0.35$~dex but their fitted radii are reliable, when compared to the radii derived from the Sonora-Bobcat evolutionary models (using these companions' bolometric luminosities and their primary stars' ages and metallicities). For another benchmark, Ross~458C, its fitted metallicity is consistent with evolutionary-based results but its fitted radii is underestimated by a factor of $\approx 1.6$. Among our sample, some objects with fitted $Z<-0.21$~dex and $R\geqslant0.75$~R$_{\rm Jup}$ might be in similar situations as HD~3651B and GJ~570D, while those with fitted $Z\geqslant-0.21$~dex and $R<0.75$~R$_{\rm Jup}$ might be in a similar situation as Ross~458C. Altogether, correcting the inaccurate atmospheric parameters of these late-T dwarfs would be the equivalent of shifting their $Z$ toward higher values and/or their $R$ toward larger values, which would alter the $Z-R$ correlation seen in Figure~\ref{fig:Rz}. The underestimation of these fitted parameters in our sample suggests the model assumptions of cloudless and chemical-equilibrium atmospheres are not adequate to fully interpret late-T dwarf spectra and should be further improved. 

To conclude, many of our late-T dwarfs with low atmospheric $Z$ might have values underestimated by as large as $0.3-0.4$~dex, and this can explain the discrepant metallicity distributions between late-T dwarfs and Hypatia stars. Thus, we cannot say more about whether the metallicities of nearby late-T dwarfs and stars are similar. More comprehensive understanding of substellar metallicities would benefit from the continuing discoveries of brown dwarfs as companions to stars or members of nearby associations. These objects can provide independent metallicities from their hosts and are thus ``metallicity benchmarks'' for calibrating forward-modeling analyses and improving our understanding of ultracool atmospheres.

\begin{figure*}[t]
\begin{center}
\includegraphics[height=4.5in]{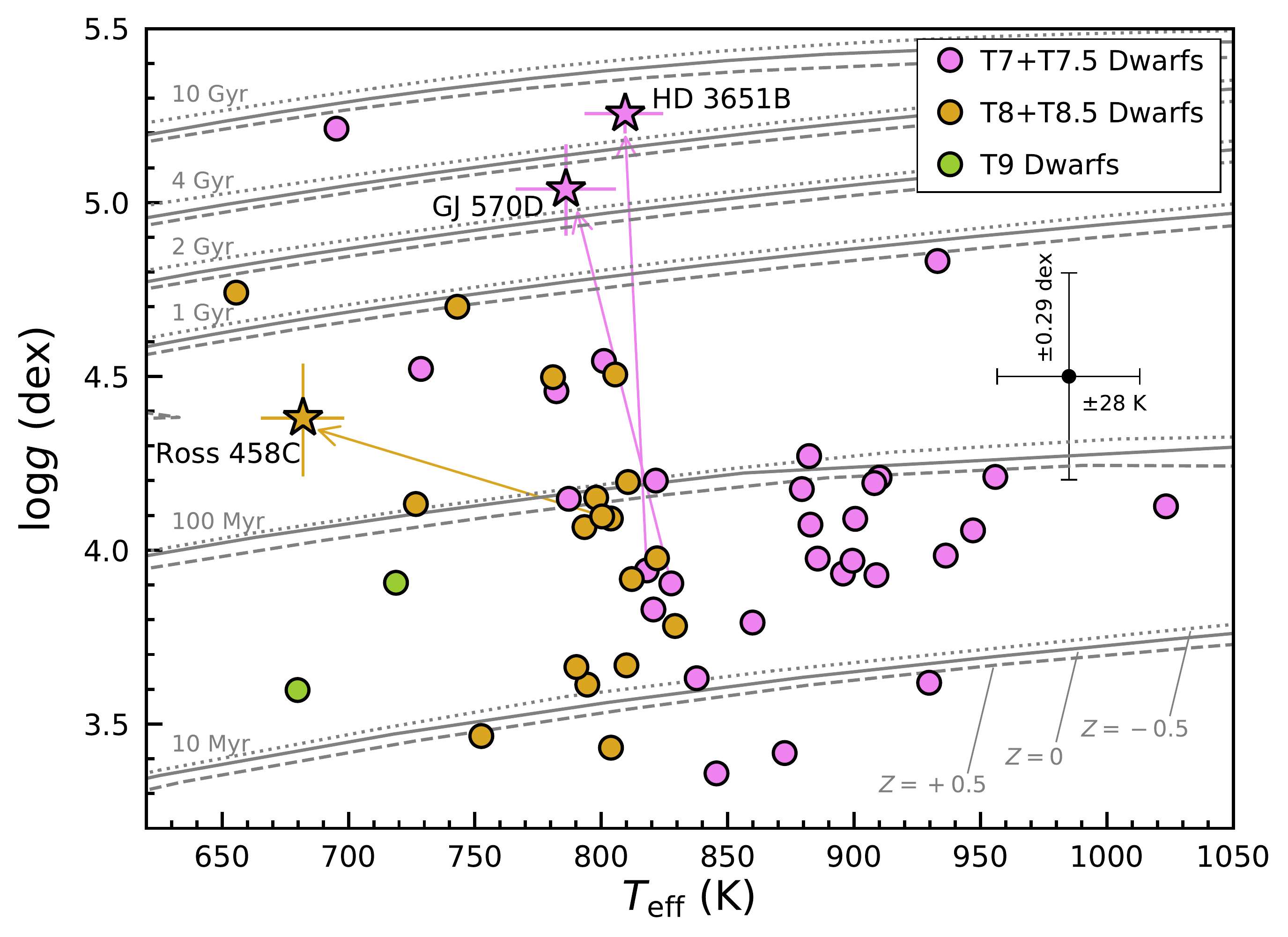}
\caption{Spectroscopically inferred $T_{\rm eff}$ and $\log{g}$ of our late-T dwarfs (binaries removed; violet for T7 and T7.5, orange for T8 and T8.5, and green for T9) and the cloudless Sonora-Bobcat isochrones (grey lines) with different ages ($10$~Myr, $100$~Myr, $1$~Gyr, $2$~Gyr, $4$~Gyr, $10$~Gyr) and metallicities ($-0.5$~dex, $0$~dex, $+0.5$~dex). Typical uncertainties for our fitted $T_{\rm eff}$ and $\log{g}$ are shown as black error bars. The evolutionary model parameters of the three benchmark companions (derived from Paper~I) are shown as stars and connected to their spectroscopically inferred parameters by arrows. Our forward-modeling analysis underestimates these benchmarks' $\log{g}$ and $Z$ or overestimates their $T_{\rm eff}$, likely due to systematics of the cloudless Sonora-Bobcat models, and consequently, their ages derived from atmospheric model parameters are implausibly younger than the ages of their primary stars. Many of our late-T dwarfs might be in similar situations as these three companions, and thus their true ages should be systematically older. }
\label{fig:teff_logg_age}
\end{center}
\end{figure*}

\subsection{Age Distribution}
\label{subsec:age}
We proceed to study ages of our late-T dwarfs. The age distribution of field brown dwarfs is essential to constrain their formation history and initial mass function \citep[e.g.,][]{2012ApJ...753..156K, 2013MNRAS.430.1171D}. However, it is challenging to age-date individual field brown dwarfs, given that their ages and masses are degenerate and thus the former cannot be well-constrained unless the latter is known independently. Previous work has estimated low-precision ages for large groups of ultracool dwarfs based on their kinematics. Such analysis is conducted in a statistical fashion and assumes the objects' space motions exhibit a larger spread with time due to dynamical evolution \citep[e.g.,][]{1977A&A....60..263W}. Comparing tangential or three-dimensional space velocities of late-M, L, and early-T dwarfs to those of earlier-type main-sequence stars, several studies have found that local ultracool dwarfs have slightly smaller velocity dispersions and perhaps younger ages \citep[$0.5-4$~Gyr; e.g.,][]{2002AJ....124.1170D, 2007AJ....133.2258S, 2007ApJ...666.1205Z}. However, based on tangential velocities of a large, volume-limited sample of M7$-$T8 objects within 20~pc, \cite{2009AJ....137....1F} suggested the kinematic age of ultracool dwarfs ($3-8$~Gyr) is indistinguishable from earlier-type stars, although the former indeed becomes younger ($2-4$~Gyr) after excluding thick-disk or halo-like objects with high tangential velocities \citep[$>100$~km~s$^{-1}$; also see][]{2007AJ....133.2258S}. 

Thanks to the long-term astrometric monitoring, many brown dwarf binaries now have measured dynamical masses. These ``mass benchmarks'' \citep[e.g.,][]{2008ApJ...689..436L} can disentangle the age-mass degeneracy of substellar objects and thus yield some of the most robust ages. Recently, \citeauthor{2017ApJS..231...15D} (\citeyear{2017ApJS..231...15D}; DL17) studied the largest sample of such benchmark binaries to date (31 systems) and derived ages from their measured dynamical masses and the \cite{2008ApJ...689.1327S} evolutionary models. They derived a substellar age distribution from a high-quality subset of 10 systems (M8$-$T5 within 30~pc) and found a median and mean of $1.3$~Gyr and $2.3$~Gyr, respectively. This distribution is systematically younger than the past population synthesis analyses of \cite{2004ApJS..155..191B} and \cite{2005ApJ...625..385A}, which found that mid-L to early-T dwarfs have ages of $\approx 2-3$~Gyr \citep[although the modeled age distribution at a given spectral type has a large spread, e.g., see Figure~8 of][]{2004ApJS..155..191B}. The DL17 age distribution is in accord with Galactic dynamic heating preferentially scattering older substellar objects out of the immediate region ($\lesssim 50$~pc) around the Earth, which has been the main focus of brown dwarf searches. 

Here we compare the DL17 age distribution to that of our late-T dwarf sample, for which we compute ages using the fitted $\{T_{\rm eff},\ \log{g},\ Z\}$ values at each step of the MCMC chains and the interpolated cloudless Sonora-Bobcat evolutionary models (Table~\ref{tab:starfish_derived_results}). Both our analysis and that of DL17 are subject to any systematics of the evolutionary models \citep[e.g.,][]{2009ApJ...692..729D, 2014ApJ...790..133D, 2018AJ....156..168B, 2018AJ....155..159B, 2018ApJ...865...28D, 2020AJ....160..196B}, but ours are also impacted by those of the atmospheric models, which are likely more uncertain. Therefore, our derived ages are expected to be less accurate than the DL17 results.

The age distribution of our late-T dwarfs is shown in Figure~\ref{fig:age}, with a median of $50$~Myr and 16th-to-84th percentiles spanning $10$~Myr to $0.4$~Gyr. This distribution is significantly younger than DL17 (with a median of $1.3$~Gyr and 16th-to-84th percentiles spanning $0.8$ to $3.0$~Gyr) and past kinematic studies. Such a young distribution is implausible, especially since the majority of our sample are not associated to any stars or clusters with such young age.\footnote{Based on the available astrometry and radial velocities of 694 T and Y dwarfs, \cite{2021ApJ...911....7Z} recently identified a number of T dwarfs as candidate members of nearby young moving groups, with their final membership assessment awaiting radial velocity measurements. Only 3 objects in our sample here are candidate young moving group members: WISE~J$024124.73-365328.0$ \citep[T7; Argus; $40-50$~Myr,][]{2019ApJ...870...27Z}, 2MASS~$1553+1532$ \citep[T7; Carina-Near; $200\pm50$~Myr,][]{2006ApJ...649L.115Z}, and WISEPC~J$225540.74-311841.8$ \citep[T8; $\beta$~Pictoris; $24\pm3$~Myr,][]{2015MNRAS.454..593B}.} The same effect is seen for the 3 benchmarks, HD~3651B, GJ~570D, and Ross~458C, whose primary stars have ages of $4.5-8.3$~Gyr, $1.4-5.2$~Gyr, and $0.15-0.8$~Gyr determined by various age-dating techniques (as summarized in Paper~I). In contrast, our spectroscopically inferred ages of these objects are much younger, $0.038^{+0.077}_{-0.025}$~Gyr, $0.031^{+0.052}_{-0.020}$~Gyr, and $0.077^{+0.314}_{-0.056}$~Gyr, respectively (Figure~\ref{fig:teff_logg_age} and Table~\ref{tab:starfish_derived_results}).

Our implausibly young age estimates represent another illustration that the assumptions made within our model atmospheres should be further improved. To derive more reliable ages of late-T dwarfs from near-infrared spectra, we could introduce more atmospheric processes into models, e.g., clouds and/or dis-equilibrium chemistry (see Section~\ref{sec:comments_sonora}).

\begin{figure*}[t]
\begin{center}
\includegraphics[height=5.5in]{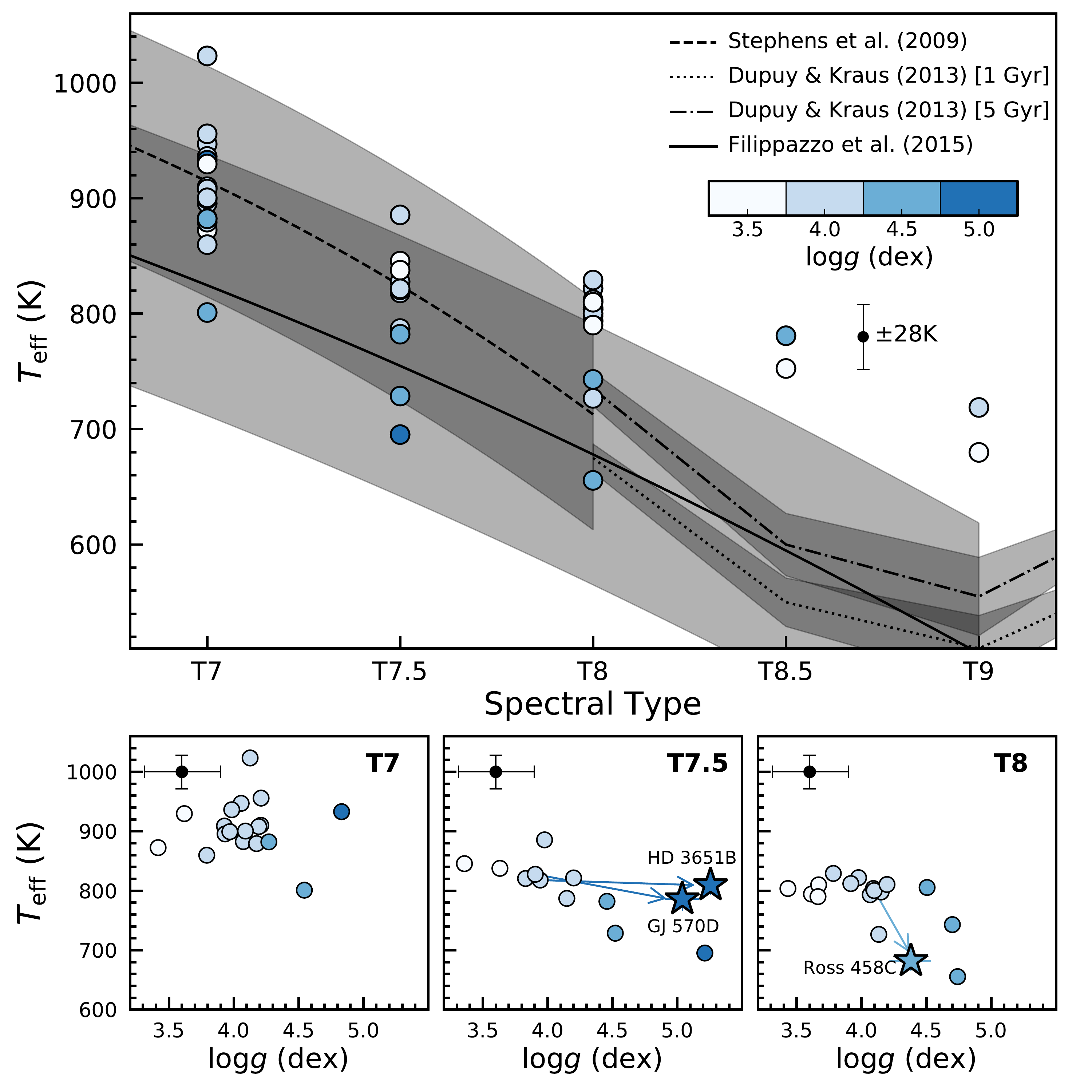}
\caption{Top plot: $T_{\rm eff}-$SpT relation of our late-T dwarfs (binaries removed), with spectroscopically inferred $\log{g}$ encoded with blue shading. The median $T_{\rm eff}$ uncertainty is shown as the black error bar. We overlay evolutionary-based effective temperature scales from \citeauthor{2009ApJ...702..154S} (\citeyear{2009ApJ...702..154S}; dashed) and \citeauthor{2015ApJ...810..158F} (\citeyear{2015ApJ...810..158F}; solid). We also plot the \cite{2013Sci...341.1492D} $T_{\rm eff}-$SpT relations corresponding to assumed ages of $1$~Gyr (dotted) and $5$~Gyr (dash-dotted). The gray shadow shows the rms of each empirical T$_{\rm eff}-$SpT relation. Bottom plots: $T_{\rm eff}$ of our T7, T7.5, and T8 dwarfs as a function of $\log{g}$. Median $T_{\rm eff}$ and $\log{g}$ uncertainties are shown as black error bars. The evolutionary model parameters of three benchmarks (derived from Paper~I) are shown as stars and connected to their spectroscopically inferred parameters by arrows.  }
\label{fig:teff_spt_logg}
\end{center}
\end{figure*}

\begin{figure*}[t]
\begin{center}
\includegraphics[height=5.5in]{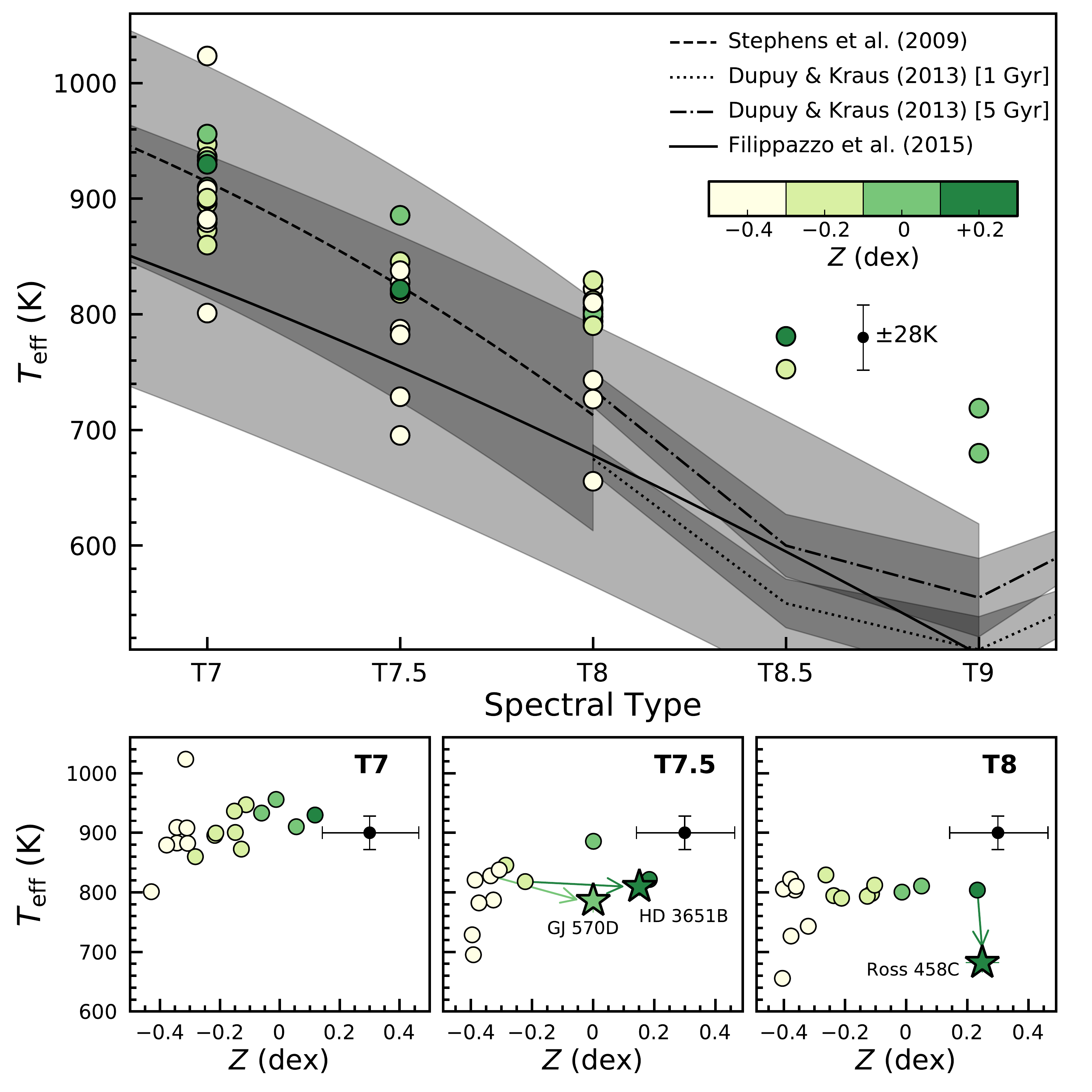}
\caption{The effect of metallicity on the $T_{\rm eff}-$SpT relation of our late-T dwarfs (binaries removed), with the same format as Figure~\ref{fig:teff_spt_logg}. Spectroscopically inferred $Z$ are encoded with green shading. Bottom plots show the objects' $T_{\rm eff}$ as a function of $Z$. }
\label{fig:teff_spt_z}
\end{center}
\end{figure*}

\begin{figure*}[t]
\begin{center}
\includegraphics[height=2in]{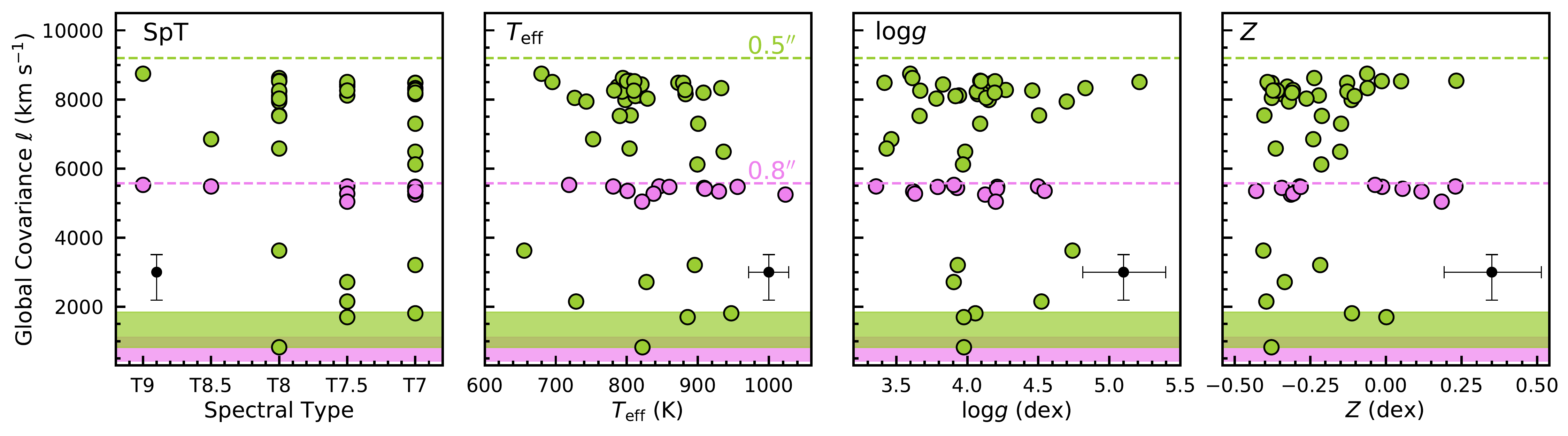}
\caption{The covariance hyper-parameter $\ell$ of our 49 late-T dwarfs (binaries removed) as a function of spectral type and fitted $\{T_{\rm eff},\ \log{g},\ Z\}$, with median uncertainties shown as black error bars. We use green and violet for $\ell$ inferred from spectra taken from the $0.5''$ and $0.8''$ slits, respectively. For a given slit, we use the shaded area to mark the $\ell$ range expected by the instrumental line spread function ($820-1840$~km~s$^{-1}$ for the $0.5''$ slit and $425-1115$~km~s$^{-1}$ for the $0.8''$ slit) and use dashed lines to mark upper boundaries of our adopted $\ell$ prior.  }
\label{fig:ell}
\end{center}
\end{figure*}

\begin{figure*}[t]
\begin{center}
\includegraphics[height=2in]{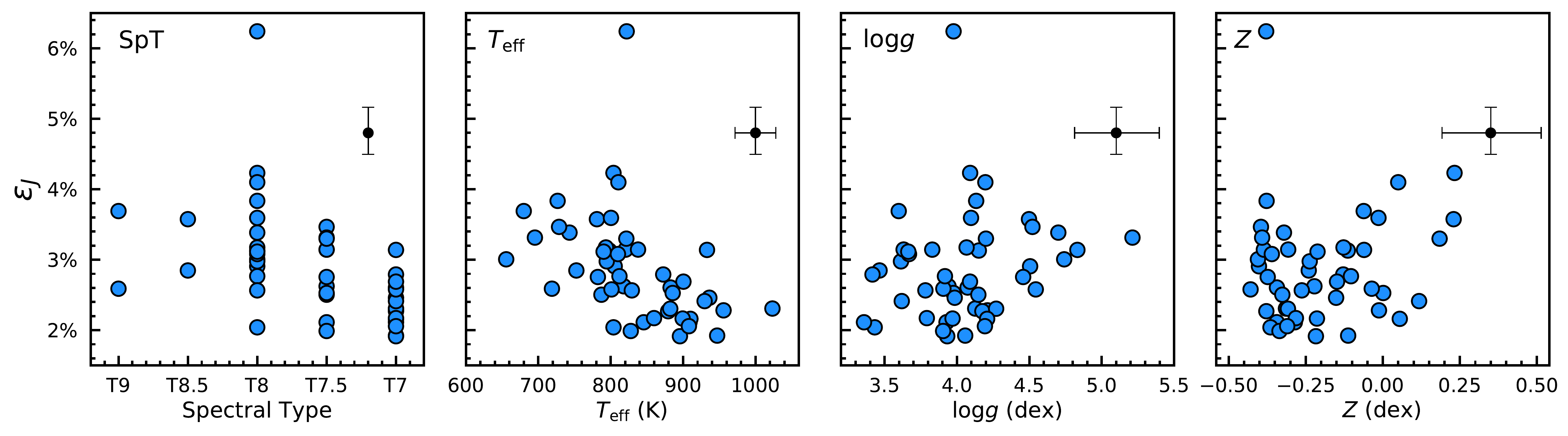}
\caption{The $\epsilon_{J}$ values of our 49 late-T dwarfs (binaries removed) computed from their covariance hyper-parameters $a_{G}$ and $a_{N}$ and the observed spectra (Section~\ref{subsec:hyper_param}). The $\epsilon_{J}$ describes the fraction that model uncertainties (i.e., systematic flux difference between the cloudless Sonora-Bobcat models and late-T dwarf spectra) comprise of the objects' observed peak $J$-band fluxes. Median uncertainties are shown as black error bars. The object with the highest $\epsilon_{J}$ among our sample is WISE~$1322-2340$. The $\epsilon_{J}$ values, namely model systematics, tend to become larger at later spectral types, cooler spectroscopically inferred $T_{\rm eff}$, and possibly higher $\log{g}$ and $Z$.   }
\label{fig:epsJ}
\end{center}
\end{figure*}

\subsection{Effective Temperature vs. Spectral Type}
\label{subsec:teff_spt}

Now we study the relation between the atmospheric-based, spectroscopically inferred $T_{\rm eff}$ and the spectral types of late-T dwarfs. First, we briefly review empirical $T_{\rm eff}$--SpT relations that have been established using evolutionary models. Similar to earlier work by \cite{2002ApJ...564..452L}, \cite{2004AJ....127.3516G} compiled parallaxes and infrared spectrophotometry of 51 M1$-$T9 dwarfs. They measured bolometric luminosities from the objects' spectral energy distributions (SEDs) and then converted these into $T_{\rm eff}$ assuming an age range of $0.1-10$~Gyr and using evolutionary models \citep{1997ApJ...491..856B, 1998A&A...337..403B, 2000ApJ...542..464C}. \cite{2008ApJ...685.1183L} and \cite{2009ApJ...702..154S} later modified this sample by removing binaries and/or adding companions discovered since the \cite{2004AJ....127.3516G} work and obtained tighter temperature scales as a function of spectral types. Focusing on the coolest substellar objects, \cite{2013Sci...341.1492D} computed $L_{\rm bol}$ for 21 T8--Y0 field dwarfs using parallaxes, broadband photometry, and bolometric corrections, and then derived $T_{\rm eff}$ using the \cite{2003AandA...402..701B} evolutionary models with age assumptions of $1$~Gyr and $5$~Gyr. The most recent $T_{\rm eff}-$SpT relation has been derived by \cite{2015ApJ...810..158F} using a much larger sample of nearly 200 M6$-$T9 dwarfs. They measured $L_{\rm bol}$ based on a homogeneous SED analysis and then estimated $T_{\rm eff}$ using the \cite{2008ApJ...689.1327S} hybrid models with an assumed age of $0.5-10$~Gyr for most objects and narrower ranges for association members and companions. The \cite{2015ApJ...810..158F} temperature scale of field dwarfs is consistent with results of \cite{2009ApJ...702..154S} and \cite{2013Sci...341.1492D}.

Figures~\ref{fig:teff_spt_logg} and \ref{fig:teff_spt_z} compare our atmospheric-based $T_{\rm eff}$ of T7$-$T9 dwarfs to the evolutionary-based temperature scales from \cite{2009ApJ...702..154S}, \cite{2013Sci...341.1492D}, and  \cite{2015ApJ...810..158F}. At a given spectral type, our spectroscopically inferred $T_{\rm eff}$ show a spread of $\pm 100$~K, similar to the empirical relations. Our results are consistent with empirical ones for T7 and T7.5 dwarfs, but appear $50-200$~K hotter at later types, likely because our derived $T_{\rm eff}$ for T8$-$T9 dwarfs are less reliable than those of T7 dwarfs. As found in Paper~I and summarized in Section~\ref{subsec:results} in this work, our spectral fits of Ross~458C (T8) predict $\approx 120$~K higher $T_{\rm eff}$ than evolutionary model analysis, while the two T7.5 benchmarks (HD~3651B and GJ~570D) have no such significant discrepancy. 

Figure~\ref{fig:teff_spt_logg} explores whether the derived $T_{\rm eff}-$SpT relation of our sample is gravity-dependent. We focus on the T7, T7.5, and T8 objects in our sample, since we have only four T8.5 and T9 dwarfs. Studying the objects' $T_{\rm eff}$ as a function of $\log{g}$, we find lower-gravity T7.5 and T8 dwarfs have on-average hotter effective temperatures at a given spectral type. Intriguingly, this phenomenon is not found in earlier T dwarfs. Based on a sample of 25 young ($\leqslant 300$~Myr) and old ($> 300$~Myr) T0--T6 benchmarks, \cite{2020ApJ...891..171Z} demonstrated that lower-gravity objects tend to have $\approx 100$~K cooler $T_{\rm eff}$ than their high-gravity counterparts at same spectral types \citep[also see][]{2006ApJ...651.1166M, 2015ApJ...810..158F, 2016ApJS..225...10F, 2016ApJ...833...96L}, which is opposite of the $T_{\rm eff}-\log{g}$ relation for late-T dwarfs seen in Figure~\ref{fig:teff_spt_logg}. For T7$-$T9 spectral types, there is only one low-gravity benchmark with direct spectroscopy, i.e., Ross~458C, for which our evolutionary model analysis derives $T_{\rm eff} = 682^{+16}_{-17}$~K (Paper~I). This temperature is consistent with those of higher-gravity field T8 dwarfs \citep[$678 \pm 113$~K based on the $T_{\rm eff}-$SpT relation of][]{2015ApJ...810..158F} and thus not in accord with the $T_{\rm eff}-\log{g}$ relation from our results, which predicts its $T_{\rm eff}$ to be much hotter. More discoveries of such young, late-type benchmarks will help investigate the gravity dependence of late-T dwarfs' effective temperatures.

In fact, the possible relation between the fitted $T_{\rm eff}$ and $\log{g}$ of our sample is likely because these parameters are over- and/or under-estimated. Some of the T7.5 dwarfs in our sample might be similar to HD~3651B and GJ~570D, whose fitted $T_{\rm eff}$ are reliable but $\log{g}$ are underestimated by $\approx 1.2$~dex when compared to the more robust evolutionary model predictions. Also, some of the T8 dwarfs might be similar to Ross~458C, whose fitted $\log{g}$ is reliable but $T_{\rm eff}$ is overestimated by $\approx 120$~K. Correcting their inaccurate atmospheric-based parameters would be the equivalent of shifting their $\log{g}$ toward higher values and/or their $T_{\rm eff}$ toward cooler values, which could alter the $T_{\rm eff}-\log{g}$ relation in Figure~\ref{fig:teff_spt_logg}.

Figure~\ref{fig:teff_spt_z} explores whether the derived $T_{\rm eff}-$SpT relation of our sample is metallicity-dependent. Studying the objects' $T_{\rm eff}$ as a function of $Z$, we find lower-metallicity T7, T7.5, and T8 dwarfs have on-average cooler effective temperatures at a given spectral type. The validity of this metallicity dependence is uncertain. For the two late-T subdwarf benchmarks, BD~$+01^{\circ}$~2920B \citep[T8; $Z = -0.38 \pm 0.06$~dex by][]{2005ApJS..159..141V} and Wolf~1130C \citep[T8; $Z = -0.70 \pm 0.12$ by][]{2018ApJ...854..145M}, previous work found their effective temperatures are similar to those of the field population, which has on average higher metallicities \citep[][]{2012MNRAS.422.1922P, 2013ApJ...777...36M}, and thus not in accord with the $T_{\rm eff}-Z$ relation seen in our results. Similar to the aforementioned $T_{\rm eff}-\log{g}$ trend, this $T_{\rm eff}-Z$ relation for late-T dwarfs in Figure~\ref{fig:teff_spt_z} might be related to modeling systematics. To further validate this potential metallicity dependence for late-T dwarfs, we need models with modified physical assumptions and more benchmarks over a wide range of metallicities.

\section{Discussion of the Cloudless Sonora-Bobcat Models}
\label{sec:comments_sonora}
Our forward-modeling analysis shows the cloudless Sonora-Bobcat models match the general spectroscopic appearance of late-T dwarfs, but the fitted physical parameters can be under- or over-estimated, suggesting the models' assumptions might be inadequate to fully interpret late-T dwarf spectra (Section~\ref{sec:atm_lateT}). Here we explore specific shortcomings, by focusing on 49 late-T dwarfs in our sample, as the other 6 objects are resolved or likely binaries (Section~\ref{sec:binaries}).

\begin{figure*}[t]
\begin{center}
\includegraphics[height=4.5in]{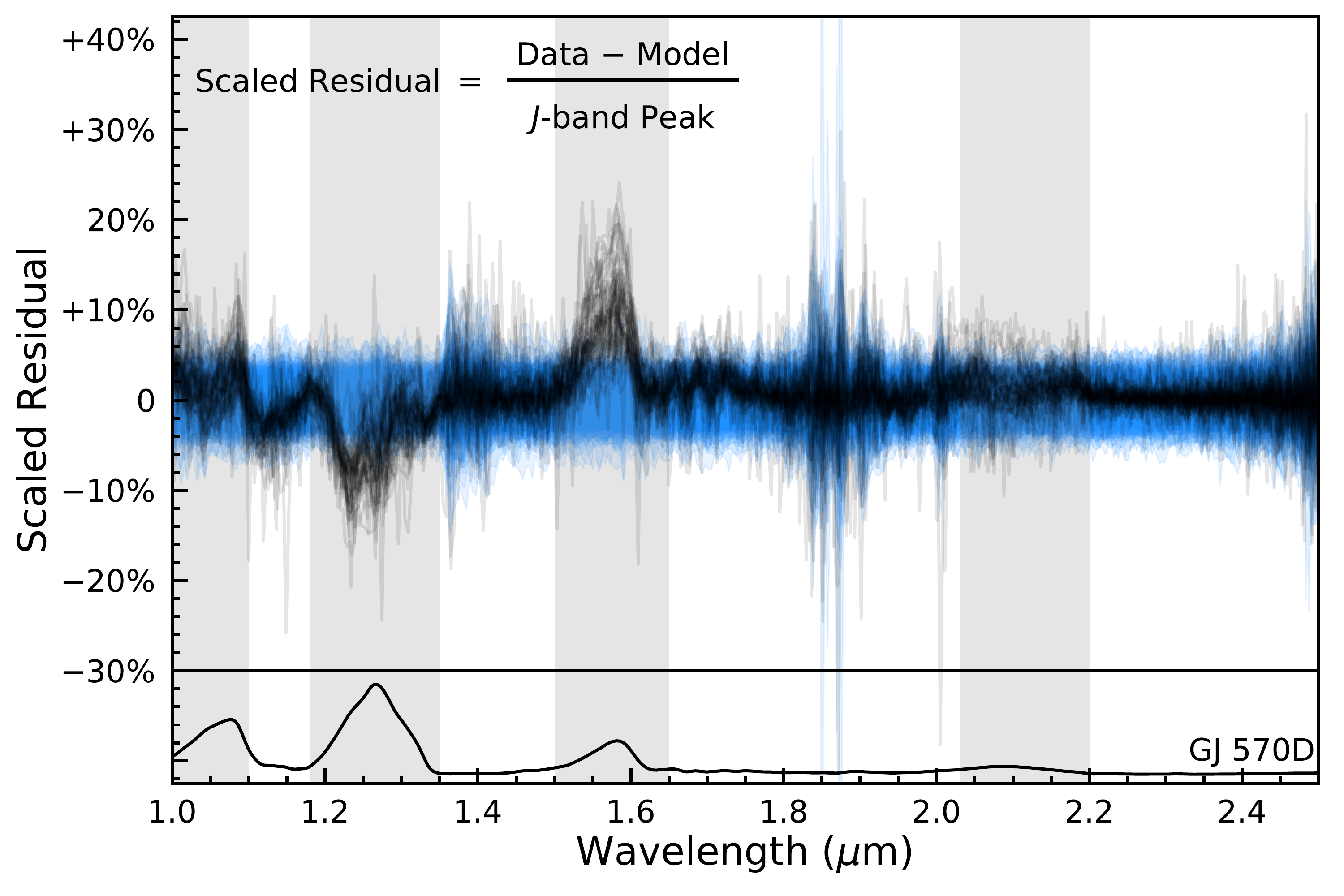}
\caption{Stacked spectral-fitting residuals of our 49 late-T dwarfs (binaries removed), with each residual (i.e., data $-$ model; black) normalized by the object's observed peak $J$-band flux. The $1\sigma$ and $2\sigma$ dispersions of $5 \times 10^{4}$ draws from Starfish's full covariance matrix are shown as deep and light blue shadows, respectively, and they are normalized by the same peak flux for each object. The resulting stack has a tight distribution at each wavelength and exhibits prominent features in $YJHK$ bands (grey shadow, with specific wavelength ranges shown in Equation~\ref{eq:Rlambda}). The spectrum of GJ~570D is plotted at the bottom as a reference. }
\label{fig:stacked_residual}
\end{center}
\end{figure*}

\begin{figure*}[t]
\begin{center}
\includegraphics[height=5.5in]{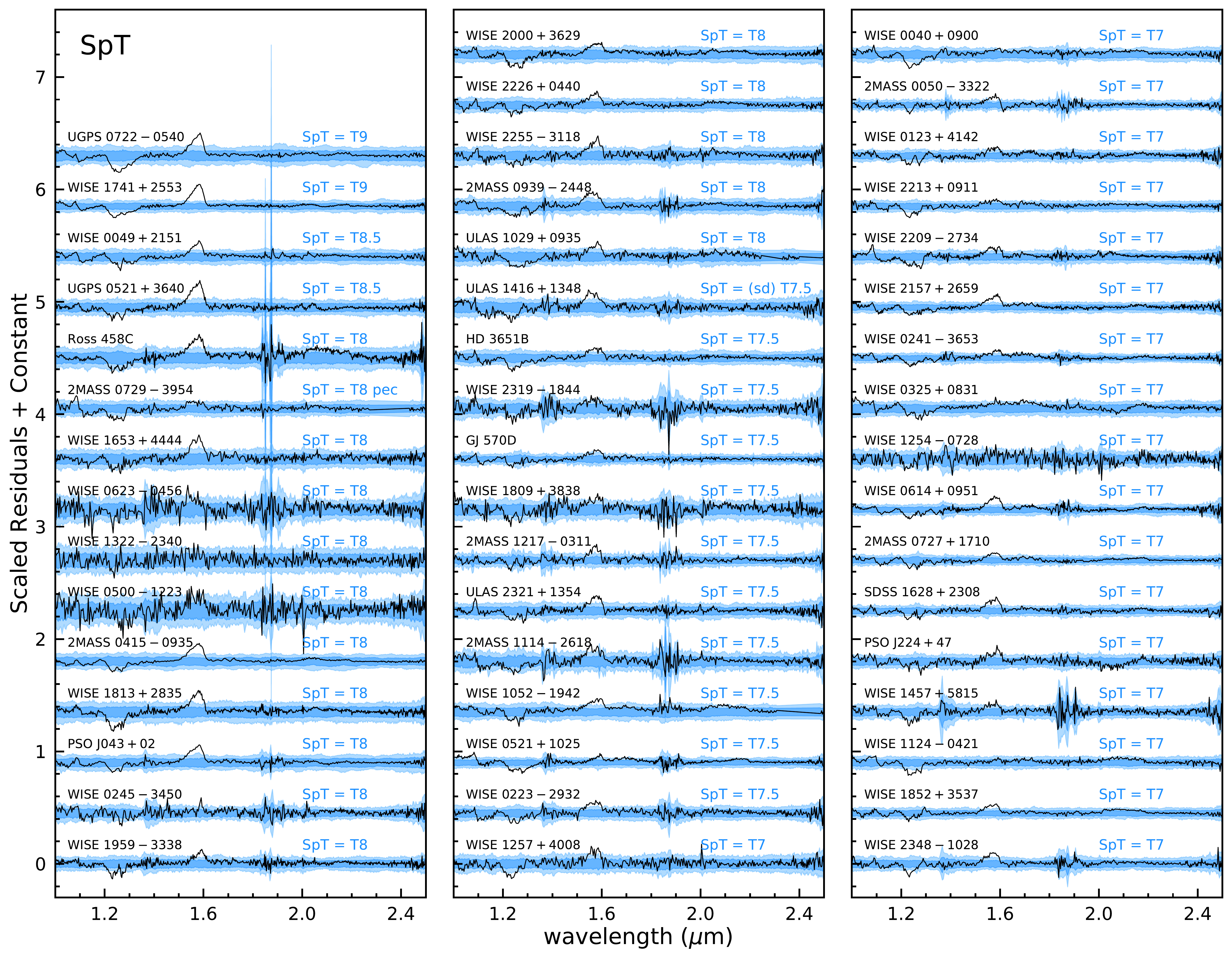}
\caption{Residuals of our 49 late-T dwarfs (binaries removed) shown in order of spectral types, and spectroscopically inferred $T_{\rm eff}$, $\log{g}$, and $Z$. We use each object's observed peak $J$-band flux to normalize its residual (black), as well as $1\sigma$ and $2\sigma$ dispersions of $5 \times 10^{4}$ draws from Starfish's full covariance matrix (blue shadows). The $J$-band and $H$-band residuals tend to be more significant with later spectral types and cooler effective temperatures. The $H$-band residual also becomes more prominent with increasing metallicities.  }
\label{fig:residual_afo_stgz}
\end{center}
\end{figure*}
\renewcommand{\thefigure}{\arabic{figure}}
\addtocounter{figure}{-1}
\begin{figure*}[t]
\begin{center}
\includegraphics[height=5.5in]{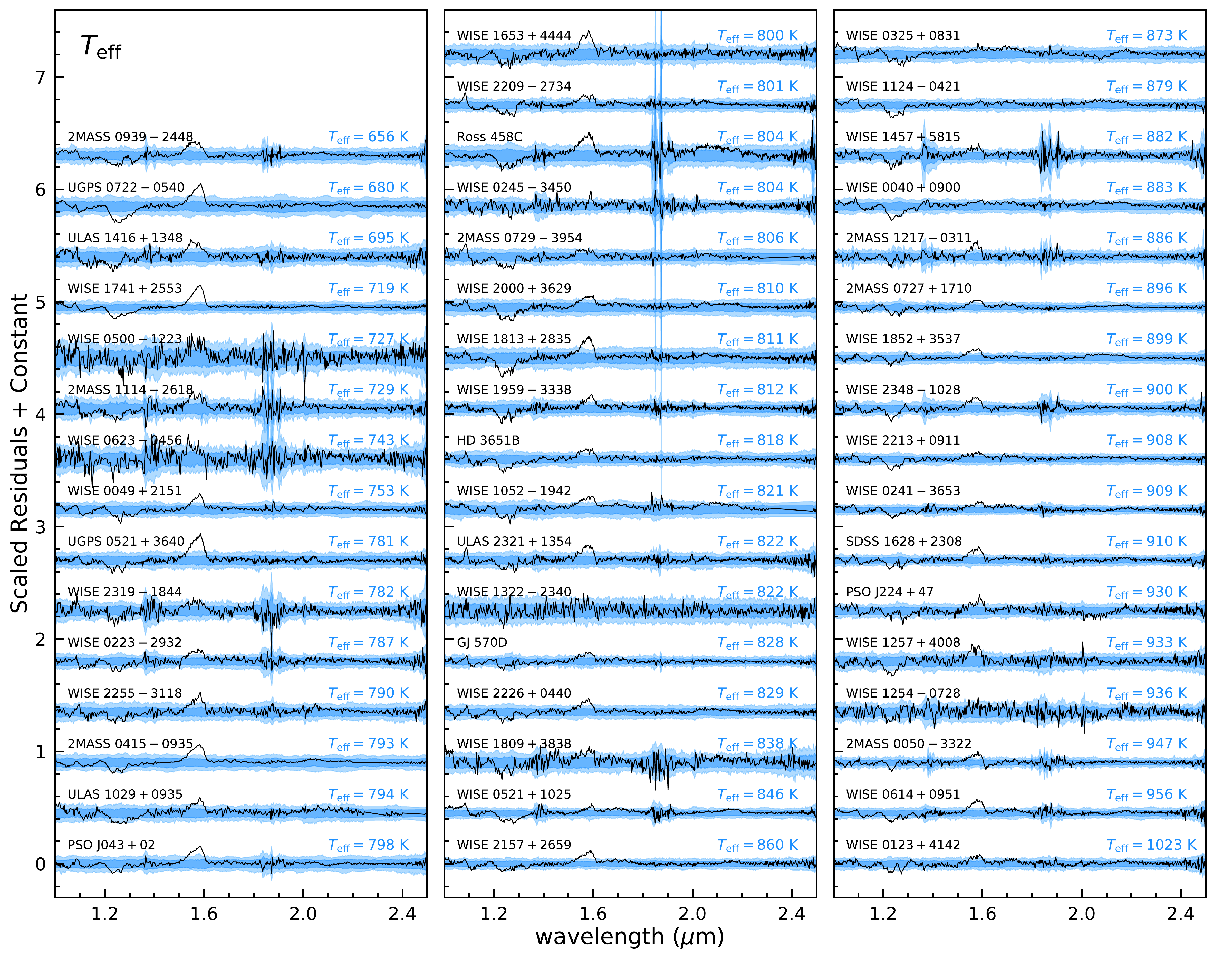}
\caption{Continued}
\label{fig:residual_afo_stgz}
\end{center}
\end{figure*}
\renewcommand{\thefigure}{\arabic{figure}}
\addtocounter{figure}{-1}
\begin{figure*}[t]
\begin{center}
\includegraphics[height=5.5in]{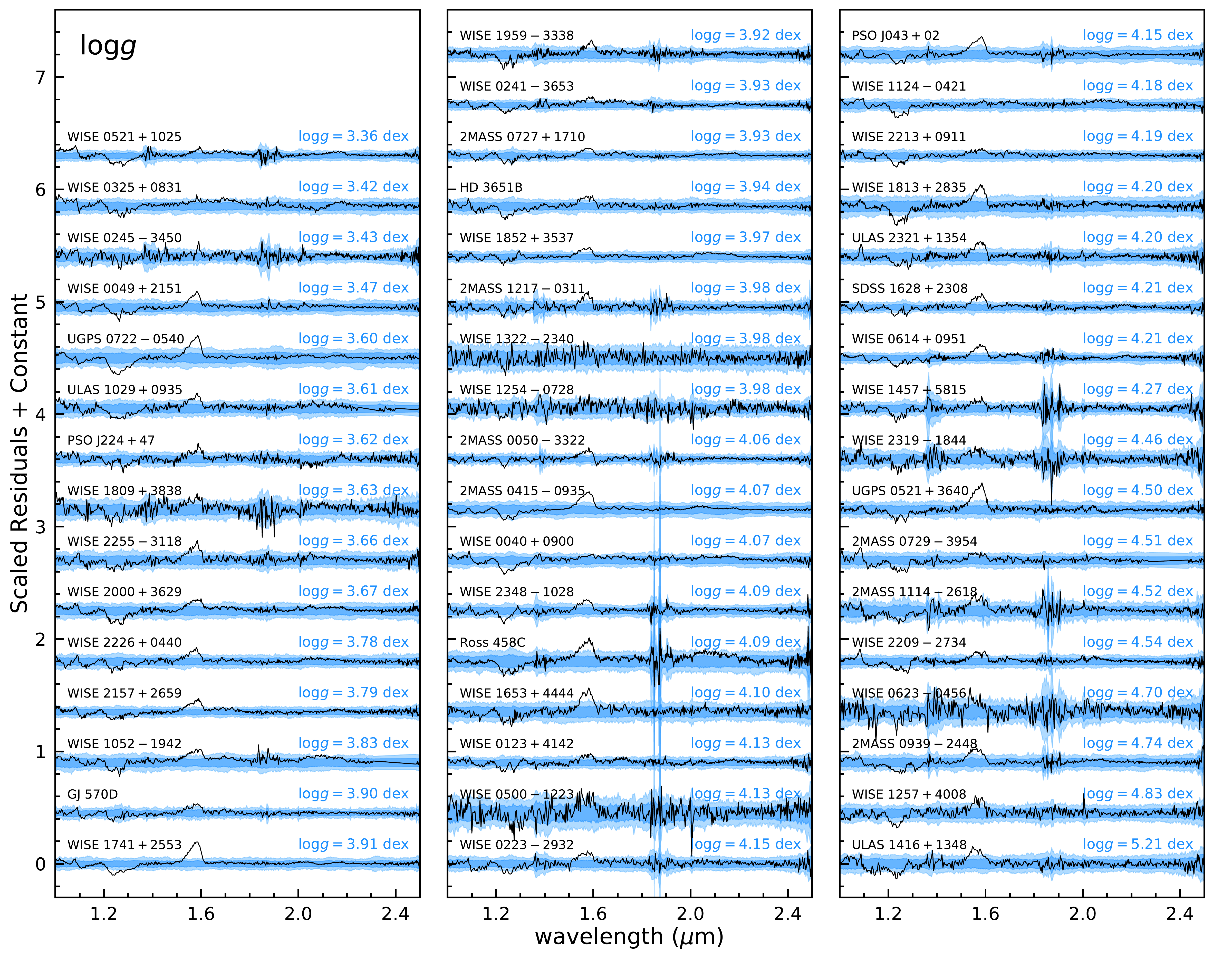}
\caption{Continued}
\label{fig:residual_afo_stgz}
\end{center}
\end{figure*}
\renewcommand{\thefigure}{\arabic{figure}}
\addtocounter{figure}{-1}
\begin{figure*}[t]
\begin{center}
\includegraphics[height=5.5in]{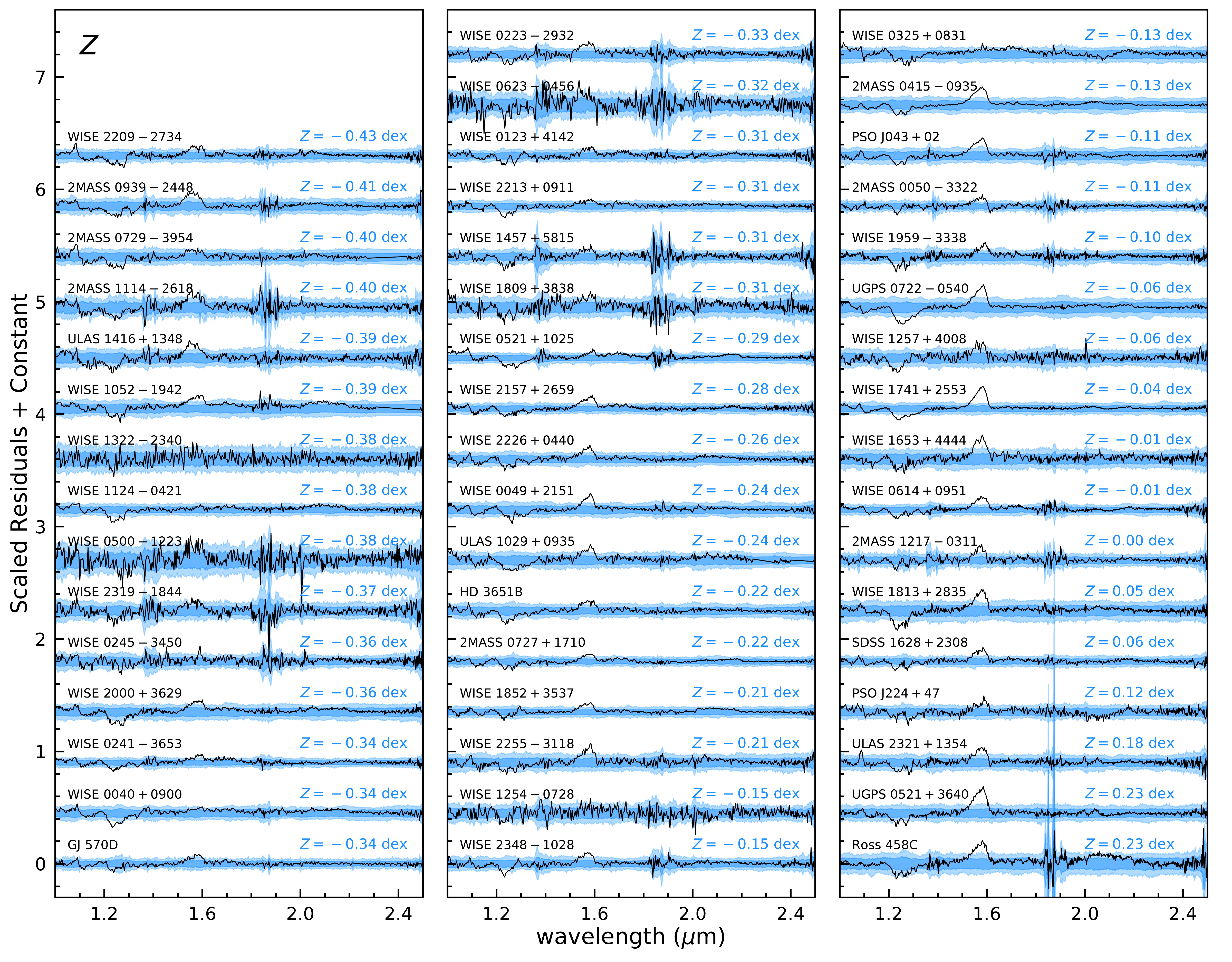}
\caption{Continued}
\label{fig:residual_afo_stgz}
\end{center}
\end{figure*}

\subsection{Starfish Covariance Hyper-Parameters}
\label{subsec:hyper_param}
As discussed in Paper~I, we can assess the systematic difference between data and models using the fitted covariance hyper-parameters $\{a_{N},\ a_{G},\ \ell\}$ of our late-T dwarfs. The parameter $\ell$ characterizes the autocorrelation wavelength of data$-$model residuals, caused by (1) the oversampled instrumental line spread function (LSF), and (2) the systematics of model assumptions. If the correlated residuals of an object are solely due to the instrumental LSF, then its fitted $\ell$ should be within $[820, 1840]$~km~s$^{-1}$ and $[425, 1115]$~km~s$^{-1}$ for spectra taken by the SpeX $0.5''$ and $0.8''$ slits, respectively (see Appendix of Paper~I). However, the spectroscopically inferred $\ell$ of almost all our late-T dwarfs are significantly higher than the range expected from the LSF, suggesting the correlated residuals of our sample do not arise from the instrument's spectral resolution but rather shortcomings of the model predictions (Figure~\ref{fig:ell}). Figure~\ref{fig:ell} also shows the inferred $\ell$ of our sample has no correlations with the objects' spectral types or fitted $\{T_{\rm eff},\ \log{g},\ Z\}$.

Given that the modeling systematics dominate the correlated residuals, we can use the other two hyper-parameters ($a_{N}$ and $a_{G}$) to quantify the modeling systematics. Specifically, we can compute $\sqrt{a_{G} / a_{N}}$ and regard it as an equivalent flux that describes the average ``model uncertainties'' (as opposed to measurement uncertainties). Normalizing this value by an object's peak $J$-band flux (Equation~7 of Paper~I), we define the normalized model uncertainty as $\epsilon_{J} = \sqrt{a_{G} / a_{N}} \big/ {\rm max}(f_{{\rm obs}, J})$, with higher values suggesting more significant data-model discrepancies. 

We summarize the derived $\epsilon_{J}$ of our sample in Table~\ref{tab:starfish_derived_results} and find the systematic difference between the cloudless Sonora-Bobcat models and late-T dwarf spectra is on average $\approx 2\%-4\%$ of the observed peak $J$-band fluxes over $1.0-2.5$~$\mu$m wavelengths\footnote{We note these values describe an average systematic difference between observed spectra and Sonora-Bobcat models over the $1.0-2.5$~$\mu$m range. As shown in Figure~\ref{fig:stacked_residual}, such systematic difference can be as high as $10\%-20\%$ of the objects' peak $J$-band fluxes over narrow wavelength ranges in $J$ and $H$ bands.}, equivalent to a S/N (the ratio of signal to model uncertainty) of $50-25$. This is in accord with our conclusion in Paper~I based on the three late-T benchmarks, which are also included in the analysis here. This result implies that model uncertainties exceed measurement uncertainties when fitting the cloudless Sonora-Bobcat models to late-T dwarf spectra with S/N $\gtrsim 50$ per pixel in $J$ band. As a consequence, increasing the S/N of observations does not necessarily improve the precision of the fitted physical parameters, as also seen in Figure~\ref{fig:comment_snr} and discussed in Section~\ref{subsubsec:comment_snr}. 

Figure~\ref{fig:epsJ} plots our objects' $\epsilon_{J}$ as a function of their spectral types and fitted physical parameters. We find the data-model difference tends to be larger toward later spectral types, cooler effective temperatures, and possibly higher $\log{g}$ and $Z$. These trends indicate some important atmospheric processes are likely missing in the model assumptions, which we discuss in the following section.

\subsection{Spectral-Fitting Residuals}
\label{subsec:stacked_residual}
Missing physical processes in model assumptions should leave footprints in our objects' spectral-fitting residuals. Thanks to our large sample of late-T dwarfs, we can investigate these residuals and assess how models deviate from the observations as a function of wavelength and atmospheric parameters. In Figure~\ref{fig:stacked_residual}, we normalize the fitting residuals for each object by its observed peak $J$-band flux and then stack them together. The resulting stack has a tight distribution as a function of wavelength and exhibits prominent features in $YJHK$ bands. 

In order to study how the residuals are correlated with atmospheric properties, we plot residuals sorted by the objects' spectral types and fitted $\{T_{\rm eff},\ \log{g},\ Z\}$ in Figure~\ref{fig:residual_afo_stgz}. We further define the following quantities $q_{Y}$, $q_{J}$, $q_{H}$, and $q_{K}$ to evaluate the data-model difference in $YJHK$ bands for each object (Table~\ref{tab:starfish_derived_results}),
\begin{equation} \label{eq:Rlambda}
\begin{aligned} 
q(\lambda\lambda) &= 1 - \ddfrac{\int_{\lambda\lambda} f_{{\rm model}, \lambda}\ d\lambda}{\int_{\lambda\lambda} f_{{\rm obs}, \lambda}\ d\lambda}   \\
{\rm such\ that} \quad q_{Y} &\equiv q([1.00\,\mu{\rm m}, 1.10\,\mu{\rm m}])   \\
\quad q_{J} &\equiv q([1.18\,\mu{\rm m}, 1.35\,\mu{\rm m}])   \\
\quad q_{H} &\equiv q([1.50\,\mu{\rm m}, 1.65\,\mu{\rm m}])   \\
\quad q_{K} &\equiv q([2.03\,\mu{\rm m}, 2.20\,\mu{\rm m}])   \\
\end{aligned} 
\end{equation}
Here $\lambda\lambda$ corresponds to the wavelength range of prominent $YJHK$-band features seen in Figure~\ref{fig:stacked_residual} and does not follow the standard definition of the filters. In addition, $f_{{\rm model}, \lambda}$ and $f_{{\rm obs}, \lambda}$ are the fitted model and the observed spectrum of the object, respectively. The sign of $q$ indicates whether the models under-predict (positive) or over-predict (negative) the data, with larger absolute values of $q$ indicating larger data-model discrepancies. By definition, model atmospheres that perfectly match the data will have $q = 0$ for all wavelength ranges.

In $Y$ band ($\approx 1.0-1.1$~$\mu$m), our fitted models slightly under-predict the spectra of most late-T dwarfs (Figure~\ref{fig:stacked_residual} and \ref{fig:residual_afo_stgz}). This is likely related to the potassium resonance doublet at $0.77$~$\mu$m \citep{1999PhRvA..60.1021A, 2000ApJ...531..438B}, whose pressure-broadened wings can extend to $Y$ and $J$ bands. The K line profile depends on the treatment of collisions between H$_{2}$ molecules and K atoms \citep[e.g.,][]{2017ApJ...850..150B, 2020A&A...637A..38P}. The cloudless Sonora-Bobcat models adopt the K line shape theory by \cite{2007A&A...465.1085A}. \cite{2016A&A...589A..21A} have improved calculations of the K$-$H$_{2}$ potential, which might predict more accurate shapes for the K~I doublet and thereby the $Y$-band fluxes \cite[e.g.,][]{2020A&A...637A..38P}. 

\begin{figure*}[t]
\begin{center}
\includegraphics[height=1.6in]{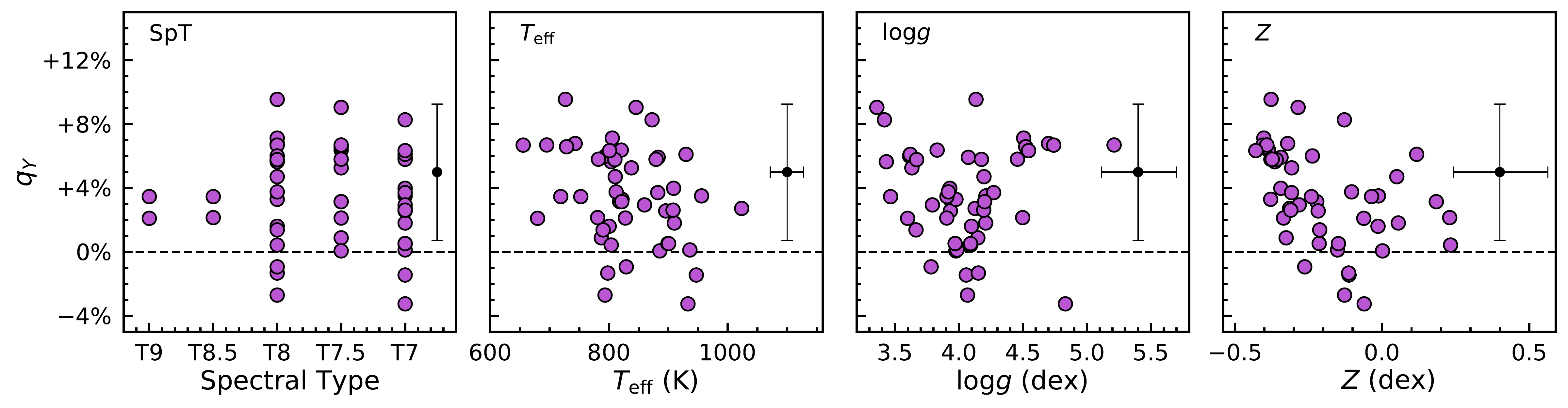}
\includegraphics[height=1.6in]{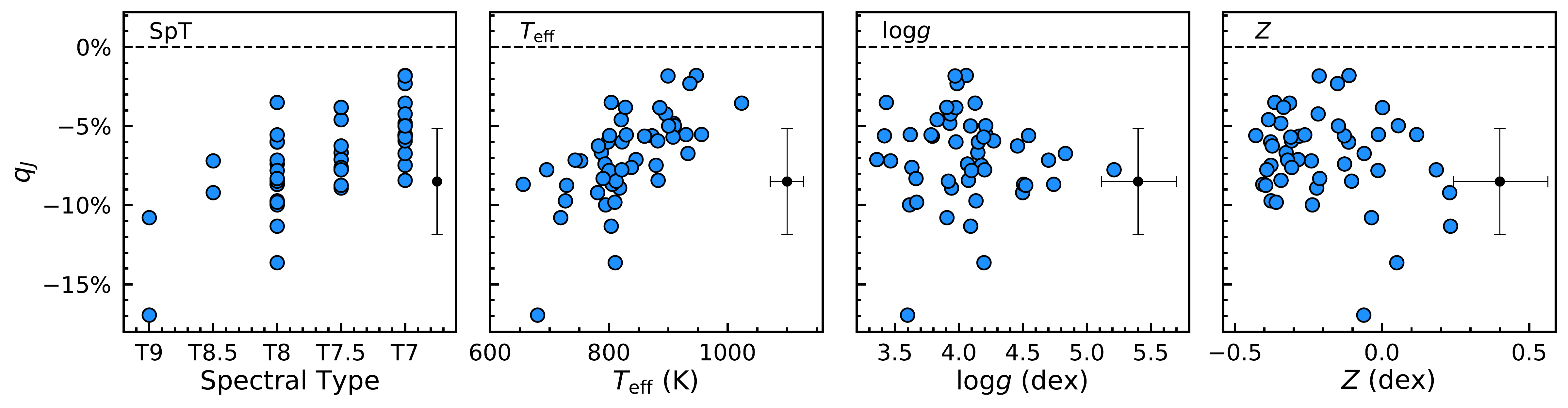}
\includegraphics[height=1.6in]{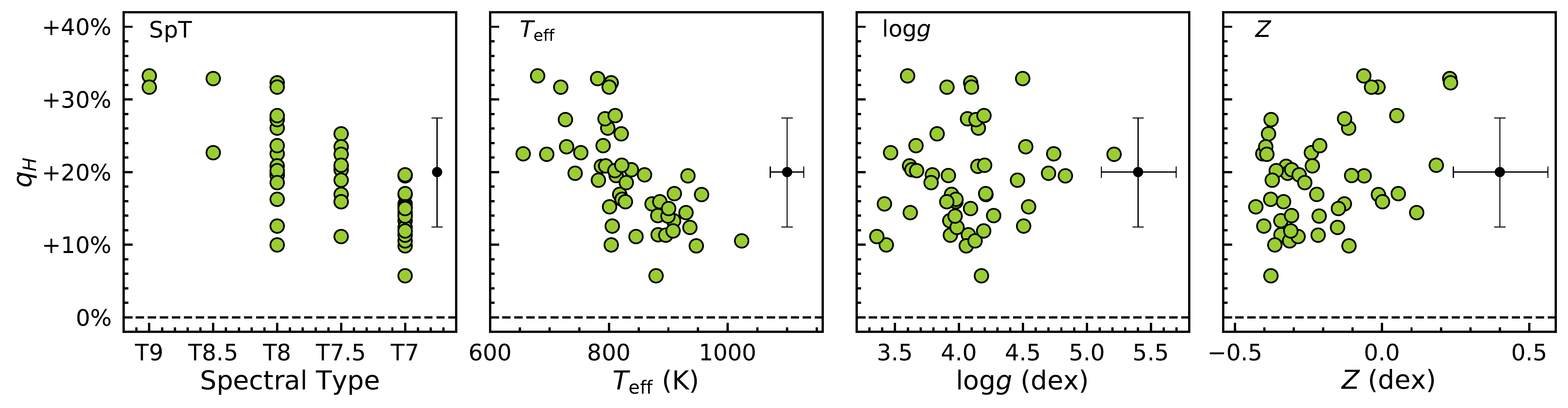}
\includegraphics[height=1.6in]{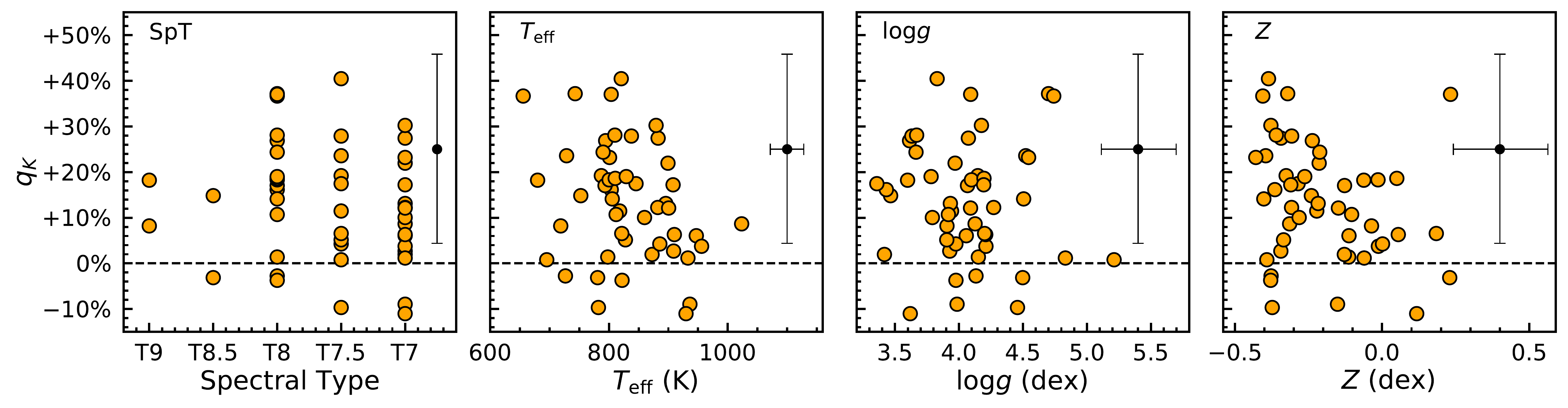}
\caption{The $q_{Y}$ (purple), $q_{J}$ (blue), $q_{H}$ (green), and $q_{K}$ (orange) of our 49 late-T dwarfs (binaries removed) as a function of spectral type and atmospheric $\{T_{\rm eff},\ \log{g},\ Z\}$, with median uncertainties shown by black error bars. The $q$ values (Equation~\ref{eq:Rlambda}) quantify the fitted residuals of the object in $YJHK$ bands. The sign of $q$ indicates whether the models under-predict (positive) or over-predict (negative) the observed spectra over a given band pass, with larger absolute values of $q$ indicating larger data-model discrepancies. By definition, perfect models should have $q = 0$ at all wavelengths (dashed horizontal lines). }
\label{fig:R_YJHK}
\end{center}
\end{figure*}

In $J$ band ($\approx 1.18-1.35$~$\mu$m), our fitted models over-predict spectra of all late-T dwarfs, and we find $q_{J}$ is correlated with spectral type and $T_{\rm eff}$ (Figure~\ref{fig:R_YJHK}), as the $J$-band residuals increase for later-type and cooler objects. This residual plausibly arises from clouds \citep[e.g.,][]{2012ApJ...756..172M}, given that fluxes at near-infrared spectral peaks are emitted from the deep atmosphere and thus more sensitive to cloud opacity.  Including clouds into Sonora-Bobcat models might therefore produce spectra that better match the observations. Also, with lower effective temperatures, condensates of various species are expected to form \citep[e.g., Na$_{2}$S, KCl, MnS, ZnS;][]{1999ApJ...519..793L, 2012ApJ...756..172M}, leading to a larger $J$-band discrepancy between data and the cloudless models. This is consistent with the $q_{J} - T_{\rm eff}$ correlation in Figure~\ref{fig:R_YJHK}.

Alternatively, the over-predicted $J$-band flux of the cloudless Sonora-Bobcat models might be related to the assumption that the deep temperature gradient in convective regions lies along an adiabatic. As demonstrated by \cite{2015ApJ...804L..17T, 2019ApJ...876..144T}, cloudless dis-equilibrium models with (1) atmospheric mixing (described by the eddy diffusion coefficients $K_{zz}$) and (2) reduced vertical temperature gradient (as compared to the adiabatic lapse rate) are more appropriate for spectra of T/Y dwarfs than cloudless, chemical-equilibrium models. While these models can explain the $J$-band residuals of our sample, they will need to explain the $q_{J}-$SpT and $q_{J}-T_{\rm eff}$ correlations as seen in Figure~\ref{fig:R_YJHK}. One implication from these correlations is that the thermo-chemical instability in the \cite{2015ApJ...804L..17T} models should be more significant with later spectral types and cooler $T_{\rm eff}$ for late-T dwarfs. To examine this hypothesis, further investigation is needed to study such instability in brown dwarf atmospheres as a function of physical properties.

In $H$ band ($\approx 1.50-1.65$~$\mu$m), our fitted models under-predict spectra of all late-T dwarfs, and we find the residuals ($q_{H}$) become more significant with later spectral types, cooler effective temperatures, and higher metallicities (Figures~\ref{fig:R_YJHK}). These can be related to the dis-equilibrium abundance of NH$_{3}$, which would be less than the amount assumed by the equilibrium chemistry within the cloudless Sonora-Bobcat models and thus lead to weaker absorption in the blue wing of $H$ band \citep[][]{2006ApJ...647..552S, 2012ApJ...750...74S, 2011ApJ...743...50C, 2014ApJ...797...41Z}. However, the NH$_{3}$ cross-section is peaked around $1.5$~$\mu$m \citep[or wavenumbers of $6600-6700$~cm$^{-1}$; e.g.,][]{2011MNRAS.413.1828Y, 2019MNRAS.490.4638C}, slightly offset from the $H$-band residuals of our late-T dwarfs which are peaked near $1.58 $~$\mu$m. Alternatively, the under-predicted $H$-band spectra could well be a consequence of the spectral-fitting procedure responding to the over-predicted $J$-band flux by choosing models that under-predict the $H$-band flux. Studying our late-T dwarf spectra using (1) cloudless models with dis-equilibrium chemistry or (2) cloudy models with equilibrium chemistry will help better understand this $H$-band residual.

In $K$ band ($\approx 2.03-2.20$~$\mu$m), our fitted models only slightly under-predict spectra of most late-T dwarfs. Similar to $J$ band, such $K$-band residuals may arise from clouds and/or reductions in the vertical temperature gradient. As shown by \citeauthor{2012ApJ...756..172M} (\citeyear{2012ApJ...756..172M}; their Figure~11) and \citeauthor{2015ApJ...804L..17T} (\citeyear{2015ApJ...804L..17T}; their Figure~1), either of these two processes can reduce fluxes in $YJ$ bands and increase the flux in $K$ band.

\section{Summary and Future Work}
\label{sec:summary}
We have conducted a forward-modeling analysis for 55 late-T (T7$-$T9) dwarfs using low-resolution ($R \approx 50-250$) near-infrared ($1.0-2.5$~$\mu$m) spectra and state-of-art, cloudless Sonora-Bobcat model atmospheres with $T_{\rm eff} = 600-1200$~K, $\log{g} = 3.25-5.5$~dex, and $Z = \{-0.5, 0, +0.5\}$~dex. Our sample contains $90\%$ of the nearby T7$-$T9 population with distances $\leqslant 25$~pc, $J$-band magnitudes $\leqslant 17.5$~mag, and declinations from $-40^{\circ}$ to $+70^{\circ}$. Our work is the largest analysis of brown dwarf spectra using multi-metallicity models to date, as well as the most systematic test of any set of ultracool model atmospheres. Our forward-modeling framework was constructed and validated in Paper~I, which uses the Bayesian inference tool Starfish \citep[][]{2015ApJ...812..128C}. Compared to traditional forward-modeling studies, our analysis produces more realistic error estimates since we account for uncertainties from model interpolation and correlated residuals due to instrumental effects and systematics of model assumptions. 

We have inferred effective temperatures ($T_{\rm eff}$), surface gravities ($\log{g}$), metallicities ($Z$), radii ($R$), masses ($M$), and bolometric luminosities ($L_{\rm bol}$) for our late-T dwarfs, with the typical resulting $\{T_{\rm eff},\ \log{g},\ Z\}$ uncertainties being $\approx 1/3-1/2$ of the Sonora-Bobcat model grid spacing. We found no difference in the precision of these physical parameters for spectra with two different spectral resolutions ($R \approx 80-250$ and $50-160$). Combining the resulting parameter posteriors of our entire sample, we found some fitted parameters are correlated, including $T_{\rm eff}$ and $R$, $R$ and $M$, and $\log{g}$ and $Z$. Correlations within the first two pairs are expected by the Stephen-Boltzmann law and the calculation of mass from spectroscopic parameters, respectively. The third correlation illustrates that $\log{g}$ and $Z$ are degenerate in the cloudless Sonora-Bobcat models, and we provide a quantification of this $\log{g}-Z$ dependence. For late-T dwarfs, the degeneracy acts such that an increase in $Z$, combined with a $3.4\times$ larger increase in $\log{g}$, results in a spectrum that has similar fitted atmospheric parameters. Consequently, using solar-metallicity model atmospheres to study late-T dwarfs whose metallicities are in fact non-solar will bias the inferred $\log{g}$, but likely not other parameters.

By virtue of having a large sample of spectra within a focused spectral type range (as opposed to a smaller sample of objects over a wide spectral type range), we can study population properties of T7--T9 dwarfs to provide useful diagnostics about our set of grid models, which assume cloudless and chemical-equilibrium atmospheres:
\begin{enumerate}
\item The spectroscopically inferred metallicities of our entire sample and a volume-limited subset within 25~pc are $0.3-0.4$~dex lower than those of nearby FGKM stars \citep[e.g.,][]{2014AJ....148...54H}. This significant discrepancy is unlikely to be a real difference between the substellar and stellar populations, but rather because our fitted metallicities are underestimated, which is also seen from our Paper~I analysis of the late-T benchmarks HD~3651B and GJ~570D.
\item The inferred ages of our sample, based on their fitted $\{T_{\rm eff},\ \log{g},\ Z\}$ and evolutionary models, have a median of $50$~Myr and a $1\sigma$ confidence interval of $10$~Myr--$0.4$~Gyr. These values are implausibly younger than the robustly determined ages of nearby M8--T5 binaries \citep[][]{2017ApJS..231...15D} and thus are likely underestimated.
\item The spectroscopically inferred effective temperatures of our sample show a similar spread ($\pm 100$~K) at a given spectral type as compared to empirical effective temperature scales, but our $T_{\rm eff}$ appear systematically hotter (by $50-200$~K) for $\geqslant$T8 dwarfs. Also, our derived $T_{\rm eff}-$SpT relation for late-T dwarfs is weakly correlated with fitted $\log{g}$ and $Z$, as objects with either lower $\log{g}$ or higher $Z$ have on average hotter $T_{\rm eff}$ at a given spectral type. The possible gravity and metallicity dependence seen in this work might be caused by the over- and/or under-estimated physical parameters from spectral fitting, but it should be further validated using more late-T benchmarks with diverse ages and metallicities.
\item The spectroscopically inferred masses of our sample are unphysically small (mostly $1-8$~M$_{\rm Jup}$), due to the underestimate of their fitted $\log{g}$ and/or $R$. 
\end{enumerate}

Using the hyper-parameters from our spectral-fitting results, we quantified that the systematic difference between the observed late-T dwarf spectra and the Sonora-Bobcat models is on average $\approx 2\%-4\%$ of the objects' peak $J$-band fluxes over the $1.0-2.5$~$\mu$m range (as high as $10-20\%$ of the objects' peak $J$-band fluxes over narrow wavelength ranges in $J$ and $H$ bands), equivalent to a S/N of $50-25$. Therefore, model uncertainties exceed measurement uncertainties when fitting the Sonora-Bobcat models to late-T dwarf spectra with S/N higher than these values. This can also explain why the fitted parameter precision of our sample does not improve with increasing S/N once it is above $\approx 50$ per pixel in $J$ band.

In order to investigate how to improve model assumptions, we stacked the spectral-fitting residuals of our entire sample and investigated these as a function of wavelength and the inferred atmospheric properties. We found common, prominent residual features in $YJHK$ bands: 
\begin{enumerate}
\item In $Y$ band, the cloudless Sonora-Bobcat models tend to under-predict the observed fluxes, which is likely related to the potassium line profiles. Further improvements of the alkali opacities might help reduce this residual \citep[e.g.,][]{2016A&A...589A..21A}.
\item In $J$ band, the cloudless Sonora-Bobcat models tend to over-predict the observed fluxes by an amount which is larger for later spectral types and cooler $T_{\rm eff}$. This effect is likely caused by missing opacity from clouds \citep[e.g.,][]{2012ApJ...756..172M} or the model assumption that the deep temperature gradient in convective regions lies along an adiabat \citep[e.g.,][]{2015ApJ...804L..17T}. Including clouds or assuming a reduced temperature gradient could result in spectra that better match the observations.
\item In $H$ band, the cloudless Sonora-Bobcat models tend to under-predict the observed fluxes by an amount which is larger for later spectral types, cooler $T_{\rm eff}$, and higher $Z$. This residual might be explained by the dis-equilibrium abundance of NH$_{3}$ \citep[e.g.,][]{2006ApJ...647..552S, 2011ApJ...743...50C} which is not included by the Sonora-Bobcat assumption of equilibrium chemistry, although our residuals do not seem to precisely coincide with the expected wavelength for NH$_{3}$ opacities. Also, this $H$-band residual could merely be a consequence of the spectral-fitting procedure responding to the over-predicted $J$-band flux.
\item In $K$ band, the cloudless Sonora-Bobcat models tend to under-predict the observed fluxes, which likely arises from the same reason that cause the $J$-band residuals.
\end{enumerate}

In future work, spectroscopic analysis of late-T dwarfs will benefit from models that include improved opacities, clouds, reduced vertical temperature gradient, and/or chemical dis-equilibrium. Such will also be essential to studying ultracool dwarfs with higher or lower effective temperatures than late-T dwarfs, for which there are already numerous spectra. {\it James Webb Space Telescope} can further extend spectroscopic observations down to cooler temperatures and wider wavelength coverage.  In addition, the analysis conducted in our work can also be extended to other sets of grid models to verify whether the physical assumptions made by those models can reproduce the observations, and what atmospheric processes might be included to improve data-model consistency. Since brown dwarfs harbor similar atmospheric processes as imaged exoplanets, models employed to interpret both classes of objects are generated from the same theoretical framework. Dedicated analyses of brown dwarf atmospheres will therefore help us to robustly characterize direct spectroscopy of exoplanets and thereby understand their appearance, formation, and evolution.

\acknowledgments
We thank Mark Phillips, Didier Saumon, Caroline Morley, Eugene Magnier, Paul Molli\`{e}re, Joe Zalesky, Trent Dupuy, Ehsan Gharib-Nezhad for insightful discussions and comments on this work. We thank Ian Czekala, Michael Gully-Santiago, and Miles Lucas for helpful discussions about Starfish. We also thank Michael Gully-Santiago for implementing Starfish for IRTF/SpeX prism data and sharing initial work on spectroscopic analysis for T dwarfs (\url{https://github.com/gully/jammer-Gl570D}). We thank Trent Dupuy for providing age posteriors of brown dwarf binaries with dynamical mass measurements. This work has benefited from The UltracoolSheet at \url{http://bit.ly/UltracoolSheet}, maintained by Will Best, Trent Dupuy, Michael Liu, Rob Siverd, and Zhoujian Zhang, and developed from compilations by \cite{2012ApJS..201...19D}, \cite{2013Sci...341.1492D}, \cite{2016ApJ...833...96L}, \cite{2018ApJS..234....1B}, and \cite{2021AJ....161...42B}. This work benefited from 2017--2019 Exoplanet Summer Program in the Other Worlds Laboratory (OWL) at the University of California, Santa Cruz, a program funded by the Heising-Simons Foundation. M.C.L. acknowledges the National Science Foundation (NSF) grant AST-1518339. This work is funded in part by the Gordon and Betty Moore Foundation through Grant GBMF8550 to M.C.L. The advanced computing resources from the University of Hawaii Information Technology Services -- Cyberinfrastructure are greatly acknowledged and Z. Z. thanks the technical support received from Curt Dodds. This research was greatly facilitated by the TOPCAT software written by Mark Taylor (http://www.starlink.ac.uk/topcat/). Finally, the authors wish to recognize and acknowledge the very significant cultural role and reverence that the summit of Maunakea has always had within the indigenous Hawaiian community.  We are most fortunate to have the opportunity to conduct observations from this mountain.

\facilities{IRTF (SpeX)}

\software{Starfish \citep{2015ApJ...812..128C}, {\it emcee} \citep[][]{2013PASP..125..306F}, Spextool \citep[][]{2004PASP..116..362C}, TOPCAT \citep[][]{2005ASPC..347...29T}, Astropy \citep{2013A&A...558A..33A, 2018AJ....156..123A}, IPython \citep{PER-GRA:2007}, Numpy \citep{numpy},  Scipy \citep{scipy}, Matplotlib \citep{Hunter:2007}.}

\appendix

\section{Bolometric Luminosities: Comparison with Literature}
\label{app:compare_lbol}
Among our sample, 15 late-T dwarfs have bolometric luminosities in the literature. \cite{2015ApJ...810..158F} constructed SEDs for 14 of our T7$-$T9 dwarfs by combining optical, near-infrared, and mid-infrared photometry and spectroscopy, and then computed bolometric luminosities by integrating their SEDs, with short-wavelength fluxes linearly extrapolated to zero wavelength and long-wavelength fluxes approximated by a Rayleigh-Jeans tail. \cite{2017ApJ...848...83L} performed a retrieval analysis for SpeX prism spectra of 11 of our T7$-$T8 dwarfs and computed $L_{\rm bol}$ by integrating the fitted models. Six objects in these two studies now have newer parallaxes with much higher precisions: HD~3651B, PSO~J043.5395+02.3995, 2MASS~J1217110$-$031113, Ross~458C, ULAS~1416+1348, and GJ~570D. We thereby scale these objects' literature $L_{\rm bol}$ values with the new parallaxes, without modifying the published uncertainties.

\begin{figure*}[t]
\begin{center}
\includegraphics[height=4.in]{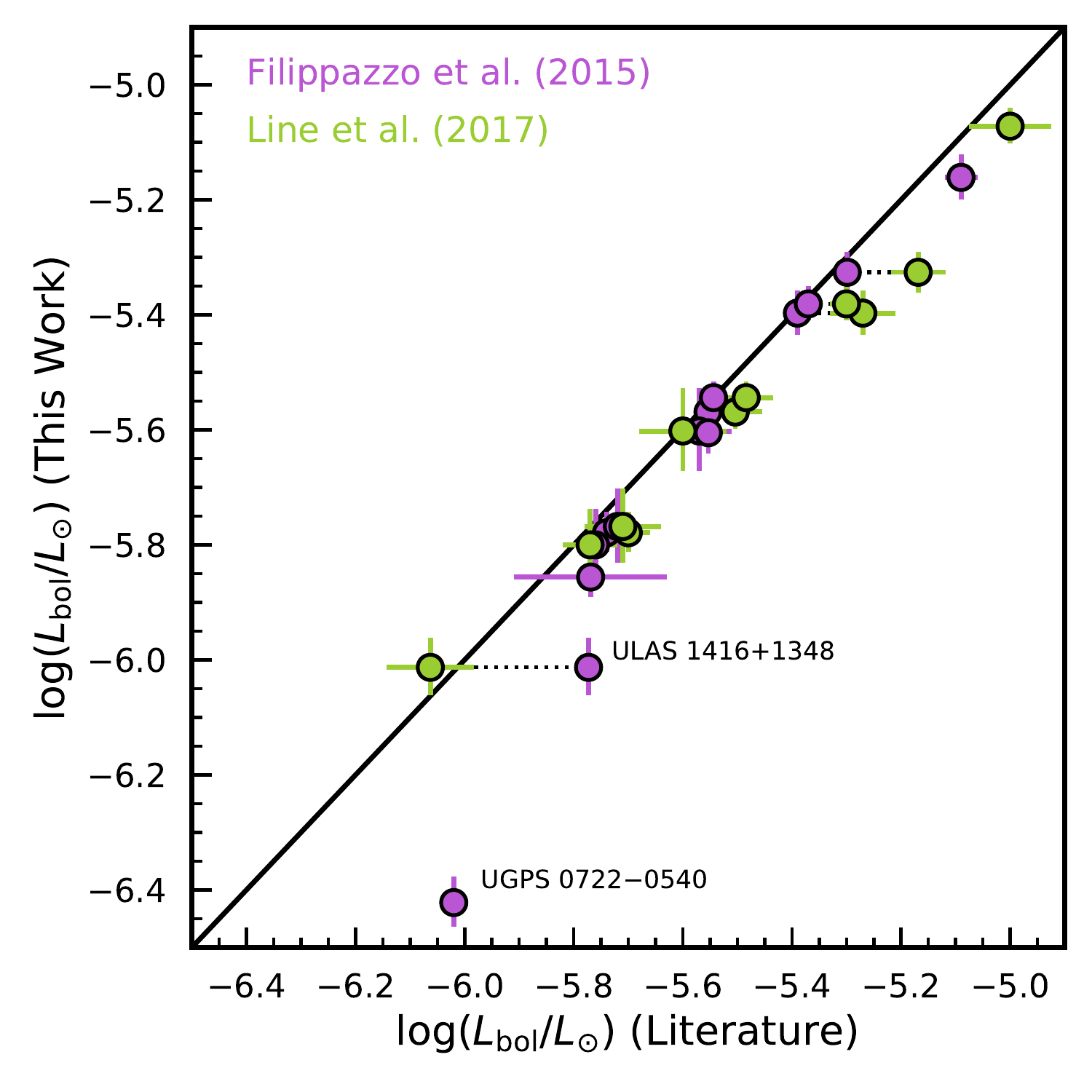}
\caption{Comparison of bolometric luminosities for 15 late-T dwarfs with measurements both by us and the literature (Appendix~\ref{app:compare_lbol}). We use dotted lines to connect results for same object. Our computed $L_{\rm bol}$'s of these objects are generally consistent with the literature values but are systematically fainter by $0.06-0.07$~dex.}
\label{fig:compare_lbol}
\end{center}
\end{figure*}

Figure~\ref{fig:compare_lbol} and Table~\ref{tab:compare_lbol} compare our $L_{\rm bol}$ values for these 15 late-T dwarfs with the literature and suggest a general consistency. Our work produces slightly fainter $L_{\rm bol}$, with the differences having a weighted mean and weighted root mean square of $0.057 \pm 0.099$~dex compared to \cite{2015ApJ...810..158F} and $0.075 \pm 0.044$~dex compared to \cite{2017ApJ...848...83L}. For UGPS~J072227.51$-$054031.2 (UGPS~0722$-$0540), our estimated $L_{\rm bol}$ is fainter than the value in \cite{2015ApJ...810..158F} by $8.4\sigma$. As shown in Figure~\ref{fig:emu_results}, our fitted cloudless models do not well match this object's observed spectrum and our spectroscopically inferred $T_{\rm eff} = 680 \pm 26$~K is much higher than the values of $500-580$~K as suggested by \cite{2012ApJ...756..172M} and \cite{2015ApJ...804L..17T}, whose models account for sulfide clouds or diabatic convection (also see Section~\ref{subsec:stacked_residual}) and can better match this object's spectrum. Our over-estimated $T_{\rm eff}$ for UGPS~0722$-$0540 might lead to a smaller ratio between the integrated fluxes in mid-infrared wavelengths (approximated by our fitted models) and in near-infrared wavelengths (indicated by our SpeX prism data), resulting in an under-estimated $L_{\rm bol}$. For ULAS~1416+1348, our derived $L_{\rm bol} = -6.013^{+0.052}_{-0.049}$~dex is $4.7\sigma$ fainter than the value in \cite{2015ApJ...810..158F}, but is consistent with that in \cite{2017ApJ...848...83L}.

\section{Bolometric Corrections}
\label{app:bc}
Table~\ref{tab:bc} presents our bolometric corrections for late-T dwarfs in $J_{\rm MKO}$, $H_{\rm MKO}$, and $K_{\rm MKO}$ bands computed using their bolometric fluxes (with $M_{\rm bol,\odot} = +4.74$). For this calculation, we exclude six objects that are resolved or candidate binaries, as well as two objects with peculiar spectra (2MASS~J0939$-$2448 and ULAS~J1416+1348; see Section~\ref{sec:binaries}). Comparison with the bolometric corrections of \cite{2010ApJ...722..311L} demonstrates the $1\sigma$ consistency. \cite{2015ApJ...810..158F} provided bolometric corrections using 2MASS bands, so we cannot compare directly to their results. Following \cite{2012ApJS..201...19D}, we also compute an estimate of the intrinsic (astrophysical) scatter in the bolometric correction at a given spectral type and band.

\begin{figure*}[t]
\begin{center}
\includegraphics[height=2.3in]{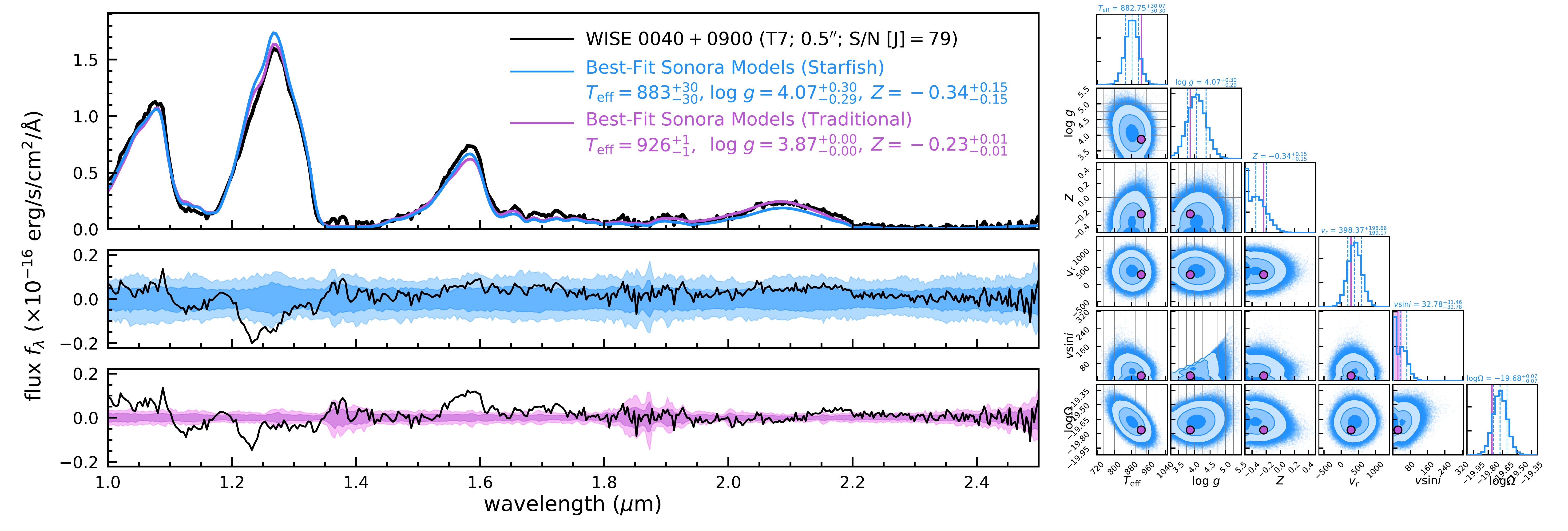}
\caption{Comparisons of results from Starfish and traditional forward-modeling analyses, with a similar format as Figure~\ref{fig:emu_results}. Left: The upper panel shows the observed spectrum (black) and the median model spectra of those interpolated at parameters drawn from the MCMC chains based on the Starfish (blue) and traditional (purple) methods. The middle and lower panel show the residual of each method (data$-$model; black). Right: Posteriors of the six physical parameters $\{T_{\rm eff},\ \log{g},\ Z,\ v_{r},\ v\sin{i},\ \log{\Omega}\}$ derived from the Starfish-based forward-modeling analysis (blue). We overlay the median values and uncertainties from the traditional method (purple), shown as vertical lines and shadows in the 1-D histograms and as circles and error bars in the 2-D histograms. We use grey vertical and horizontal lines to mark the $\{T_{\rm eff},\ \log{g},\ Z\}$ grids pionts of the cloudless Sonora-Bobcat models. Figures of spectral-fitting results for our entire sample (55 late-T dwarfs with 57 spectra) are accessible online. }
\label{fig:emulin_results}
\end{center}
\end{figure*}

\section{Forward-Modeling Analysis with the Traditional Approach}
\label{app:traditional}
We also conduct the forward-modeling analysis for our late-T dwarfs following the traditional approach described in Paper~I. We use the cloudless Sonora-Bobcat model atmospheres over their entire parameter space of $[200, 2400]$~K in $T_{\rm eff}$, $[3.25, 5.5]$~dex in $\log{g}$, and $[-0.5, +0.5]$~dex in $Z$, and use linear interpolation to synthesize model spectra with an arbitrary set of grid parameters. We determine six physical parameters $\{T_{\rm eff},\ \log{g},\ Z,\ v_{r},\ v\sin{i},\ \log{\Omega}\}$ with same priors as our Starfish-based analysis (Section~\ref{subsec:method}), and construct our covariance matrix by simply placing squared flux uncertainties of spectra along its diagonal axis.

We use {\it emcee} to fit our $1.0-2.5$~$\mu$m spectra with 24 walkers and terminate the fitting process with $3 \times 10^{4}$ iterations given that such number of iterations exceeds 50 times the autocorrelation length of all the fitted parameters. We incorporate the systematic error of $180$~${\rm km~s}^{-1}$ into the inferred radial velocity to account for the uncertainty in the wavelength calibration of the SpeX prism data (Section~\ref{subsec:method}). We also incorporate the systematic error of $0.4\sigma_{H_{\rm MKO}}$ into the inferred $\log{\Omega}$ to account for the uncertainty in flux calibration (Section~\ref{subsec:method}), where $\sigma_{H_{\rm MKO}}$ is the photometric error in an object's $H$-band magnitude. We compute the objects' radii ($R$) and masses ($M$) using their parallaxes and $\log{g}$ and $\log{\Omega}$ posteriors and also compute ages by interpolating the Sonora-Bobcat evolutionary models using the fitted $\{T_{\rm eff},\ \log{g},\ Z\}$. We summarize all inferred parameters and their uncertainties in Table~\ref{tab:traditional_results}.

\begin{figure*}[t]
\begin{center}
\includegraphics[height=5.3in]{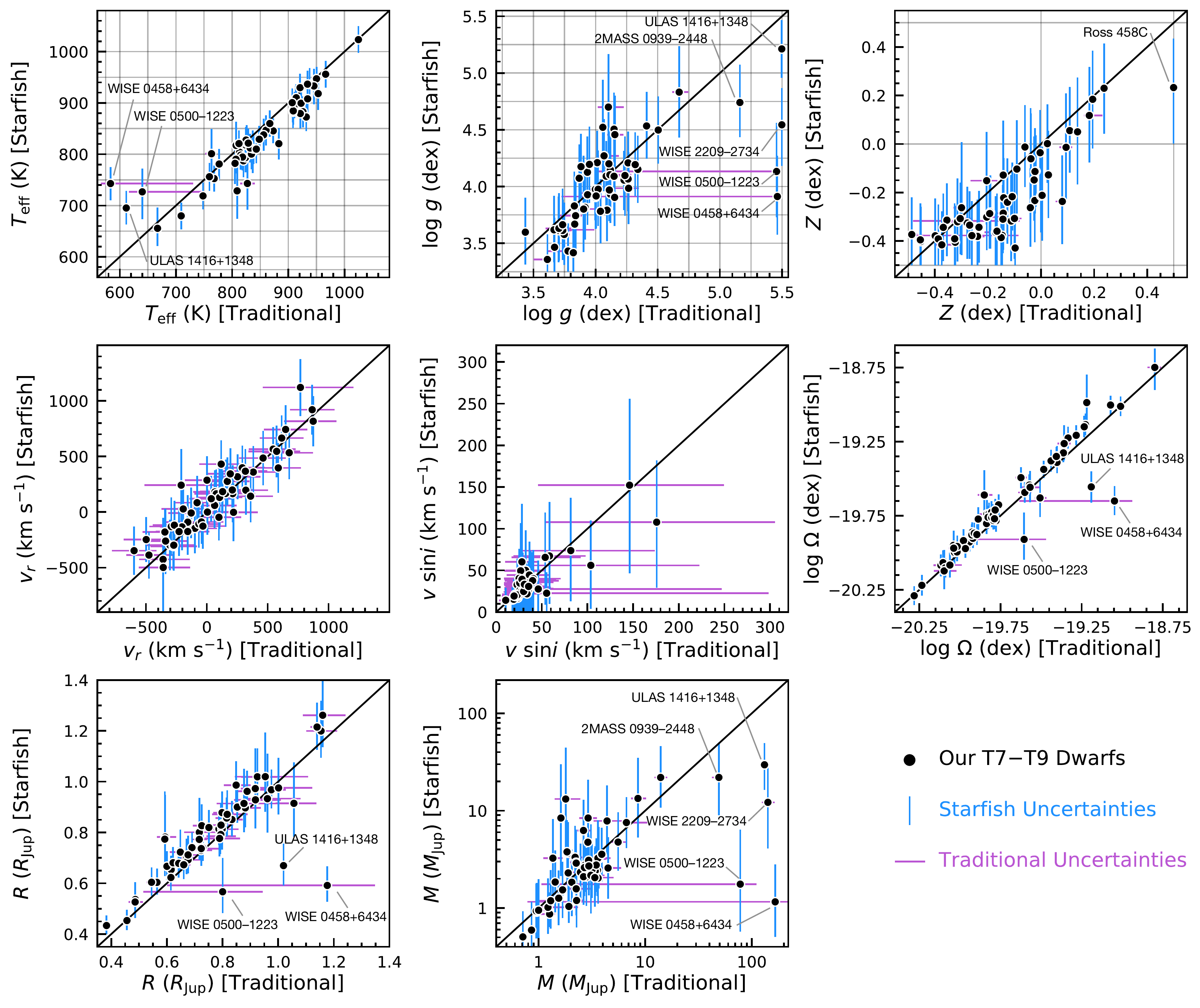}
\caption{Comparison of eight physical parameters of our late-T dwarfs $\{T_{\rm eff},\ \log{g},\ Z,\ v_{r},\ v\sin{i},\ \log{\Omega},\ R,\ M\}$ as inferred from Starfish and traditional forward-modeling analysis (masses are compared in the logarithmic scale). We use grey vertical and horizontal lines to mark the $\{T_{\rm eff},\ \log{g},\ Z\}$ grids of the cloudless Sonora-Bobcat models. The results from the two methods are generally consistent within the uncertainties, although there are systematic differences in almost all parameters except for $v_{r}$ and $v\sin i$. We label the objects that have significantly different parameters from the two methods and discuss them in Appendix~\ref{app:traditional}. }
\label{fig:emulin_compare}
\end{center}
\end{figure*}

Figure~\ref{fig:emulin_results} compares the parameter posteriors and the fitted model spectra (interpolated at parameters drawn from the MCMC samples) derived from the traditional approach to those from the Starfish analysis. While the fitted model spectra by these two methods both match the data, the residuals from the traditional approach significantly exceed measurement uncertainties in some wavelength ranges, indicating that the more sophisticated Starfish covariance matrix better describes the difference between the data and models \citep[also see Figure~5 of][]{2015ApJ...812..128C}. Also, the fitted models from the two methods are not always identical at several wavelengths, primarily because of their vastly different covariance matrices, which weight residuals and compute the likelihoods differently given the same set of physical parameters.

Figure~\ref{fig:emulin_compare} compares eight physical parameters $\{T_{\rm eff},\ \log{g},\ Z,\ v_{r},\ v\sin{i},\ \log{\Omega},\ R,\ M\}$ of late-T dwarfs inferred from the two approaches. The weighted mean and weighted root mean square of the Starfish$-$Traditional parameter differences are $-16 \pm 23$~K for $T_{\rm eff}$, $-0.09 \pm 0.23$~dex for $\log{g}$, $-0.09 \pm 0.10$~dex for $Z$, $45 \pm 125$~km~s$^{-1}$ for $v_{r}$, $-1 \pm 6$~km~s$^{-1}$ for $v\sin{i}$, $0.040 \pm 0.061$~dex for $\log{\Omega}$, $0.03 \pm 0.06$~dex for $R$, and $-0.17 \pm 1.14$~dex for $M$. Therefore, the results from these two spectral-fitting methods are generally consistent. Several objects have significantly different properties based on the two methods: WISE~$2209-2734$, ULAS~$1416+1348$, WISE~$0500-1223$, 2MASS~$0939-2448$, and WISE~$0458+6434$. 

For WISE~$2209-2734$, the Starfish analysis derives $\log{g}$ and $Z$ that are smaller than the traditional method by $0.95 \pm 0.34$~dex and $0.33 \pm 0.10$~dex, respectively. However, the fitted model spectra from the two methods are similar, given that these parameter offsets follow the $\log{g}-Z$ degeneracy of the cloudless Sonora-Bobcat models that we find based on the stacked posteriors of our late-T dwarf sample (Section~\ref{subsubsec:param_corr}). The different results from the two methods are thus likely caused by the different covariance matrices used to compute the likelihood. 

ULAS~$1416+1348$ is a wide-orbit companion to a low-metallicity late-L dwarf (see Section~\ref{subsec:candidate}). Its $\{T_{\rm eff},\ \log{g},\ \log{\Omega},\ R,\ M\}$ are different between the two methods by $1-6\sigma$, but we find the fitted model spectra from Starfish better match the data especially for the blue wing of $Y$ band (Figure~\ref{fig:emulin_results}). 

Among the remaining three objects, WISE~$0458+6434$ is a resolved $0.51''$ T8.5$+$T9 binary (Section~\ref{subsec:resolved}). Binarity can result in spectral peculiarity and cause the two forward-modeling approaches to derive different parameters. The binarity of WISE~$0500-1223$ and 2MASS~$0939-2448$ are uncertain (Section~\ref{subsec:candidate}). In addition, parameter posteriors of WISE~$0500-1223$ inferred from the traditional method show two peaks, with one peak consistent with the Starfish results. This anomaly might also be related to the low S/N of data ($\approx 13$ in $J$ band).

Table~\ref{tab:typical_error} compares typical parameter uncertainties between the two methods. The traditional forward-modeling approach produces artificially small parameter errors by factors of $2-15$ in $\{T_{\rm eff},\ \log{g},\ Z,\ \log{\Omega},\ R,\ M\}$ and factors of $1-2$ in $v_{r}$ and $v\sin i$ (also see Figure~\ref{fig:comment_snr}). We note that the larger error estimates from Starfish are more realistic, given that Starfish accounts for uncertainties from model interpolation and correlated residuals.

\end{CJK*}

\clearpage
\bibliographystyle{aasjournal}
\bibliography{ms}


\clearpage
\tabletypesize{\scriptsize}

\global\pdfpageattr\expandafter{\the\pdfpageattr/Rotate 90} 
\begin{longrotatetable} 
\begin{deluxetable}{lcccccccccccc} 
\tablewidth{0pc} 
\setlength{\tabcolsep}{0.05in} 
\tablecaption{Sample of T7$-$T9 Dwarfs: Spectroscopy \label{tab:spectroscopy}} 
\tablehead{ \multicolumn{1}{l}{}  &  \multicolumn{1}{c}{}  &  \multicolumn{1}{c}{}  &  \multicolumn{1}{c}{}  &  \multicolumn{4}{c}{IRTF/SpeX Spectra}  &  \multicolumn{1}{c}{}  &  \multicolumn{4}{c}{References}   \\ 
\cline{5-8} \cline{10-13} 
\multicolumn{1}{l}{Object}  &  \multicolumn{1}{c}{Type\tablenotemark{a}}  &  \multicolumn{1}{c}{SpT}  &  \multicolumn{1}{c}{}  &  \multicolumn{1}{c}{Slit}  &  \multicolumn{1}{c}{$J$-band S/N\tablenotemark{b}}  &  \multicolumn{1}{c}{Obs. Date\tablenotemark{c}}  &  \multicolumn{1}{c}{A0V Star\tablenotemark{c}}  &  \multicolumn{1}{c}{}  &  \multicolumn{1}{c}{Discovery}  &  \multicolumn{1}{c}{Type}  &  \multicolumn{1}{c}{SpT}  &  \multicolumn{1}{c}{Spectra}  \\ 
\multicolumn{1}{l}{}  &  \multicolumn{1}{c}{}  &  \multicolumn{1}{c}{}  &  \multicolumn{1}{c}{}  &  \multicolumn{1}{c}{($''$)}  &  \multicolumn{1}{c}{}  &  \multicolumn{1}{c}{(UT)}  &  \multicolumn{1}{c}{}  &  \multicolumn{1}{c}{}  &  \multicolumn{1}{c}{}  &  \multicolumn{1}{c}{}  &  \multicolumn{1}{c}{}  &  \multicolumn{1}{c}{}   } 
\startdata 
HD~$3651$B  &  C  &  T7.5  &    &  0.5  &  73  &  --  &  --  &  & 10 & 10 & 12 & 11   \\ 
WISE~J$004024.88+090054.8$  &    &  T7  &    &  0.5  &  80  &  --  &  --  &  & 34 & ... & 34 & 38   \\ 
WISE~J$004945.61+215120.0$  &    &  T8.5  &    &  0.5  &  78  &  --  &  --  &  & 34 & ... & 34 & 38   \\ 
  &    &    &    &  0.8  &  35  &  2012-01-20  &  HD~$9711$  &  & 34 & ... & 34 & 40   \\ 
2MASS~J$00501994-3322402$  &    &  T7  &    &  0.5  &  61  &  --  &  --  &  & 5 & ... & 6 & 8   \\ 
WISEPA~J$012333.21+414203.9$  &    &  T7  &    &  0.8  &  60  &  2017-10-22  &  HD~$1561$  &  & 25 & ... & 25 & 40   \\ 
WISEPC~J$022322.39-293258.1$  &    &  T7.5  &    &  0.5  &  50  &  --  &  --  &  & 25 & ... & 25 & 38   \\ 
WISE~J$024124.73-365328.0$  &    &  T7  &    &  0.8  &  57  &  2017-11-15  &  HD~$18546$  &  & 29 & ... & 39 & 40   \\
\enddata 
\tablenotetext{a}{``B'': Resolved binaries; ``B?'': Binarity is likely but not confirmed (Section~\ref{subsec:candidate}); ``C'': Companions to stars or brown dwarfs. Remaining objects are single field dwarfs.}  
\tablenotetext{b}{Signal-to-noise ratio around the $J$-band peak ($1.2-1.3$~$\mu$m) of the object's spectra. }  
\tablenotetext{c}{For spectra taken from our own observations, we show the observation date and the A0V telluric standard star.}  
\tablerefs{(1) \cite{1999ApJ...522L..65B}, (2) \cite{2000ApJ...531L..57B}, (3) \cite{2002ApJ...564..421B}, (4) \cite{2004AJ....127.2856B}, (5) \cite{2005AJ....130.2326T}, (6) \cite{2006ApJ...637.1067B}, (7) \cite{2006ApJS..166..585B}, (8) \cite{2006ApJ...639.1095B}, (9) \cite{2006AJ....131.2722C}, (10) \cite{2006MNRAS.373L..31M}, (11) \cite{2007ApJ...658..617B}, (12) \cite{2007ApJ...654..570L}, (13) \cite{2007AJ....134.1162L}, (14) \cite{2010AJ....139.2448B}, (15) \cite{2010MNRAS.406.1885B}, (16) \cite{2010ApJ...710.1142B}, (17) \cite{2010MNRAS.404.1952B}, (18) \cite{2010MNRAS.405.1140G}, (19) \cite{2010MNRAS.408L..56L}, (20) \cite{2010AandA...515A..92S}, (21) \cite{2010AandA...510L...8S}, (22) \cite{2011MNRAS.414L..90B}, (23) \cite{2011ApJ...743...50C}, (24) \cite{2011AJ....142...57G}, (25) \cite{2011ApJS..197...19K}, (26) \cite{2011ApJ...740L..32L}, (27) \cite{2011ApJ...726...30M}, (28) \cite{2011AandA...532L...5S}, (29) \cite{2012ApJ...753..156K}, (30) \cite{2012ApJ...760..152L}, (31) \cite{2012ApJ...758...57L}, (32) \cite{2013MNRAS.433..457B}, (33) \cite{2013AandA...557A..43B}, (34) \cite{2013ApJS..205....6M}, (35) \cite{2013PASP..125..809T}, (36) \cite{2014AJ....147..113C}, (37) \cite{2015ApJ...814..118B}, (38) \cite{2017ASInC..14....7B}, (39) \cite{2018ApJS..236...28T}, (40) This work} 
\end{deluxetable} 
\end{longrotatetable}

\global\pdfpageattr\expandafter{\the\pdfpageattr/Rotate 90} 
\begin{longrotatetable} 
\begin{deluxetable}{lcccccccc} 
\tablewidth{0pc} 
\setlength{\tabcolsep}{0.05in} 
\tablecaption{Sample of T7$-$T9 Dwarfs: Astrometry \label{tab:astrometry}} 
\tablehead{ \multicolumn{1}{l}{Object}  &  \multicolumn{1}{c}{Type\tablenotemark{a}}  &  \multicolumn{1}{c}{SpT}  &  \multicolumn{1}{c}{R.A.\tablenotemark{b}}  &  \multicolumn{1}{c}{Decl.\tablenotemark{b}}  &  \multicolumn{1}{c}{$\mu_{\alpha}\cos{\delta}$}  &  \multicolumn{1}{c}{$\mu_{\delta}$}  &  \multicolumn{1}{c}{Parallax}  &  \multicolumn{1}{c}{References}   \\ 
\multicolumn{1}{l}{}  &  \multicolumn{1}{c}{}  &  \multicolumn{1}{c}{}  &  \multicolumn{1}{c}{(hh:mm:ss.ss)}  &  \multicolumn{1}{c}{(dd:mm:ss.s)}  &  \multicolumn{1}{c}{(mas yr$^{-1}$)}  &  \multicolumn{1}{c}{(mas yr$^{-1}$)}  &  \multicolumn{1}{c}{(mas)}  &  \multicolumn{1}{c}{}   } 
\startdata 
HD~$3651$B  &  C  &  T7.5  &  00:39:18.91  &  $\mathbin{\phantom{-}}$21:15:16.8  &  $-462.06 \pm 0.11$  &  $-369.81 \pm 0.06$  &  $89.79 \pm 0.06$  &  6,7 \\ 
WISE~J$004024.88+090054.8$  &    &  T7  &  00:40:24.89  &  $\mathbin{\phantom{-}}$09:00:54.5  &  $-53.90 \pm 2.80$  &  $-62.60 \pm 2.30$  &  $71.40 \pm 2.70$  &  9 \\ 
WISE~J$004945.61+215120.0$  &    &  T8.5  &  00:49:45.65  &  $\mathbin{\phantom{-}}$21:51:19.7  &  $-483.20 \pm 1.40$  &  $-46.50 \pm 1.40$  &  $139.90 \pm 2.50$  &  8 \\ 
2MASS~J$00501994-3322402$  &    &  T7  &  00:50:21.05  &  $-$33:22:28.9  &  $1150.50 \pm 2.20$  &  $939.10 \pm 2.10$  &  $94.60 \pm 2.40$  &  3 \\ 
WISEPA~J$012333.21+414203.9$  &    &  T7  &  01:23:33.25  &  $\mathbin{\phantom{-}}$41:42:03.9  &  $601.10 \pm 1.30$  &  $88.80 \pm 1.30$  &  $39.40 \pm 2.40$  &  9 \\ 
WISEPC~J$022322.39-293258.1$  &    &  T7.5  &  02:23:22.38  &  $-$29:32:57.3  &  $780.20 \pm 1.10$  &  $-535.10 \pm 1.50$  &  $80.70 \pm 2.60$  &  8 \\ 
WISE~J$024124.73-365328.0$  &    &  T7  &  02:41:24.75  &  $-$36:53:28.1  &  $242.00 \pm 1.50$  &  $148.40 \pm 1.40$  &  $52.40 \pm 2.70$  &  8 \\ 
WISE~J$024512.62-345047.8$  &    &  T8  &  02:45:12.62  &  $-$34:50:47.8  &  --  &  --  &  --  &  ... \\
\enddata 
\tablenotetext{a}{``B'': Resolved binaries; ``B?'': Binarity is likely but not confirmed (Section~\ref{subsec:candidate}); ``C'': Companions to stars or brown dwarfs. Remaining objects are single field dwarfs.}  
\tablenotetext{b}{Coordinates are provided at epoch J2000 with equinox J2000.} 
\tablerefs{(1) \cite{2003AJ....126..975T}, (2) \cite{2008ApJ...689L..53B}, (3) \cite{2012ApJS..201...19D}, (4) \cite{2012ApJ...752...56F}, (5) \cite{2012ApJ...748...74L}, (6) \cite{2016AandA...595A...1G}, (7) \cite{2018AandA...616A...1G}, (8) \cite{2019ApJS..240...19K}, (9) \cite{2020AJ....159..257B}} 
\end{deluxetable} 
\end{longrotatetable}

\global\pdfpageattr\expandafter{\the\pdfpageattr/Rotate 90} 
\begin{longrotatetable} 
\begin{deluxetable}{lcccccccccccccccc} 
\tablewidth{0pc} 
\setlength{\tabcolsep}{0.05in} 
\tablecaption{Sample of T7$-$T9 Dwarfs: Photometry \label{tab:photometry}} 
\tablehead{ \multicolumn{1}{l}{}  &  \multicolumn{1}{c}{}  &  \multicolumn{1}{c}{}  &  \multicolumn{1}{c}{}  &  \multicolumn{5}{c}{Near-Infrared MKO Photometry}  &  \multicolumn{1}{c}{}  &  \multicolumn{3}{c}{AllWISE Photometry}  &  \multicolumn{1}{c}{}  &  \multicolumn{3}{c}{{\it Spitzer}/IRAC Photometry}   \\ 
\cline{5-9} \cline{11-13} \cline{15-17} 
\multicolumn{1}{l}{Object}  &  \multicolumn{1}{c}{Type\tablenotemark{a}}  &  \multicolumn{1}{c}{SpT}  &  \multicolumn{1}{c}{}  &  \multicolumn{1}{c}{$Y_{\rm MKO}$}  &  \multicolumn{1}{c}{$J_{\rm MKO}$}  &  \multicolumn{1}{c}{$H_{\rm MKO}$}  &  \multicolumn{1}{c}{$K_{\rm MKO}$}  &  \multicolumn{1}{c}{References}  &  \multicolumn{1}{c}{}  &  \multicolumn{1}{c}{$W1$}  &  \multicolumn{1}{c}{$W2$}  &  \multicolumn{1}{c}{References}  &  \multicolumn{1}{c}{}  &  \multicolumn{1}{c}{$[3.6]$}  &  \multicolumn{1}{c}{$[4.5]$}  &  \multicolumn{1}{c}{References}  \\ 
\multicolumn{1}{l}{}  &  \multicolumn{1}{c}{}  &  \multicolumn{1}{c}{}  &  \multicolumn{1}{c}{}  &  \multicolumn{1}{c}{(mag)}  &  \multicolumn{1}{c}{(mag)}  &  \multicolumn{1}{c}{(mag)}  &  \multicolumn{1}{c}{(mag)}  &  \multicolumn{1}{c}{}  &  \multicolumn{1}{c}{}  &  \multicolumn{1}{c}{(mag)}  &  \multicolumn{1}{c}{(mag)}  &  \multicolumn{1}{c}{}  &  \multicolumn{1}{c}{}  &  \multicolumn{1}{c}{(mag)}  &  \multicolumn{1}{c}{(mag)}  &  \multicolumn{1}{c}{}   } 
\startdata 
HD~$3651$B  &  C  &  T7.5  &  &  $17.12 \pm 0.06$  &  $16.16 \pm 0.03$  &  $16.68 \pm 0.04$  &  $16.87 \pm 0.05$  &  4,26  &  &  --  &  --  &  ...  &  &  $15.38 \pm 0.04$  &  $13.62 \pm 0.02$  &  8  \\ 
WISE~J$004024.88+090054.8$  &    &  T7  &  &  $17.15 \pm 0.02$  &  $16.13 \pm 0.01$  &  $16.56 \pm 0.02$  &  $16.55 \pm 0.05$  &  12  &  &  $15.99 \pm 0.06$  &  $13.83 \pm 0.05$  &  19  &  &  $15.07 \pm 0.02$  &  $13.76 \pm 0.02$  &  17  \\ 
WISE~J$004945.61+215120.0$  &    &  T8.5  &  &  $17.29 \pm 0.17$  &  $16.44 \pm 0.16$  &  $16.72 \pm 0.02$  &  $16.80 \pm 0.17$  &  24,26  &  &  $15.92 \pm 0.06$  &  $13.03 \pm 0.03$  &  19  &  &  $15.01 \pm 0.02$  &  $13.04 \pm 0.02$  &  17  \\ 
2MASS~J$00501994-3322402$  &    &  T7  &  &  $16.85 \pm 0.10$  &  $15.65 \pm 0.10$  &  $16.04 \pm 0.10$  &  $15.91 \pm 0.10$  &  8  &  &  $15.60 \pm 0.04$  &  $13.58 \pm 0.03$  &  19  &  &  $14.87 \pm 0.02$  &  $13.59 \pm 0.02$  &  24  \\ 
WISEPA~J$012333.21+414203.9$  &    &  T7  &  &  $18.10 \pm 0.05$  &  $16.99 \pm 0.02$  &  $17.25 \pm 0.05$  &  $17.25 \pm 0.06$  &  25,26  &  &  $16.84 \pm 0.08$  &  $14.95 \pm 0.07$  &  19  &  &  $16.12 \pm 0.03$  &  $14.84 \pm 0.02$  &  24  \\ 
WISEPC~J$022322.39-293258.1$  &    &  T7.5  &  &  $18.08 \pm 0.07$  &  $17.10 \pm 0.05$  &  $17.30 \pm 0.11$  &  $17.59 \pm 0.08$  &  10,26  &  &  $16.87 \pm 0.08$  &  $14.02 \pm 0.04$  &  19  &  &  $15.81 \pm 0.03$  &  $14.01 \pm 0.02$  &  10  \\ 
WISE~J$024124.73-365328.0$\tablenotemark{b}  &    &  T7  &  &  $17.73 \pm 0.05$  &  $16.72 \pm 0.01$  &  $17.04 \pm 0.07$  &  $17.27 \pm 0.05$  &  16,24,27  &  &  $16.86 \pm 0.08$  &  $14.35 \pm 0.04$  &  19  &  &  $15.74 \pm 0.03$  &  $14.35 \pm 0.02$  &  24  \\ 
WISE~J$024512.62-345047.8$  &    &  T8  &  &  $18.65 \pm 0.11$  &  $17.77 \pm 0.09$  &  $18.27 \pm 0.10$  &  $18.05 \pm 0.11$  &  17,26  &  &  $17.03 \pm 0.10$  &  $14.50 \pm 0.05$  &  19  &  &  $16.05 \pm 0.03$  &  $14.59 \pm 0.02$  &  17  \\
\enddata 
\tablenotetext{a}{``B'': Resolved binaries; ``B?'': Binarity is likely but not confirmed (Section~\ref{subsec:candidate}); ``C'': Companions to stars or brown dwarfs. Remaining objects are single field dwarfs.}  
\tablenotetext{b}{We calculate synthetic photometry for WISE~J$024124.73-365328.0$ ($Y_{\rm MKO}$ and $K_{\rm MKO}$) and UGPS~J$052127.27+364048.6$ ($Y_{\rm MKO}$) using their SpeX prism spectra and the approach described in \cite{2021AJ....161...42B}.}  
\tablerefs{(1) \cite{2002ApJ...564..452L}, (2) \cite{2004AJ....127.3553K}, (3) \cite{2006AJ....131.2722C}, (4) \cite{2007ApJ...654..570L}, (5) \cite{2009ApJ...695.1517L}, (6) \cite{2010MNRAS.406.1885B}, (7) \cite{2010MNRAS.408L..56L}, (8) \cite{2010ApJ...710.1627L}, (9) \cite{2011AJ....142...57G}, (10) \cite{2011ApJS..197...19K}, (11) \cite{2012ApJS..201...19D}, (12) \cite{2012yCat.2314....0L}, (13) \cite{2012yCat.2316....0U}, (14) \cite{2013MNRAS.433..457B}, (15) \cite{2013Sci...341.1492D}, (16) \cite{2013Msngr.154...35M}, (17) \cite{2013ApJS..205....6M}, (18) \cite{2013PASP..125..809T}, (19) \cite{2014yCat.2328....0C}, (20) \cite{2014AJ....147..113C}, (21) \cite{2015ApJ...814..118B}, (22) \cite{2016yCat.2343....0E}, (23) \cite{2017ApJ...842..118L}, (24) \cite{2019ApJS..240...19K}, (25) \cite{2020AJ....159..257B}, (26) \cite{2021AJ....161...42B}, (27) This work} 
\end{deluxetable} 
\end{longrotatetable}

\begin{longrotatetable} 
\begin{deluxetable}{lllllllllllclcl} 
\tablewidth{0pc} 
\setlength{\tabcolsep}{0.05in} 
\tablecaption{Starfish-Based Forward-Modeling Analysis: Fitted parameters of T7--T9 Dwarfs \label{tab:starfish_fitting_results}} 
\tablehead{ \multicolumn{1}{l}{}  &  \multicolumn{1}{l}{}  &  \multicolumn{1}{l}{}  &  \multicolumn{1}{l}{}  &  \multicolumn{1}{l}{}  &  \multicolumn{1}{l}{}  &  \multicolumn{1}{l}{}  &  \multicolumn{1}{l}{}  &  \multicolumn{1}{l}{}  &  \multicolumn{1}{l}{}  &  \multicolumn{1}{l}{}  &  \multicolumn{1}{c}{}  &  \multicolumn{3}{c}{additive $\log{\Omega}$ systematics\tablenotemark{a}}   \\ 
\cline{13-15} 
\multicolumn{1}{l}{Object}  &  \multicolumn{1}{l}{Slit}  &  \multicolumn{1}{l}{$T_{\rm eff}$}  &  \multicolumn{1}{l}{$\log{g}$}  &  \multicolumn{1}{l}{$Z$}  &  \multicolumn{1}{l}{$v_{r}$}  &  \multicolumn{1}{l}{$v\sin{i}$}  &  \multicolumn{1}{l}{$\log{\Omega}$}  &  \multicolumn{1}{l}{$a_{N}$}  &  \multicolumn{1}{l}{$\log{a_{G}}$}  &  \multicolumn{1}{l}{$\ell$}  &  \multicolumn{1}{c}{}  &  \multicolumn{1}{l}{$J$ band}  &  \multicolumn{1}{c}{}  &  \multicolumn{1}{l}{$K$ band}   \\ 
\multicolumn{1}{l}{}  &  \multicolumn{1}{l}{($''$)}  &  \multicolumn{1}{l}{(K)}  &  \multicolumn{1}{l}{(dex)}  &  \multicolumn{1}{l}{(dex)}  &  \multicolumn{1}{l}{(km~s$^{-1}$)}  &  \multicolumn{1}{l}{(km~s$^{-1}$)}  &  \multicolumn{1}{l}{(dex)}  &  \multicolumn{1}{l}{}  &  \multicolumn{1}{l}{(dex)}  &  \multicolumn{1}{l}{(km~s$^{-1}$)}  &  \multicolumn{1}{c}{}  &  \multicolumn{1}{l}{(dex)}  &  \multicolumn{1}{c}{}  &  \multicolumn{1}{l}{(dex)}   } 
\startdata 
${\rm HD\ 3651B}$  &  $0.5''$  &  $818^{+28\ (19)}_{-28\ (19)}$  &  $3.94^{+0.29\ (0.20)}_{-0.28\ (0.20)}$  &  $-0.22^{+0.16\ (+0.11)}_{-0.16\ (-0.10)}$  &  $\mathbin{\phantom{-}}318^{+202\ (95)}_{-202\ (96)}$  &  $27^{+27\ (20)}_{-27\ (18)}$  &  $-19.543^{+0.072\ (0.051)}_{-0.070\ (0.046)}$  &  $1.06^{+0.04}_{-0.04}$  &  $-34.64^{+0.11}_{-0.11}$  &  $8114^{+751}_{-1015}$  &  &  $+0.049$  &  &  $-0.029$ \\ 
${\rm WISE\ 0040+0900}$  &  $0.5''$  &  $883^{+30\ (23)}_{-30\ (22)}$  &  $4.07^{+0.30\ (0.22)}_{-0.29\ (0.22)}$  &  $-0.34^{+0.15\ (+0.09)}_{-0.15\ (-0.09)}$  &  $\mathbin{\phantom{-}}398^{+199\ (88)}_{-199\ (85)}$  &  $33^{+31\ (26)}_{-33\ (22)}$  &  $-19.676^{+0.072\ (0.052)}_{-0.070\ (0.047)}$  &  $1.03^{+0.04}_{-0.04}$  &  $-34.68^{+0.11}_{-0.10}$  &  $8153^{+725}_{-1020}$  &  &  $+0.038$  &  &  $-0.015$ \\ 
${\rm WISE\ 0049+2151}$  &  $0.5''$  &  $753^{+26\ (16)}_{-26\ (16)}$  &  $3.47^{+0.26\ (0.19)}_{-0.22\ (0.14)}$  &  $-0.24^{+0.18\ (+0.14)}_{-0.17\ (-0.12)}$  &  $-180^{+209\ (107)}_{-209\ (110)}$  &  $15^{+15\ (10)}_{-15\ (10)}$  &  $-19.348^{+0.069\ (0.045)}_{-0.069\ (0.048)}$  &  $1.04^{+0.04}_{-0.04}$  &  $-34.50^{+0.12}_{-0.11}$  &  $6848^{+1065}_{-907}$  &  &  $-0.067$  &  &  $-0.016$ \\ 
  &  $0.8''$  &  $756^{+27\ (19)}_{-28\ (19)}$  &  $3.58^{+0.28\ (0.21)}_{-0.27\ (0.19)}$  &  $-0.23^{+0.18\ (+0.14)}_{-0.17\ (-0.12)}$  &  $-91^{+243\ (166)}_{-245\ (168)}$  &  $16^{+17\ (12)}_{-16\ (11)}$  &  $-19.390^{+0.073\ (0.051)}_{-0.072\ (0.052)}$  &  $1.05^{+0.04}_{-0.04}$  &  $-34.66^{+0.08}_{-0.07}$  &  $5454^{+90}_{-190}$  &  &  $-0.035$  &  &  $-0.055$ \\
\enddata 
\tablecomments{We report the median and $1\sigma$ uncertainties for each fitted parameter with systematic errors incorporated (Section~\ref{subsec:method}). The formal $1\sigma$ errors of parameters directly obtained from the spectral fitting process are shown in parentheses.} 
\tablenotetext{a}{The small shifts to be added to our fitted $\log{\Omega}$ if the objects' observed spectra are flux-calibrated by their $J_{\rm MKO}$ or $K_{\rm MKO}$ photometry rather than $H_{\rm MKO}$ used in this work (see Section~\ref{subsec:method}).}  
\end{deluxetable} 
\end{longrotatetable} 

\begin{longrotatetable} 
\begin{deluxetable}{llllllllllll} 
\tablewidth{0pc} 
\setlength{\tabcolsep}{0.05in} 
\tablecaption{Starfish-Based Forward-Modeling Analysis: Derived properties of T7--T9 Dwarfs \label{tab:starfish_derived_results}} 
\tablehead{   \multicolumn{1}{l}{}  &  \multicolumn{1}{l}{}  &  \multicolumn{1}{l}{}  &  \multicolumn{1}{l}{}  &  \multicolumn{1}{l}{}  &  \multicolumn{1}{l}{}  &  \multicolumn{1}{l}{}  &  \multicolumn{1}{l}{}  &  \multicolumn{4}{c}{$q_{(\lambda\lambda)}$ quantities}   \\ 
\cline{9-12} 
\multicolumn{1}{l}{Object}  &  \multicolumn{1}{l}{Slit}  &  \multicolumn{1}{l}{$R$}  &  \multicolumn{1}{l}{$M$}  &  \multicolumn{1}{l}{$\log{(L_{\rm bol}/L_{\odot})}$}  &  \multicolumn{1}{l}{Age}  &  \multicolumn{1}{l}{$\epsilon_{J}$}  &  \multicolumn{1}{l}{}  &  \multicolumn{1}{l}{$q_{Y}$}  &  \multicolumn{1}{l}{$q_{J}$}  &  \multicolumn{1}{l}{$q_{H}$}  &  \multicolumn{1}{l}{$q_{K}$}   \\ 
\multicolumn{1}{l}{}  &  \multicolumn{1}{l}{($''$)}  &  \multicolumn{1}{l}{($R_{\rm Jup}$)}  &  \multicolumn{1}{l}{($M_{\rm Jup}$)}  &  \multicolumn{1}{l}{(dex)}  &  \multicolumn{1}{l}{(Myr)}  &  \multicolumn{1}{l}{($\%$)}  &  \multicolumn{1}{l}{}  &  \multicolumn{1}{l}{($\%$)}  &  \multicolumn{1}{l}{($\%$)}  &  \multicolumn{1}{l}{($\%$)}  &  \multicolumn{1}{l}{($\%$)}   } 
\startdata 
${\rm HD\ 3651B}$  &  $0.5''$  & $0.81^{+0.07}_{-0.06}$  &  $2.33^{+2.27}_{-1.14}$  &  $-5.569^{+0.026}_{-0.029}$  &  $38^{+76}_{-25}$  &  $2.6^{+0.4}_{-0.3}$  &  &  $3 \pm 4$  &  $-9 \pm 4$  &  $17 \pm 8$  &  $11 \pm 24$  \\ 
${\rm WISE\ 0040+0900}$  &  $0.5''$  & $0.88^{+0.08}_{-0.08}$  &  $3.67^{+3.95}_{-1.86}$  &  $-5.347^{+0.045}_{-0.043}$  &  $50^{+195}_{-34}$  &  $2.6^{+0.3}_{-0.3}$  &  &  $6 \pm 4$  &  $-8 \pm 3$  &  $11 \pm 8$  &  $27 \pm 18$  \\ 
${\rm WISE\ 0049+2151}$  &  $0.5''$  & $0.65^{+0.06}_{-0.05}$  &  $0.51^{+0.42}_{-0.19}$  &  $-5.895^{+0.039}_{-0.039}$  &  $8^{+13}_{-4}$  &  $2.8^{+0.4}_{-0.3}$  &  &  $3 \pm 4$  &  $-7 \pm 4$  &  $23 \pm 9$  &  $15 \pm 26$  \\ 
  &  $0.8''$  & $0.62^{+0.06}_{-0.05}$  &  $0.59^{+0.57}_{-0.27}$  &  $-5.918^{+0.041}_{-0.039}$  &  $12^{+23}_{-7}$  &  $2.6^{+0.3}_{-0.2}$  &  &  $6 \pm 4$  &  $-8 \pm 3$  &  $27 \pm 7$  &  $24 \pm 19$  \\ 
${\rm 2MASS\ 0050-3322}$  &  $0.5''$  & $0.73^{+0.06}_{-0.06}$  &  $2.45^{+2.04}_{-1.13}$  &  $-5.397^{+0.039}_{-0.038}$  &  $40^{+105}_{-25}$  &  $1.9^{+0.2}_{-0.1}$  &  &  $-1 \pm 2$  &  $-2 \pm 1$  &  $10 \pm 3$  &  $6 \pm 7$  \\ 
${\rm WISE\ 0123+4142}$  &  $0.8''$  & $0.79^{+0.08}_{-0.07}$  &  $3.32^{+3.67}_{-1.72}$  &  $-5.191^{+0.060}_{-0.058}$  &  $41^{+168}_{-28}$  &  $2.3^{+0.2}_{-0.2}$  &  &  $3 \pm 3$  &  $-4 \pm 2$  &  $11 \pm 4$  &  $9 \pm 12$  \\ 
${\rm WISE\ 0223-2932}$  &  $0.5''$  & $0.72^{+0.09}_{-0.07}$  &  $2.95^{+3.43}_{-1.55}$  &  $-5.718^{+0.054}_{-0.051}$  &  $91^{+318}_{-63}$  &  $2.5^{+0.3}_{-0.3}$  &  &  $1 \pm 4$  &  $-7 \pm 3$  &  $21 \pm 8$  &  $19 \pm 28$  \\ 
${\rm WISE\ 0241-3653}$  &  $0.8''$  & $0.90^{+0.09}_{-0.08}$  &  $2.75^{+2.44}_{-1.29}$  &  $-5.287^{+0.057}_{-0.054}$  &  $26^{+47}_{-17}$  &  $2.1^{+0.2}_{-0.2}$  &  &  $4 \pm 2$  &  $-5 \pm 2$  &  $13 \pm 4$  &  $3 \pm 13$  \\
\enddata 
\tablecomments{Radii ($R$) is derived from the objects' parallaxes and spectroscopically inferred $\log{\Omega}$. Mass ($M$) is derived from $R$ and fitted $\log{g}$. Bolometric luminosity ($\log{(L_{\rm bol}/L_{\odot})}$ is derived from integrating the objects' $1.0-2.5$~$\mu$m SpeX spectra and fitted model spectra to shorter and longer wavelengths spanning $0.4-50$~$\mu$m, with spectra scaled by measured parallaxes (Section~\ref{subsec:results}). Age is derived from the spectroscopically inferred $T_{\rm eff}$, $\log{g}$, $Z$, and the cloudless Sonora-Bobcat evolutionary models (Section~\ref{subsec:age}). The measure of the data-model discrepancy ($\epsilon_{J}$) is derived from the fitted hyper-parameters $a_{N}$ and $a_{G}$, as well as the objects' $J$-band peak fluxes (Section~\ref{sec:comments_sonora}). The measure of spectral-fitting residuals in $YJHK$ bands, $q(\lambda\lambda)$, is defined in Equation~\ref{eq:Rlambda} and computed using the objects' observed spectra and their fitted model spectra (Section~\ref{subsec:stacked_residual}). } 
\tablenotetext{a}{WISE~$0245-3450$ lacks a parallax and thus has no radii, mass, or bolometric luminosity results.  }  
\end{deluxetable} 
\end{longrotatetable}

\begin{longrotatetable} 
\begin{deluxetable}{lcccccccc} 
\tablewidth{0pt}
\setlength{\tabcolsep}{0.15in} 
\tablecaption{Typical Parameter Uncertainties for Our Forward-Modeling Analysis \label{tab:typical_error}} 
\tablehead{ \multicolumn{1}{l}{Method}  &  \multicolumn{1}{c}{$T_{\rm eff}$}  &  \multicolumn{1}{c}{$\log\ g$}  &  \multicolumn{1}{c}{$Z$}  &  \multicolumn{1}{c}{$v_{r}$}  &  \multicolumn{1}{c}{$v\ \sin\ i$}  &  \multicolumn{1}{c}{$\log\ \Omega$}  &  \multicolumn{1}{c}{$R$}  &  \multicolumn{1}{c}{$M$}   \\ 
\multicolumn{1}{l}{}  &  \multicolumn{1}{c}{(K)}  &  \multicolumn{1}{c}{(dex)}  &  \multicolumn{1}{c}{(dex)}  &  \multicolumn{1}{c}{(km~s$^{-1}$)}  &  \multicolumn{1}{c}{(km~s$^{-1}$)}  &  \multicolumn{1}{c}{(dex)}  &  \multicolumn{1}{c}{(R$_{\rm Jup}$)}  &  \multicolumn{1}{c}{(M$_{\rm Jup}$)}   } 
\startdata 
{\it Starfish} & & & & & & & & \\ 
formal & $20$ & $0.22$ & $0.102$ & $127$ & $21$ & $0.049$ & $0.06$ & $1.3$   \\ 
adopted & $28$ & $0.29$ & $0.154$ & $220$ & $29$ & $0.073$ & $0.08$ & $1.9$   \\ 
\cline{1-9} 
{\it Traditional} & & & & & & & & \\ 
formal & $3$ & $0.02$ & $0.011$ & $29$ & $20$ & $0.006$ & $0.03$ & $0.2$   \\ 
adopted & $3$ & $0.02$ & $0.011$ & $183$ & $20$ & $0.024$ & $0.04$ & $0.3$   \\ 
\enddata 
\tablecomments{Formal uncertainties are obtained directly from the spectral fitting process and the adopted uncertainties are those with systematic errors incorporated (Section~\ref{subsec:method} and Appendix~\ref{app:traditional}). We only show the formal uncertainties if they are different from the adopted ones. } 
\end{deluxetable} 
\end{longrotatetable}

\global\pdfpageattr\expandafter{\the\pdfpageattr/Rotate 90} 
\begin{longrotatetable} 
\begin{deluxetable}{lccccccccccc} 
\tablewidth{0pc} 
\tablecaption{Bolometric Luminosities from This Work, \cite{2015ApJ...810..158F}, and \cite{2017ApJ...848...83L} \label{tab:compare_lbol}} 
\tablehead{ \multicolumn{1}{l}{}  &  \multicolumn{1}{c}{}  &  \multicolumn{2}{c}{This Work}  &  \multicolumn{1}{c}{}  &  \multicolumn{7}{c}{Literature Values\tablenotemark{a}}   \\ 
\cline{3-4}  \cline{6-12}
\multicolumn{1}{l}{}  &  \multicolumn{1}{c}{}  &  \multicolumn{1}{c}{}  &  \multicolumn{1}{c}{}  &  \multicolumn{1}{c}{}  &  \multicolumn{1}{c}{}  &  \multicolumn{1}{c}{}  &  \multicolumn{2}{c}{\cite{2015ApJ...810..158F} $\log{(L_{\rm bol}/L_{\odot})}$}  &  \multicolumn{1}{c}{}  &  \multicolumn{2}{c}{\cite{2017ApJ...848...83L} $\log{(L_{\rm bol}/L_{\odot})}$}  \\ 
\cline{8-9} \cline{11-12} 
\multicolumn{1}{l}{Object}  &  \multicolumn{1}{c}{SpT}  &  \multicolumn{1}{c}{Parallax}  &  \multicolumn{1}{c}{$\log{(L_{\rm bol}/L_{\odot})}$}  &  \multicolumn{1}{c}{}  &  \multicolumn{1}{c}{Parallax}  &  \multicolumn{1}{c}{Par. Ref.}  &  \multicolumn{1}{c}{original}  &  \multicolumn{1}{c}{modified}  &  \multicolumn{1}{c}{}  &  \multicolumn{1}{c}{original}  &  \multicolumn{1}{c}{modified}  \\ 
\multicolumn{1}{l}{}  &  \multicolumn{1}{c}{}  &  \multicolumn{1}{c}{(mas)}  &  \multicolumn{1}{c}{(dex)}  &  \multicolumn{1}{c}{}  &  \multicolumn{1}{c}{(mas)}  &  \multicolumn{1}{c}{}  &  \multicolumn{1}{c}{(dex)}  &  \multicolumn{1}{c}{(dex)}  &  \multicolumn{1}{c}{}  &  \multicolumn{1}{c}{(dex)}  &  \multicolumn{1}{c}{(dex)}  } 
\startdata 
HD~$3651$B  &  T7.5  &  $89.79 \pm 0.06$  &  $-5.569^{+0.026}_{-0.029}$  &  &  $90.42 \pm 0.32$  &  3  &  $-5.56 \pm 0.01$  &  $-5.55 \pm 0.01$  &  &  $-5.51 \pm 0.05$  &  $-5.50 \pm 0.05$  \\ 
2MASS~J$00501994-3322402$  &  T7  &  $94.60 \pm 2.40$  &  $-5.397^{+0.039}_{-0.038}$  &  &  $94.60 \pm 2.40$  &  5  &  $-5.39 \pm 0.02$  &  $-5.39 \pm 0.02$  &  &  $-5.27 \pm 0.06$  &  $-5.27 \pm 0.06$  \\ 
PSO~J$043.5395+02.3995$  &  T8  &  $146.10 \pm 1.50$  &  $-5.856^{+0.034}_{-0.034}$  &  &  $166.00 \pm 26.00$  &  7  &  $-5.88 \pm 0.14$  &  $-5.77 \pm 0.14$  &  &  --  &  --  \\ 
2MASSI~J$0415195-093506$  &  T8  &  $175.20 \pm 1.70$  &  $-5.779^{+0.036}_{-0.033}$  &  &  $175.20 \pm 1.70$  &  5  &  $-5.74 \pm 0.01$  &  $-5.74 \pm 0.01$  &  &  $-5.70 \pm 0.04$  &  $-5.70 \pm 0.04$  \\ 
UGPS~J$072227.51-054031.2$  &  T9  &  $242.80 \pm 2.40$  &  $-6.422^{+0.045}_{-0.042}$  &  &  $242.80 \pm 2.40$  &  8  &  $-6.02 \pm 0.02$  &  $-6.02 \pm 0.02$  &  &  --  &  --  \\ 
2MASSI~J$0727182+171001$  &  T7  &  $112.50 \pm 0.90$  &  $-5.381^{+0.031}_{-0.029}$  &  &  $112.50 \pm 0.90$  &  5  &  $-5.37 \pm 0.01$  &  $-5.37 \pm 0.01$  &  &  $-5.30 \pm 0.03$  &  $-5.30 \pm 0.03$  \\ 
2MASS~J$07290002-3954043$  &  T8 pec  &  $126.30 \pm 8.30$  &  $-5.602^{+0.075}_{-0.069}$  &  &  $126.30 \pm 8.30$  &  6  &  $-5.57 \pm 0.06$  &  $-5.57 \pm 0.06$  &  &  $-5.60 \pm 0.08$  &  $-5.60 \pm 0.08$  \\ 
2MASS~J$09393548-2448279$  &  T8  &  $187.30 \pm 4.60$  &  $-5.768^{+0.066}_{-0.063}$  &  &  $187.30 \pm 4.60$  &  4  &  $-5.72 \pm 0.02$  &  $-5.72 \pm 0.02$  &  &  $-5.71 \pm 0.07$  &  $-5.71 \pm 0.07$  \\
\enddata 
\tablenotetext{a}{We compute the modified bolometric luminosities by scaling the original literature $L_{\rm bol}$'s to the more precise parallaxes adopted in this work without modifying the uncertainties provided by the literature.}  
\tablerefs{(1) \cite{2003AJ....126..975T}, (2) \cite{2007AandA...474..653V}, (3) \cite{2007AandA...474..653V}, (4) \cite{2008ApJ...689L..53B}, (5) \cite{2012ApJS..201...19D}, (6) \cite{2012ApJ...752...56F}, (7) \cite{2012ApJ...753..156K}, (8) \cite{2012ApJ...748...74L}} 
\end{deluxetable} 
\end{longrotatetable}

\global\pdfpageattr\expandafter{\the\pdfpageattr/Rotate 90} 
\begin{longrotatetable} 
\begin{deluxetable}{lcccccccccccccccc} 
\tablewidth{0pc} 
\tablecaption{Bolometric Corrections for T7$-$T9 Dwarfs \label{tab:bc}} 
\tablehead{ \multicolumn{1}{l}{}  &  \multicolumn{1}{c}{}  &  \multicolumn{1}{c}{}  &  \multicolumn{4}{c}{BC$_{J_{\rm MKO}}$}   &  \multicolumn{1}{c}{}  &  \multicolumn{4}{c}{BC$_{H_{\rm MKO}}$}  &  \multicolumn{1}{c}{}  &  \multicolumn{4}{c}{BC$_{K_{\rm MKO}}$}  \\ 
\cline{4-7}  \cline{9-12}  \cline{14-17}
\multicolumn{1}{l}{Spectral Type}  &  \multicolumn{1}{c}{N$_{\rm obj}$}  &  \multicolumn{1}{c}{}  &  \multicolumn{1}{c}{mean}  &  \multicolumn{1}{c}{error}  &  \multicolumn{1}{c}{$\sigma_{\rm add}$}  &  \multicolumn{1}{c}{rms}  &  \multicolumn{1}{c}{}  &  \multicolumn{1}{c}{mean}  &  \multicolumn{1}{c}{error}  &  \multicolumn{1}{c}{$\sigma_{\rm add}$}  &  \multicolumn{1}{c}{rms}  &  \multicolumn{1}{c}{}  &  \multicolumn{1}{c}{mean}  &  \multicolumn{1}{c}{error}  &  \multicolumn{1}{c}{$\sigma_{\rm add}$}  &  \multicolumn{1}{c}{rms} \\
\multicolumn{1}{l}{}  &  \multicolumn{1}{c}{}  &  \multicolumn{1}{c}{}  &  \multicolumn{1}{c}{(mag)}  &  \multicolumn{1}{c}{(mag)}  &  \multicolumn{1}{c}{(mag)}  &  \multicolumn{1}{c}{(mag)}  &  \multicolumn{1}{c}{}  &  \multicolumn{1}{c}{(mag)}  &  \multicolumn{1}{c}{(mag)}  &  \multicolumn{1}{c}{(mag)}  &  \multicolumn{1}{c}{(mag)}  &  \multicolumn{1}{c}{}  &  \multicolumn{1}{c}{(mag)}  &  \multicolumn{1}{c}{(mag)}  &  \multicolumn{1}{c}{(mag)}  &  \multicolumn{1}{c}{(mag)} }
\startdata
$[$T7, T7.5)   &   18   &   &   2.67   &    0.02   &  0.09  &   0.10    &   &   2.32   &    0.02   &  0.00  &   0.07   &  &    2.35   &    0.02   &  0.17  &   0.17      \\
$[$T7.5, T8)   &   10   &   &   2.58   &    0.03   &  0.07  &   0.10    &   &   2.21   &    0.03   &  0.00  &   0.06   &  &    2.17   &    0.03   &  0.32  &   0.26      \\
$[$T8, T8.5)   &   14   &   &   2.57   &    0.03   &  0.15  &   0.16    &   &   2.20   &    0.03   &  0.00  &   0.08   &  &    2.21   &    0.03   &  0.29  &   0.29      \\
$[$T8.5, T9)   &   3     &   &   2.59   &    0.07   &  0.23  &   0.18    &   &   2.16   &    0.05   &  0.20  &   0.15   &  &    2.17   &    0.09   &  0.20  &   0.18      \\
$[$T9, T9.5)   &   2     &   &   2.29   &    0.07   &  0.02  &   0.05    &   &   2.04   &    0.08   &  0.11  &   0.07   &  &    1.69   &    0.11   &  0.37  &   0.17     \\
\enddata
\tablecomments{For each spectral type bin and photometric band, we tabulate the weighted average, weighted error, and weighted root mean square (rms) of the bolometric corrections, by excluding resolved or candidate binaries and two objects with peculiar spectra (Appendix~\ref{app:bc}). The $\sigma_{\rm add}$ column gives the additional uncertainty needed to make the reduced $\chi^{2} \approx$ 1 for the weighted average, as an estimate of the intrinsic (astrophysical) variations. When using these bolometric corrections, we recommend adopting the larger of the weighted error and $\sigma_{\rm add}$ for the uncertainty on the correction.}  
\end{deluxetable} 
\end{longrotatetable}

\global\pdfpageattr\expandafter{\the\pdfpageattr/Rotate 90} 
\begin{longrotatetable} 
\begin{deluxetable}{lccllllllclll} 
\tablewidth{0pc} 
\setlength{\tabcolsep}{0.1in} 
\tablecaption{Traditional Forward-Modeling Analysis: Fitted and Derived Properties of T7--T9 Dwarfs  \label{tab:traditional_results}} 
\tablehead{ \multicolumn{1}{l}{}  &  \multicolumn{1}{c}{}  &  \multicolumn{1}{c}{}  &  \multicolumn{6}{c}{Fitted Parameters\tablenotemark{a}}   &   \multicolumn{1}{c}{}  &  \multicolumn{3}{c}{Derived Parameters\tablenotemark{b}}   \\ 
\cline{4-9} \cline{11-13} 
\multicolumn{1}{l}{Object}  &  \multicolumn{1}{l}{Slit}  &  \multicolumn{1}{c}{}  &  \multicolumn{1}{l}{$T_{\rm eff}$}  &  \multicolumn{1}{l}{$\log{g}$}  &  \multicolumn{1}{l}{$Z$}  &  \multicolumn{1}{l}{$v_{r}$}  &  \multicolumn{1}{l}{$v\sin{i}$}  &  \multicolumn{1}{l}{$\log{\Omega}$}  &  \multicolumn{1}{c}{}  &  \multicolumn{1}{l}{$R$}  &  \multicolumn{1}{l}{$M$}  &  \multicolumn{1}{l}{Age}   \\ 
\multicolumn{1}{l}{}  &  \multicolumn{1}{l}{($''$)}  &  \multicolumn{1}{c}{}  &  \multicolumn{1}{l}{(K)}  &  \multicolumn{1}{l}{(dex)}  &  \multicolumn{1}{l}{(dex)}  &  \multicolumn{1}{l}{(km~s$^{-1}$)}  &  \multicolumn{1}{l}{(km~s$^{-1}$)}  &  \multicolumn{1}{l}{(dex)}  &  \multicolumn{1}{c}{}  &  \multicolumn{1}{l}{($R_{\rm Jup}$)}  &  \multicolumn{1}{l}{($M_{\rm Jup}$)}  &  \multicolumn{1}{l}{(Myr)}   } 
\startdata 
${\rm HD\ 3651B}$  &  $0.5''$  &  & $824^{+2}_{-3}$  &  $4.00^{+0.01}_{-0.12}$  &  $-0.14^{+0.02}_{-0.02}$  &  $\mathbin{\phantom{-}}252^{+188\ (23)}_{-197\ (29)}$  &  $28^{+17}_{-18}$  &  $-19.572^{+0.021\ (0.006)}_{-0.018\ (0.005)}$  &   &  $0.79^{+0.02}_{-0.02}$  &  $2.47^{+0.16}_{-0.54}$  &  $47^{+1}_{-17}$     \\ 
${\rm WISE\ 0040+0900}$  &  $0.5''$  &  & $926^{+1}_{-1}$  &  $3.87^{+0.00}_{-0.00}$  &  $-0.23^{+0.01}_{-0.01}$  &  $\mathbin{\phantom{-}}291^{+188\ (20)}_{-202\ (24)}$  &  $23^{+18}_{-14}$  &  $-19.762^{+0.010\ (0.002)}_{-0.009\ (0.001)}$  &   &  $0.80^{+0.04}_{-0.03}$  &  $1.90^{+0.20}_{-0.15}$  &  $20^{+0}_{-0}$     \\ 
${\rm WISE\ 0049+2151}$  &  $0.5''$  &  & $768^{+2}_{-2}$  &  $3.67^{+0.02}_{-0.02}$  &  $-0.13^{+0.01}_{-0.01}$  &  $-345^{+193\ (19)}_{-184\ (18)}$  &  $18^{+12}_{-12}$  &  $-19.401^{+0.010\ (0.005)}_{-0.009\ (0.004)}$  &   &  $0.62^{+0.01}_{-0.01}$  &  $0.71^{+0.04}_{-0.04}$  &  $16^{+1}_{-1}$     \\ 
  &  $0.8''$  &  & $760^{+2}_{-2}$  &  $3.75^{+0.04}_{-0.03}$  &  $-0.02^{+0.02}_{-0.02}$  &  $-157^{+181\ (29)}_{-181\ (29)}$  &  $19^{+13}_{-13}$  &  $-19.402^{+0.011\ (0.007)}_{-0.011\ (0.007)}$  &   &  $0.61^{+0.01}_{-0.01}$  &  $0.86^{+0.09}_{-0.07}$  &  $23^{+4}_{-2}$     \\ 
${\rm 2MASS\ 0050-3322}$  &  $0.5''$  &  & $950^{+1}_{-1}$  &  $4.23^{+0.01}_{-0.02}$  &  $-0.02^{+0.01}_{-0.01}$  &  $\mathbin{\phantom{-}}615^{+180\ (18)}_{-180\ (19)}$  &  $35^{+25}_{-24}$  &  $-19.600^{+0.040\ (0.002)}_{-0.040\ (0.003)}$  &   &  $0.72^{+0.04}_{-0.04}$  &  $3.56^{+0.42}_{-0.37}$  &  $79^{+5}_{-6}$     \\ 
${\rm WISE\ 0123+4142}$  &  $0.8''$  &  & $1025^{+1}_{-1}$  &  $3.93^{+0.01}_{-0.01}$  &  $-0.35^{+0.01}_{-0.01}$  &  $-97^{+182\ (27)}_{-181\ (25)}$  &  $28^{+17}_{-19}$  &  $-20.281^{+0.024\ (0.003)}_{-0.024\ (0.002)}$  &   &  $0.79^{+0.06}_{-0.05}$  &  $2.20^{+0.37}_{-0.28}$  &  $19^{+1}_{-0}$     \\
\enddata 
\tablenotetext{a}{We report the median and $1\sigma$ errors of each parameter with the systematic errors in $v_{r}$ and $\log{\Omega}$ incorporated (Section~\ref{subsec:method}). The formal $1\sigma$ erros of $v_{r}$ and $\log{\Omega}$ directly from the spectral fitting process are shown in parentheses. }  
\tablenotetext{b}{Parameters are derived in the same approach as those in Table~\ref{tab:starfish_derived_results}. WISE~$0245-3450$ lacks a parallax and thus has no radii or mass results. Ages $>12$~Gyr are shown as ``--''.}  
\end{deluxetable} 
\end{longrotatetable}

\vfill
\eject
\end{document}